\documentclass[referee]{aa} % for a referee version
\usepackage{graphicx}
\usepackage{txfonts}
\usepackage{pdflscape}
\usepackage{amsmath} % For multiple line equations
\usepackage{natbib,twoopt}
\usepackage[breaklinks=true]{hyperref} %% to avoid \citeads line fills
\bibpunct{(}{)}{;}{a}{}{,} %% natbib format for A&A and ApJ

\makeatletter
\newcommandtwoopt{\citeads}[3][][]{\href{http://adsabs.harvard.edu/abs/#3}%
{\def\hyper@linkstart##1##2{}%
\let\hyper@linkend\@empty\citealp[#1][#2]{#3}}}
\newcommandtwoopt{\citepads}[3][][]{\href{http://adsabs.harvard.edu/abs/#3}%
{\def\hyper@linkstart##1##2{}%
\let\hyper@linkend\@empty\citep[#1][#2]{#3}}}
\newcommandtwoopt{\citetads}[3][][]{\href{http://adsabs.harvard.edu/abs/#3}%
{\def\hyper@linkstart##1##2{}%
\let\hyper@linkend\@empty\citet[#1][#2]{#3}}}
\newcommandtwoopt{\citeyearads}[3][][]%
{\href{http://adsabs.harvard.edu/abs/#3}
{\def\hyper@linkstart##1##2{}%
\let\hyper@linkend\@empty\citeyear[#1][#2]{#3}}}
\makeatother

\let\oldhref\href
\renewcommand{\href}[2]{\oldhref{#1}{\hbox{#2}}}

\usepackage{color}
\definecolor{mygreen}{RGB}{0,128,0}

\hypersetup{colorlinks=true,linkcolor=blue,citecolor=blue,urlcolor=blue}

\newcommand{\muas}{$\mu$as}
\newcommand{\muasyr}{$\mu$as\,yr$^{-1}$}
\newcommand{\gaia} {\textit{Gaia}}

\newcommand{\qsos} {QSOs}
\newcommand\gdrone{\gaia~DR1}
\newcommand\gdrtwo{\gaia~DR2}
\newcommand{\gcrf}{\textit{Gaia}-CRF}
\newcommand{\gcrftwo}{\textit{Gaia}-CRF2}
\newcommand{\icrf}{ICRF3-prototype}

\newcommand\figref[1]{Fig.~\ref{#1}}

\newcommand\tabref[1]{Table~\ref{#1}}

\begin{document}

   \title{\textit{Gaia} Data Release 2 \\ The Celestial reference frame (\textit{Gaia}-CRF2)}

%   \subtitle{The Gaia CRF2 }

\author{
{\it Gaia} Collaboration
\and F.        ~Mignard                       \inst{\ref{inst:0001}}
\and S.A.      ~Klioner                       \inst{\ref{inst:0002}}
\and L.        ~Lindegren                     \inst{\ref{inst:0003}}
\and J.        ~Hern\'{a}ndez                 \inst{\ref{inst:0004}}
\and U.        ~Bastian                       \inst{\ref{inst:0005}}
\and A.        ~Bombrun                       \inst{\ref{inst:0006}}
\and D.        ~Hobbs                         \inst{\ref{inst:0003}}
\and U.        ~Lammers                       \inst{\ref{inst:0004}}
\and D.        ~Michalik                      \inst{\ref{inst:0003}}
\and M.        ~Ramos-Lerate                  \inst{\ref{inst:0010}}
\and M.        ~Biermann                      \inst{\ref{inst:0005}}
\and J.        ~Fern\'{a}ndez-Hern\'{a}ndez   \inst{\ref{inst:0012}}
\and R.        ~Geyer                         \inst{\ref{inst:0002}}
\and T.        ~Hilger                        \inst{\ref{inst:0002}}
\and H.I.      ~Siddiqui                      \inst{\ref{inst:0015}}
\and H.        ~Steidelm\"{ u}ller            \inst{\ref{inst:0002}}
\and C.        ~Babusiaux                     \inst{\ref{inst:0017},\ref{inst:0018}}
\and C.        ~Barache                       \inst{\ref{inst:0019}}
\and S.        ~Lambert                       \inst{\ref{inst:0019}}
\and A.H.      ~Andrei                        \inst{\ref{inst:0021},\ref{inst:0022},\ref{inst:0019}}
\and G.        ~Bourda                        \inst{\ref{inst:0024}}
\and P.        ~Charlot                       \inst{\ref{inst:0024}}
\and A.G.A.    ~Brown                         \inst{\ref{inst:0026}}
\and A.        ~Vallenari                     \inst{\ref{inst:0027}}
\and T.        ~Prusti                        \inst{\ref{inst:0028}}
\and J.H.J.    ~de Bruijne                    \inst{\ref{inst:0028}}
\and C.A.L.    ~Bailer-Jones                  \inst{\ref{inst:0030}}
\and D.W.      ~Evans                         \inst{\ref{inst:0031}}
\and L.        ~Eyer                          \inst{\ref{inst:0032}}
\and F.        ~Jansen                        \inst{\ref{inst:0033}}
\and C.        ~Jordi                         \inst{\ref{inst:0034}}
\and X.        ~Luri                          \inst{\ref{inst:0034}}
\and C.        ~Panem                         \inst{\ref{inst:0036}}
\and D.        ~Pourbaix                      \inst{\ref{inst:0037},\ref{inst:0038}}
\and S.        ~Randich                       \inst{\ref{inst:0039}}
\and P.        ~Sartoretti                    \inst{\ref{inst:0017}}
\and C.        ~Soubiran                      \inst{\ref{inst:0024}}
\and F.        ~van Leeuwen                   \inst{\ref{inst:0031}}
\and N.A.      ~Walton                        \inst{\ref{inst:0031}}
\and F.        ~Arenou                        \inst{\ref{inst:0017}}
\and M.        ~Cropper                       \inst{\ref{inst:0045}}
\and R.        ~Drimmel                       \inst{\ref{inst:0046}}
\and D.        ~Katz                          \inst{\ref{inst:0017}}
\and M.G.      ~Lattanzi                      \inst{\ref{inst:0046}}
\and J.        ~Bakker                        \inst{\ref{inst:0004}}
\and C.        ~Cacciari                      \inst{\ref{inst:0050}}
\and J.        ~Casta\~{n}eda                 \inst{\ref{inst:0034}}
\and L.        ~Chaoul                        \inst{\ref{inst:0036}}
\and N.        ~Cheek                         \inst{\ref{inst:0053}}
\and F.        ~De Angeli                     \inst{\ref{inst:0031}}
\and C.        ~Fabricius                     \inst{\ref{inst:0034}}
\and R.        ~Guerra                        \inst{\ref{inst:0004}}
\and B.        ~Holl                          \inst{\ref{inst:0032}}
\and E.        ~Masana                        \inst{\ref{inst:0034}}
\and R.        ~Messineo                      \inst{\ref{inst:0059}}
\and N.        ~Mowlavi                       \inst{\ref{inst:0032}}
\and K.        ~Nienartowicz                  \inst{\ref{inst:0061}}
\and P.        ~Panuzzo                       \inst{\ref{inst:0017}}
\and J.        ~Portell                       \inst{\ref{inst:0034}}
\and M.        ~Riello                        \inst{\ref{inst:0031}}
\and G.M.      ~Seabroke                      \inst{\ref{inst:0045}}
\and P.        ~Tanga                         \inst{\ref{inst:0001}}
\and F.        ~Th\'{e}venin                  \inst{\ref{inst:0001}}
\and G.        ~Gracia-Abril                  \inst{\ref{inst:0068},\ref{inst:0005}}
\and G.        ~Comoretto                     \inst{\ref{inst:0015}}
\and M.        ~Garcia-Reinaldos              \inst{\ref{inst:0004}}
\and D.        ~Teyssier                      \inst{\ref{inst:0015}}
\and M.        ~Altmann                       \inst{\ref{inst:0005},\ref{inst:0019}}
\and R.        ~Andrae                        \inst{\ref{inst:0030}}
\and M.        ~Audard                        \inst{\ref{inst:0032}}
\and I.        ~Bellas-Velidis                \inst{\ref{inst:0077}}
\and K.        ~Benson                        \inst{\ref{inst:0045}}
\and J.        ~Berthier                      \inst{\ref{inst:0079}}
\and R.        ~Blomme                        \inst{\ref{inst:0080}}
\and P.        ~Burgess                       \inst{\ref{inst:0031}}
\and G.        ~Busso                         \inst{\ref{inst:0031}}
\and B.        ~Carry                         \inst{\ref{inst:0001},\ref{inst:0079}}
\and A.        ~Cellino                       \inst{\ref{inst:0046}}
\and G.        ~Clementini                    \inst{\ref{inst:0050}}
\and M.        ~Clotet                        \inst{\ref{inst:0034}}
\and O.        ~Creevey                       \inst{\ref{inst:0001}}
\and M.        ~Davidson                      \inst{\ref{inst:0089}}
\and J.        ~De Ridder                     \inst{\ref{inst:0090}}
\and L.        ~Delchambre                    \inst{\ref{inst:0091}}
\and A.        ~Dell'Oro                      \inst{\ref{inst:0039}}
\and C.        ~Ducourant                     \inst{\ref{inst:0024}}
\and M.        ~Fouesneau                     \inst{\ref{inst:0030}}
\and Y.        ~Fr\'{e}mat                    \inst{\ref{inst:0080}}
\and L.        ~Galluccio                     \inst{\ref{inst:0001}}
\and M.        ~Garc\'{i}a-Torres             \inst{\ref{inst:0097}}
\and J.        ~Gonz\'{a}lez-N\'{u}\~{n}ez    \inst{\ref{inst:0053},\ref{inst:0099}}
\and J.J.      ~Gonz\'{a}lez-Vidal            \inst{\ref{inst:0034}}
\and E.        ~Gosset                        \inst{\ref{inst:0091},\ref{inst:0038}}
\and L.P.      ~Guy                           \inst{\ref{inst:0061},\ref{inst:0104}}
\and J.-L.     ~Halbwachs                     \inst{\ref{inst:0105}}
\and N.C.      ~Hambly                        \inst{\ref{inst:0089}}
\and D.L.      ~Harrison                      \inst{\ref{inst:0031},\ref{inst:0108}}
\and D.        ~Hestroffer                    \inst{\ref{inst:0079}}
\and S.T.      ~Hodgkin                       \inst{\ref{inst:0031}}
\and A.        ~Hutton                        \inst{\ref{inst:0111}}
\and G.        ~Jasniewicz                    \inst{\ref{inst:0112}}
\and A.        ~Jean-Antoine-Piccolo          \inst{\ref{inst:0036}}
\and S.        ~Jordan                        \inst{\ref{inst:0005}}
\and A.J.      ~Korn                          \inst{\ref{inst:0115}}
\and A.        ~Krone-Martins                 \inst{\ref{inst:0116}}
\and A.C.      ~Lanzafame                     \inst{\ref{inst:0117},\ref{inst:0118}}
\and T.        ~Lebzelter                     \inst{\ref{inst:0119}}
\and W.        ~L\"{ o}ffler                  \inst{\ref{inst:0005}}
\and M.        ~Manteiga                      \inst{\ref{inst:0121},\ref{inst:0122}}
\and P.M.      ~Marrese                       \inst{\ref{inst:0123},\ref{inst:0124}}
\and J.M.      ~Mart\'{i}n-Fleitas            \inst{\ref{inst:0111}}
\and A.        ~Moitinho                      \inst{\ref{inst:0116}}
\and A.        ~Mora                          \inst{\ref{inst:0111}}
\and K.        ~Muinonen                      \inst{\ref{inst:0128},\ref{inst:0129}}
\and J.        ~Osinde                        \inst{\ref{inst:0130}}
\and E.        ~Pancino                       \inst{\ref{inst:0039},\ref{inst:0124}}
\and T.        ~Pauwels                       \inst{\ref{inst:0080}}
\and J.-M.     ~Petit                         \inst{\ref{inst:0134}}
\and A.        ~Recio-Blanco                  \inst{\ref{inst:0001}}
\and P.J.      ~Richards                      \inst{\ref{inst:0136}}
\and L.        ~Rimoldini                     \inst{\ref{inst:0061}}
\and A.C.      ~Robin                         \inst{\ref{inst:0134}}
\and L.M.      ~Sarro                         \inst{\ref{inst:0139}}
\and C.        ~Siopis                        \inst{\ref{inst:0037}}
\and M.        ~Smith                         \inst{\ref{inst:0045}}
\and A.        ~Sozzetti                      \inst{\ref{inst:0046}}
\and M.        ~S\"{ u}veges                  \inst{\ref{inst:0030}}
\and J.        ~Torra                         \inst{\ref{inst:0034}}
\and W.        ~van Reeven                    \inst{\ref{inst:0111}}
\and U.        ~Abbas                         \inst{\ref{inst:0046}}
\and A.        ~Abreu Aramburu                \inst{\ref{inst:0147}}
\and S.        ~Accart                        \inst{\ref{inst:0148}}
\and C.        ~Aerts                         \inst{\ref{inst:0090},\ref{inst:0150}}
\and G.        ~Altavilla                     \inst{\ref{inst:0123},\ref{inst:0124},\ref{inst:0050}}
\and M.A.      ~\'{A}lvarez                   \inst{\ref{inst:0121}}
\and R.        ~Alvarez                       \inst{\ref{inst:0004}}
\and J.        ~Alves                         \inst{\ref{inst:0119}}
\and R.I.      ~Anderson                      \inst{\ref{inst:0157},\ref{inst:0032}}
\and E.        ~Anglada Varela                \inst{\ref{inst:0012}}
\and E.        ~Antiche                       \inst{\ref{inst:0034}}
\and T.        ~Antoja                        \inst{\ref{inst:0028},\ref{inst:0034}}
\and B.        ~Arcay                         \inst{\ref{inst:0121}}
\and T.L.      ~Astraatmadja                  \inst{\ref{inst:0030},\ref{inst:0165}}
\and N.        ~Bach                          \inst{\ref{inst:0111}}
\and S.G.      ~Baker                         \inst{\ref{inst:0045}}
\and L.        ~Balaguer-N\'{u}\~{n}ez        \inst{\ref{inst:0034}}
\and P.        ~Balm                          \inst{\ref{inst:0015}}
\and C.        ~Barata                        \inst{\ref{inst:0116}}
\and D.        ~Barbato                       \inst{\ref{inst:0171},\ref{inst:0046}}
\and F.        ~Barblan                       \inst{\ref{inst:0032}}
\and P.S.      ~Barklem                       \inst{\ref{inst:0115}}
\and D.        ~Barrado                       \inst{\ref{inst:0175}}
\and M.        ~Barros                        \inst{\ref{inst:0116}}
\and M.A.      ~Barstow                       \inst{\ref{inst:0177}}
\and S.        ~Bartholom\'{e} Mu\~{n}oz      \inst{\ref{inst:0034}}
\and J.-L.     ~Bassilana                     \inst{\ref{inst:0148}}
\and U.        ~Becciani                      \inst{\ref{inst:0118}}
\and M.        ~Bellazzini                    \inst{\ref{inst:0050}}
\and A.        ~Berihuete                     \inst{\ref{inst:0182}}
\and S.        ~Bertone                       \inst{\ref{inst:0046},\ref{inst:0019},\ref{inst:0185}}
\and L.        ~Bianchi                       \inst{\ref{inst:0186}}
\and O.        ~Bienaym\'{e}                  \inst{\ref{inst:0105}}
\and S.        ~Blanco-Cuaresma               \inst{\ref{inst:0032},\ref{inst:0024},\ref{inst:0190}}
\and T.        ~Boch                          \inst{\ref{inst:0105}}
\and C.        ~Boeche                        \inst{\ref{inst:0027}}
\and R.        ~Borrachero                    \inst{\ref{inst:0034}}
\and D.        ~Bossini                       \inst{\ref{inst:0027}}
\and S.        ~Bouquillon                    \inst{\ref{inst:0019}}
\and A.        ~Bragaglia                     \inst{\ref{inst:0050}}
\and L.        ~Bramante                      \inst{\ref{inst:0059}}
\and M.A.      ~Breddels                      \inst{\ref{inst:0198}}
\and A.        ~Bressan                       \inst{\ref{inst:0199}}
\and N.        ~Brouillet                     \inst{\ref{inst:0024}}
\and T.        ~Br\"{ u}semeister             \inst{\ref{inst:0005}}
\and E.        ~Brugaletta                    \inst{\ref{inst:0118}}
\and B.        ~Bucciarelli                   \inst{\ref{inst:0046}}
\and A.        ~Burlacu                       \inst{\ref{inst:0036}}
\and D.        ~Busonero                      \inst{\ref{inst:0046}}
\and A.G.      ~Butkevich                     \inst{\ref{inst:0002}}
\and R.        ~Buzzi                         \inst{\ref{inst:0046}}
\and E.        ~Caffau                        \inst{\ref{inst:0017}}
\and R.        ~Cancelliere                   \inst{\ref{inst:0209}}
\and G.        ~Cannizzaro                    \inst{\ref{inst:0210},\ref{inst:0150}}
\and T.        ~Cantat-Gaudin                 \inst{\ref{inst:0027},\ref{inst:0034}}
\and R.        ~Carballo                      \inst{\ref{inst:0214}}
\and T.        ~Carlucci                      \inst{\ref{inst:0019}}
\and J.M.      ~Carrasco                      \inst{\ref{inst:0034}}
\and L.        ~Casamiquela                   \inst{\ref{inst:0034}}
\and M.        ~Castellani                    \inst{\ref{inst:0123}}
\and A.        ~Castro-Ginard                 \inst{\ref{inst:0034}}
\and L.        ~Chemin                        \inst{\ref{inst:0220}}
\and A.        ~Chiavassa                     \inst{\ref{inst:0001}}
\and G.        ~Cocozza                       \inst{\ref{inst:0050}}
\and G.        ~Costigan                      \inst{\ref{inst:0026}}
\and S.        ~Cowell                        \inst{\ref{inst:0031}}
\and F.        ~Crifo                         \inst{\ref{inst:0017}}
\and M.        ~Crosta                        \inst{\ref{inst:0046}}
\and C.        ~Crowley                       \inst{\ref{inst:0006}}
\and J.        ~Cuypers$^\dagger$             \inst{\ref{inst:0080}}
\and C.        ~Dafonte                       \inst{\ref{inst:0121}}
\and Y.        ~Damerdji                      \inst{\ref{inst:0091},\ref{inst:0231}}
\and A.        ~Dapergolas                    \inst{\ref{inst:0077}}
\and P.        ~David                         \inst{\ref{inst:0079}}
\and M.        ~David                         \inst{\ref{inst:0234}}
\and P.        ~de Laverny                    \inst{\ref{inst:0001}}
\and F.        ~De Luise                      \inst{\ref{inst:0236}}
\and R.        ~De March                      \inst{\ref{inst:0059}}
\and R.        ~de Souza                      \inst{\ref{inst:0238}}
\and A.        ~de Torres                     \inst{\ref{inst:0006}}
\and J.        ~Debosscher                    \inst{\ref{inst:0090}}
\and E.        ~del Pozo                      \inst{\ref{inst:0111}}
\and M.        ~Delbo                         \inst{\ref{inst:0001}}
\and A.        ~Delgado                       \inst{\ref{inst:0031}}
\and H.E.      ~Delgado                       \inst{\ref{inst:0139}}
\and S.        ~Diakite                       \inst{\ref{inst:0134}}
\and C.        ~Diener                        \inst{\ref{inst:0031}}
\and E.        ~Distefano                     \inst{\ref{inst:0118}}
\and C.        ~Dolding                       \inst{\ref{inst:0045}}
\and P.        ~Drazinos                      \inst{\ref{inst:0249}}
\and J.        ~Dur\'{a}n                     \inst{\ref{inst:0130}}
\and B.        ~Edvardsson                    \inst{\ref{inst:0115}}
\and H.        ~Enke                          \inst{\ref{inst:0252}}
\and K.        ~Eriksson                      \inst{\ref{inst:0115}}
\and P.        ~Esquej                        \inst{\ref{inst:0254}}
\and G.        ~Eynard Bontemps               \inst{\ref{inst:0036}}
\and C.        ~Fabre                         \inst{\ref{inst:0256}}
\and M.        ~Fabrizio                      \inst{\ref{inst:0123},\ref{inst:0124}}
\and S.        ~Faigler                       \inst{\ref{inst:0259}}
\and A.J.      ~Falc\~{a}o                    \inst{\ref{inst:0260}}
\and M.        ~Farr\`{a}s Casas              \inst{\ref{inst:0034}}
\and L.        ~Federici                      \inst{\ref{inst:0050}}
\and G.        ~Fedorets                      \inst{\ref{inst:0128}}
\and P.        ~Fernique                      \inst{\ref{inst:0105}}
\and F.        ~Figueras                      \inst{\ref{inst:0034}}
\and F.        ~Filippi                       \inst{\ref{inst:0059}}
\and K.        ~Findeisen                     \inst{\ref{inst:0017}}
\and A.        ~Fonti                         \inst{\ref{inst:0059}}
\and E.        ~Fraile                        \inst{\ref{inst:0254}}
\and M.        ~Fraser                        \inst{\ref{inst:0031},\ref{inst:0271}}
\and B.        ~Fr\'{e}zouls                  \inst{\ref{inst:0036}}
\and M.        ~Gai                           \inst{\ref{inst:0046}}
\and S.        ~Galleti                       \inst{\ref{inst:0050}}
\and D.        ~Garabato                      \inst{\ref{inst:0121}}
\and F.        ~Garc\'{i}a-Sedano             \inst{\ref{inst:0139}}
\and A.        ~Garofalo                      \inst{\ref{inst:0277},\ref{inst:0050}}
\and N.        ~Garralda                      \inst{\ref{inst:0034}}
\and A.        ~Gavel                         \inst{\ref{inst:0115}}
\and P.        ~Gavras                        \inst{\ref{inst:0017},\ref{inst:0077},\ref{inst:0249}}
\and J.        ~Gerssen                       \inst{\ref{inst:0252}}
\and P.        ~Giacobbe                      \inst{\ref{inst:0046}}
\and G.        ~Gilmore                       \inst{\ref{inst:0031}}
\and S.        ~Girona                        \inst{\ref{inst:0287}}
\and G.        ~Giuffrida                     \inst{\ref{inst:0124},\ref{inst:0123}}
\and F.        ~Glass                         \inst{\ref{inst:0032}}
\and M.        ~Gomes                         \inst{\ref{inst:0116}}
\and M.        ~Granvik                       \inst{\ref{inst:0128},\ref{inst:0293}}
\and A.        ~Gueguen                       \inst{\ref{inst:0017},\ref{inst:0295}}
\and A.        ~Guerrier                      \inst{\ref{inst:0148}}
\and J.        ~Guiraud                       \inst{\ref{inst:0036}}
\and R.        ~Guti\'{e}rrez-S\'{a}nchez     \inst{\ref{inst:0015}}
\and R.        ~Haigron                       \inst{\ref{inst:0017}}
\and D.        ~Hatzidimitriou                \inst{\ref{inst:0249},\ref{inst:0077}}
\and M.        ~Hauser                        \inst{\ref{inst:0005},\ref{inst:0030}}
\and M.        ~Haywood                       \inst{\ref{inst:0017}}
\and U.        ~Heiter                        \inst{\ref{inst:0115}}
\and A.        ~Helmi                         \inst{\ref{inst:0198}}
\and J.        ~Heu                           \inst{\ref{inst:0017}}
\and W.        ~Hofmann                       \inst{\ref{inst:0005}}
\and G.        ~Holland                       \inst{\ref{inst:0031}}
\and H.E.      ~Huckle                        \inst{\ref{inst:0045}}
\and A.        ~Hypki                         \inst{\ref{inst:0026},\ref{inst:0312}}
\and V.        ~Icardi                        \inst{\ref{inst:0059}}
\and K.        ~Jan{\ss}en                    \inst{\ref{inst:0252}}
\and G.        ~Jevardat de Fombelle          \inst{\ref{inst:0061}}
\and P.G.      ~Jonker                        \inst{\ref{inst:0210},\ref{inst:0150}}
\and \'{A}.L.  ~Juh\'{a}sz                    \inst{\ref{inst:0318},\ref{inst:0319}}
\and F.        ~Julbe                         \inst{\ref{inst:0034}}
\and A.        ~Karampelas                    \inst{\ref{inst:0249},\ref{inst:0322}}
\and A.        ~Kewley                        \inst{\ref{inst:0031}}
\and J.        ~Klar                          \inst{\ref{inst:0252}}
\and A.        ~Kochoska                      \inst{\ref{inst:0325},\ref{inst:0326}}
\and R.        ~Kohley                        \inst{\ref{inst:0004}}
\and K.        ~Kolenberg                     \inst{\ref{inst:0328},\ref{inst:0090},\ref{inst:0190}}
\and M.        ~Kontizas                      \inst{\ref{inst:0249}}
\and E.        ~Kontizas                      \inst{\ref{inst:0077}}
\and S.E.      ~Koposov                       \inst{\ref{inst:0031},\ref{inst:0334}}
\and G.        ~Kordopatis                    \inst{\ref{inst:0001}}
\and Z.        ~Kostrzewa-Rutkowska           \inst{\ref{inst:0210},\ref{inst:0150}}
\and P.        ~Koubsky                       \inst{\ref{inst:0338}}
\and A.F.      ~Lanza                         \inst{\ref{inst:0118}}
\and Y.        ~Lasne                         \inst{\ref{inst:0148}}
\and J.-B.     ~Lavigne                       \inst{\ref{inst:0148}}
\and Y.        ~Le Fustec                     \inst{\ref{inst:0342}}
\and C.        ~Le Poncin-Lafitte             \inst{\ref{inst:0019}}
\and Y.        ~Lebreton                      \inst{\ref{inst:0017},\ref{inst:0345}}
\and S.        ~Leccia                        \inst{\ref{inst:0346}}
\and N.        ~Leclerc                       \inst{\ref{inst:0017}}
\and I.        ~Lecoeur-Taibi                 \inst{\ref{inst:0061}}
\and H.        ~Lenhardt                      \inst{\ref{inst:0005}}
\and F.        ~Leroux                        \inst{\ref{inst:0148}}
\and S.        ~Liao                          \inst{\ref{inst:0046},\ref{inst:0352},\ref{inst:0353}}
\and E.        ~Licata                        \inst{\ref{inst:0186}}
\and H.E.P.    ~Lindstr{\o}m                  \inst{\ref{inst:0355},\ref{inst:0356}}
\and T.A.      ~Lister                        \inst{\ref{inst:0357}}
\and E.        ~Livanou                       \inst{\ref{inst:0249}}
\and A.        ~Lobel                         \inst{\ref{inst:0080}}
\and M.        ~L\'{o}pez                     \inst{\ref{inst:0175}}
\and S.        ~Managau                       \inst{\ref{inst:0148}}
\and R.G.      ~Mann                          \inst{\ref{inst:0089}}
\and G.        ~Mantelet                      \inst{\ref{inst:0005}}
\and O.        ~Marchal                       \inst{\ref{inst:0017}}
\and J.M.      ~Marchant                      \inst{\ref{inst:0365}}
\and M.        ~Marconi                       \inst{\ref{inst:0346}}
\and S.        ~Marinoni                      \inst{\ref{inst:0123},\ref{inst:0124}}
\and G.        ~Marschalk\'{o}                \inst{\ref{inst:0318},\ref{inst:0370}}
\and D.J.      ~Marshall                      \inst{\ref{inst:0371}}
\and M.        ~Martino                       \inst{\ref{inst:0059}}
\and G.        ~Marton                        \inst{\ref{inst:0318}}
\and N.        ~Mary                          \inst{\ref{inst:0148}}
\and D.        ~Massari                       \inst{\ref{inst:0198}}
\and G.        ~Matijevi\v{c}                 \inst{\ref{inst:0252}}
\and T.        ~Mazeh                         \inst{\ref{inst:0259}}
\and P.J.      ~McMillan                      \inst{\ref{inst:0003}}
\and S.        ~Messina                       \inst{\ref{inst:0118}}
\and N.R.      ~Millar                        \inst{\ref{inst:0031}}
\and D.        ~Molina                        \inst{\ref{inst:0034}}
\and R.        ~Molinaro                      \inst{\ref{inst:0346}}
\and L.        ~Moln\'{a}r                    \inst{\ref{inst:0318}}
\and P.        ~Montegriffo                   \inst{\ref{inst:0050}}
\and R.        ~Mor                           \inst{\ref{inst:0034}}
\and R.        ~Morbidelli                    \inst{\ref{inst:0046}}
\and T.        ~Morel                         \inst{\ref{inst:0091}}
\and D.        ~Morris                        \inst{\ref{inst:0089}}
\and A.F.      ~Mulone                        \inst{\ref{inst:0059}}
\and T.        ~Muraveva                      \inst{\ref{inst:0050}}
\and I.        ~Musella                       \inst{\ref{inst:0346}}
\and G.        ~Nelemans                      \inst{\ref{inst:0150},\ref{inst:0090}}
\and L.        ~Nicastro                      \inst{\ref{inst:0050}}
\and L.        ~Noval                         \inst{\ref{inst:0148}}
\and W.        ~O'Mullane                     \inst{\ref{inst:0004},\ref{inst:0104}}
\and C.        ~Ord\'{e}novic                 \inst{\ref{inst:0001}}
\and D.        ~Ord\'{o}\~{n}ez-Blanco        \inst{\ref{inst:0061}}
\and P.        ~Osborne                       \inst{\ref{inst:0031}}
\and C.        ~Pagani                        \inst{\ref{inst:0177}}
\and I.        ~Pagano                        \inst{\ref{inst:0118}}
\and F.        ~Pailler                       \inst{\ref{inst:0036}}
\and H.        ~Palacin                       \inst{\ref{inst:0148}}
\and L.        ~Palaversa                     \inst{\ref{inst:0031},\ref{inst:0032}}
\and A.        ~Panahi                        \inst{\ref{inst:0259}}
\and M.        ~Pawlak                        \inst{\ref{inst:0408},\ref{inst:0409}}
\and A.M.      ~Piersimoni                    \inst{\ref{inst:0236}}
\and F.-X.     ~Pineau                        \inst{\ref{inst:0105}}
\and E.        ~Plachy                        \inst{\ref{inst:0318}}
\and G.        ~Plum                          \inst{\ref{inst:0017}}
\and E.        ~Poggio                        \inst{\ref{inst:0171},\ref{inst:0046}}
\and E.        ~Poujoulet                     \inst{\ref{inst:0416}}
\and A.        ~Pr\v{s}a                      \inst{\ref{inst:0326}}
\and L.        ~Pulone                        \inst{\ref{inst:0123}}
\and E.        ~Racero                        \inst{\ref{inst:0053}}
\and S.        ~Ragaini                       \inst{\ref{inst:0050}}
\and N.        ~Rambaux                       \inst{\ref{inst:0079}}
\and S.        ~Regibo                        \inst{\ref{inst:0090}}
\and C.        ~Reyl\'{e}                     \inst{\ref{inst:0134}}
\and F.        ~Riclet                        \inst{\ref{inst:0036}}
\and V.        ~Ripepi                        \inst{\ref{inst:0346}}
\and A.        ~Riva                          \inst{\ref{inst:0046}}
\and A.        ~Rivard                        \inst{\ref{inst:0148}}
\and G.        ~Rixon                         \inst{\ref{inst:0031}}
\and T.        ~Roegiers                      \inst{\ref{inst:0429}}
\and M.        ~Roelens                       \inst{\ref{inst:0032}}
\and M.        ~Romero-G\'{o}mez              \inst{\ref{inst:0034}}
\and N.        ~Rowell                        \inst{\ref{inst:0089}}
\and F.        ~Royer                         \inst{\ref{inst:0017}}
\and L.        ~Ruiz-Dern                     \inst{\ref{inst:0017}}
\and G.        ~Sadowski                      \inst{\ref{inst:0037}}
\and T.        ~Sagrist\`{a} Sell\'{e}s       \inst{\ref{inst:0005}}
\and J.        ~Sahlmann                      \inst{\ref{inst:0004},\ref{inst:0438}}
\and J.        ~Salgado                       \inst{\ref{inst:0439}}
\and E.        ~Salguero                      \inst{\ref{inst:0012}}
\and N.        ~Sanna                         \inst{\ref{inst:0039}}
\and T.        ~Santana-Ros                   \inst{\ref{inst:0312}}
\and M.        ~Sarasso                       \inst{\ref{inst:0046}}
\and H.        ~Savietto                      \inst{\ref{inst:0444}}
\and M.        ~Schultheis                    \inst{\ref{inst:0001}}
\and E.        ~Sciacca                       \inst{\ref{inst:0118}}
\and M.        ~Segol                         \inst{\ref{inst:0447}}
\and J.C.      ~Segovia                       \inst{\ref{inst:0053}}
\and D.        ~S\'{e}gransan                 \inst{\ref{inst:0032}}
\and I-C.      ~Shih                          \inst{\ref{inst:0017}}
\and L.        ~Siltala                       \inst{\ref{inst:0128},\ref{inst:0452}}
\and A.F.      ~Silva                         \inst{\ref{inst:0116}}
\and R.L.      ~Smart                         \inst{\ref{inst:0046}}
\and K.W.      ~Smith                         \inst{\ref{inst:0030}}
\and E.        ~Solano                        \inst{\ref{inst:0175},\ref{inst:0457}}
\and F.        ~Solitro                       \inst{\ref{inst:0059}}
\and R.        ~Sordo                         \inst{\ref{inst:0027}}
\and S.        ~Soria Nieto                   \inst{\ref{inst:0034}}
\and J.        ~Souchay                       \inst{\ref{inst:0019}}
\and A.        ~Spagna                        \inst{\ref{inst:0046}}
\and F.        ~Spoto                         \inst{\ref{inst:0001},\ref{inst:0079}}
\and U.        ~Stampa                        \inst{\ref{inst:0005}}
\and I.A.      ~Steele                        \inst{\ref{inst:0365}}
\and C.A.      ~Stephenson                    \inst{\ref{inst:0015}}
\and H.        ~Stoev                         \inst{\ref{inst:0468}}
\and F.F.      ~Suess                         \inst{\ref{inst:0031}}
\and J.        ~Surdej                        \inst{\ref{inst:0091}}
\and L.        ~Szabados                      \inst{\ref{inst:0318}}
\and E.        ~Szegedi-Elek                  \inst{\ref{inst:0318}}
\and D.        ~Tapiador                      \inst{\ref{inst:0473},\ref{inst:0474}}
\and F.        ~Taris                         \inst{\ref{inst:0019}}
\and G.        ~Tauran                        \inst{\ref{inst:0148}}
\and M.B.      ~Taylor                        \inst{\ref{inst:0477}}
\and R.        ~Teixeira                      \inst{\ref{inst:0238}}
\and D.        ~Terrett                       \inst{\ref{inst:0136}}
\and P.        ~Teyssandier                   \inst{\ref{inst:0019}}
\and W.        ~Thuillot                      \inst{\ref{inst:0079}}
\and A.        ~Titarenko                     \inst{\ref{inst:0001}}
\and F.        ~Torra Clotet                  \inst{\ref{inst:0483}}
\and C.        ~Turon                         \inst{\ref{inst:0017}}
\and A.        ~Ulla                          \inst{\ref{inst:0485}}
\and E.        ~Utrilla                       \inst{\ref{inst:0111}}
\and S.        ~Uzzi                          \inst{\ref{inst:0059}}
\and M.        ~Vaillant                      \inst{\ref{inst:0148}}
\and G.        ~Valentini                     \inst{\ref{inst:0236}}
\and V.        ~Valette                       \inst{\ref{inst:0036}}
\and A.        ~van Elteren                   \inst{\ref{inst:0026}}
\and E.        ~Van Hemelryck                 \inst{\ref{inst:0080}}
\and M.        ~van Leeuwen                   \inst{\ref{inst:0031}}
\and M.        ~Vaschetto                     \inst{\ref{inst:0059}}
\and A.        ~Vecchiato                     \inst{\ref{inst:0046}}
\and J.        ~Veljanoski                    \inst{\ref{inst:0198}}
\and Y.        ~Viala                         \inst{\ref{inst:0017}}
\and D.        ~Vicente                       \inst{\ref{inst:0287}}
\and S.        ~Vogt                          \inst{\ref{inst:0429}}
\and C.        ~von Essen                     \inst{\ref{inst:0500}}
\and H.        ~Voss                          \inst{\ref{inst:0034}}
\and V.        ~Votruba                       \inst{\ref{inst:0338}}
\and S.        ~Voutsinas                     \inst{\ref{inst:0089}}
\and G.        ~Walmsley                      \inst{\ref{inst:0036}}
\and M.        ~Weiler                        \inst{\ref{inst:0034}}
\and O.        ~Wertz                         \inst{\ref{inst:0506}}
\and T.        ~Wevers                        \inst{\ref{inst:0031},\ref{inst:0150}}
\and \L{}.     ~Wyrzykowski                   \inst{\ref{inst:0031},\ref{inst:0408}}
\and A.        ~Yoldas                        \inst{\ref{inst:0031}}
\and M.        ~\v{Z}erjal                    \inst{\ref{inst:0325},\ref{inst:0513}}
\and H.        ~Ziaeepour                     \inst{\ref{inst:0134}}
\and J.        ~Zorec                         \inst{\ref{inst:0515}}
\and S.        ~Zschocke                      \inst{\ref{inst:0002}}
\and S.        ~Zucker                        \inst{\ref{inst:0517}}
\and C.        ~Zurbach                       \inst{\ref{inst:0112}}
\and T.        ~Zwitter                       \inst{\ref{inst:0325}}
}
\institute{
     Universit\'{e} C\^{o}te d'Azur, Observatoire de la C\^{o}te d'Azur, CNRS, Laboratoire Lagrange, Bd de l'Observatoire, CS 34229, 06304 Nice Cedex 4, France\relax                                        \label{inst:0001}
\and Lohrmann Observatory, Technische Universit\"{ a}t Dresden, Mommsenstra{\ss}e 13, 01062 Dresden, Germany\relax                                                                                           \label{inst:0002}
\and Lund Observatory, Department of Astronomy and Theoretical Physics, Lund University, Box 43, 22100 Lund, Sweden\relax                                                                                    \label{inst:0003}
\and European Space Astronomy Centre (ESA/ESAC), Camino bajo del Castillo, s/n, Urbanizacion Villafranca del Castillo, Villanueva de la Ca\~{n}ada, 28692 Madrid, Spain\relax                                \label{inst:0004}
\and Astronomisches Rechen-Institut, Zentrum f\"{ u}r Astronomie der Universit\"{ a}t Heidelberg, M\"{ o}nchhofstr. 12-14, 69120 Heidelberg, Germany\relax                                                   \label{inst:0005}
\and HE Space Operations BV for ESA/ESAC, Camino bajo del Castillo, s/n, Urbanizacion Villafranca del Castillo, Villanueva de la Ca\~{n}ada, 28692 Madrid, Spain\relax                                       \label{inst:0006}
\and Vitrociset Belgium for ESA/ESAC, Camino bajo del Castillo, s/n, Urbanizacion Villafranca del Castillo, Villanueva de la Ca\~{n}ada, 28692 Madrid, Spain\relax                                           \label{inst:0010}
\and ATG Europe for ESA/ESAC, Camino bajo del Castillo, s/n, Urbanizacion Villafranca del Castillo, Villanueva de la Ca\~{n}ada, 28692 Madrid, Spain\relax                                                   \label{inst:0012}
\and Telespazio Vega UK Ltd for ESA/ESAC, Camino bajo del Castillo, s/n, Urbanizacion Villafranca del Castillo, Villanueva de la Ca\~{n}ada, 28692 Madrid, Spain\relax                                       \label{inst:0015}
\and GEPI, Observatoire de Paris, Universit\'{e} PSL, CNRS, 5 Place Jules Janssen, 92190 Meudon, France\relax                                                                                                \label{inst:0017}
\and Univ. Grenoble Alpes, CNRS, IPAG, 38000 Grenoble, France\relax                                                                                                                                          \label{inst:0018}
\and SYRTE, Observatoire de Paris, Universit\'{e} PSL, CNRS,  Sorbonne Universit\'{e}, LNE, 61 avenue de l’Observatoire 75014 Paris, France\relax                                                          \label{inst:0019}
\and ON/MCTI-BR, Rua Gal. Jos\'{e} Cristino 77, Rio de Janeiro, CEP 20921-400, RJ,  Brazil\relax                                                                                                             \label{inst:0021}
\and OV/UFRJ-BR, Ladeira Pedro Ant\^{o}nio 43, Rio de Janeiro, CEP 20080-090, RJ, Brazil\relax                                                                                                               \label{inst:0022}
\and Laboratoire d'astrophysique de Bordeaux, Univ. Bordeaux, CNRS, B18N, all{\'e}e Geoffroy Saint-Hilaire, 33615 Pessac, France\relax                                                                       \label{inst:0024}
\and Leiden Observatory, Leiden University, Niels Bohrweg 2, 2333 CA Leiden, The Netherlands\relax                                                                                                           \label{inst:0026}
\and INAF - Osservatorio astronomico di Padova, Vicolo Osservatorio 5, 35122 Padova, Italy\relax                                                                                                             \label{inst:0027}
\and Science Support Office, Directorate of Science, European Space Research and Technology Centre (ESA/ESTEC), Keplerlaan 1, 2201AZ, Noordwijk, The Netherlands\relax                                       \label{inst:0028}
\and Max Planck Institute for Astronomy, K\"{ o}nigstuhl 17, 69117 Heidelberg, Germany\relax                                                                                                                 \label{inst:0030}
\and Institute of Astronomy, University of Cambridge, Madingley Road, Cambridge CB3 0HA, United Kingdom\relax                                                                                                \label{inst:0031}
\and Department of Astronomy, University of Geneva, Chemin des Maillettes 51, 1290 Versoix, Switzerland\relax                                                                                                \label{inst:0032}
\and Mission Operations Division, Operations Department, Directorate of Science, European Space Research and Technology Centre (ESA/ESTEC), Keplerlaan 1, 2201 AZ, Noordwijk, The Netherlands\relax          \label{inst:0033}
\and Institut de Ci\`{e}ncies del Cosmos, Universitat  de  Barcelona  (IEEC-UB), Mart\'{i} i  Franqu\`{e}s  1, 08028 Barcelona, Spain\relax                                                                  \label{inst:0034}
\and CNES Centre Spatial de Toulouse, 18 avenue Edouard Belin, 31401 Toulouse Cedex 9, France\relax                                                                                                          \label{inst:0036}
\and Institut d'Astronomie et d'Astrophysique, Universit\'{e} Libre de Bruxelles CP 226, Boulevard du Triomphe, 1050 Brussels, Belgium\relax                                                                 \label{inst:0037}
\and F.R.S.-FNRS, Rue d'Egmont 5, 1000 Brussels, Belgium\relax                                                                                                                                               \label{inst:0038}
\and INAF - Osservatorio Astrofisico di Arcetri, Largo Enrico Fermi 5, 50125 Firenze, Italy\relax                                                                                                            \label{inst:0039}
\and Mullard Space Science Laboratory, University College London, Holmbury St Mary, Dorking, Surrey RH5 6NT, United Kingdom\relax                                                                            \label{inst:0045}
\and INAF - Osservatorio Astrofisico di Torino, via Osservatorio 20, 10025 Pino Torinese (TO), Italy\relax                                                                                                   \label{inst:0046}
\and INAF - Osservatorio di Astrofisica e Scienza dello Spazio di Bologna, via Piero Gobetti 93/3, 40129 Bologna, Italy\relax                                                                                \label{inst:0050}
\and Serco Gesti\'{o}n de Negocios for ESA/ESAC, Camino bajo del Castillo, s/n, Urbanizacion Villafranca del Castillo, Villanueva de la Ca\~{n}ada, 28692 Madrid, Spain\relax                                \label{inst:0053}
\and ALTEC S.p.a, Corso Marche, 79,10146 Torino, Italy\relax                                                                                                                                                 \label{inst:0059}
\and Department of Astronomy, University of Geneva, Chemin d'Ecogia 16, 1290 Versoix, Switzerland\relax                                                                                                      \label{inst:0061}
\and Gaia DPAC Project Office, ESAC, Camino bajo del Castillo, s/n, Urbanizacion Villafranca del Castillo, Villanueva de la Ca\~{n}ada, 28692 Madrid, Spain\relax                                            \label{inst:0068}
\and National Observatory of Athens, I. Metaxa and Vas. Pavlou, Palaia Penteli, 15236 Athens, Greece\relax                                                                                                   \label{inst:0077}
\and IMCCE, Observatoire de Paris, Universit\'{e} PSL, CNRS,  Sorbonne Universit\'{e}, Univ. Lille, 77 av. Denfert-Rochereau, 75014 Paris, France\relax                                                      \label{inst:0079}
\and Royal Observatory of Belgium, Ringlaan 3, 1180 Brussels, Belgium\relax                                                                                                                                  \label{inst:0080}
\and Institute for Astronomy, University of Edinburgh, Royal Observatory, Blackford Hill, Edinburgh EH9 3HJ, United Kingdom\relax                                                                            \label{inst:0089}
\and Instituut voor Sterrenkunde, KU Leuven, Celestijnenlaan 200D, 3001 Leuven, Belgium\relax                                                                                                                \label{inst:0090}
\and Institut d'Astrophysique et de G\'{e}ophysique, Universit\'{e} de Li\`{e}ge, 19c, All\'{e}e du 6 Ao\^{u}t, B-4000 Li\`{e}ge, Belgium\relax                                                              \label{inst:0091}
\and \'{A}rea de Lenguajes y Sistemas Inform\'{a}ticos, Universidad Pablo de Olavide, Ctra. de Utrera, km 1. 41013, Sevilla, Spain\relax                                                                     \label{inst:0097}
\and ETSE Telecomunicaci\'{o}n, Universidade de Vigo, Campus Lagoas-Marcosende, 36310 Vigo, Galicia, Spain\relax                                                                                             \label{inst:0099}
\and Large Synoptic Survey Telescope, 950 N. Cherry Avenue, Tucson, AZ 85719, USA\relax                                                                                                                      \label{inst:0104}
\and Observatoire Astronomique de Strasbourg, Universit\'{e} de Strasbourg, CNRS, UMR 7550, 11 rue de l'Universit\'{e}, 67000 Strasbourg, France\relax                                                       \label{inst:0105}
\and Kavli Institute for Cosmology, University of Cambridge, Madingley Road, Cambride CB3 0HA, United Kingdom\relax                                                                                          \label{inst:0108}
\and Aurora Technology for ESA/ESAC, Camino bajo del Castillo, s/n, Urbanizacion Villafranca del Castillo, Villanueva de la Ca\~{n}ada, 28692 Madrid, Spain\relax                                            \label{inst:0111}
\and Laboratoire Univers et Particules de Montpellier, Universit\'{e} Montpellier, Place Eug\`{e}ne Bataillon, CC72, 34095 Montpellier Cedex 05, France\relax                                                \label{inst:0112}
\and Department of Physics and Astronomy, Division of Astronomy and Space Physics, Uppsala University, Box 516, 75120 Uppsala, Sweden\relax                                                                  \label{inst:0115}
\and CENTRA, Universidade de Lisboa, FCUL, Campo Grande, Edif. C8, 1749-016 Lisboa, Portugal\relax                                                                                                           \label{inst:0116}
\and Universit\`{a} di Catania, Dipartimento di Fisica e Astronomia, Sezione Astrofisica, Via S. Sofia 78, 95123 Catania, Italy\relax                                                                        \label{inst:0117}
\and INAF - Osservatorio Astrofisico di Catania, via S. Sofia 78, 95123 Catania, Italy\relax                                                                                                                 \label{inst:0118}
\and University of Vienna, Department of Astrophysics, T\"{ u}rkenschanzstra{\ss}e 17, A1180 Vienna, Austria\relax                                                                                           \label{inst:0119}
\and CITIC – Department of Computer Science, University of A Coru\~{n}a, Campus de Elvi\~{n}a S/N, 15071, A Coru\~{n}a, Spain\relax                                                                        \label{inst:0121}
\and CITIC – Astronomy and Astrophysics, University of A Coru\~{n}a, Campus de Elvi\~{n}a S/N, 15071, A Coru\~{n}a, Spain\relax                                                                            \label{inst:0122}
\and INAF - Osservatorio Astronomico di Roma, Via di Frascati 33, 00078 Monte Porzio Catone (Roma), Italy\relax                                                                                              \label{inst:0123}
\and Space Science Data Center - ASI, Via del Politecnico SNC, 00133 Roma, Italy\relax                                                                                                                       \label{inst:0124}
\and University of Helsinki, Department of Physics, P.O. Box 64, 00014 Helsinki, Finland\relax                                                                                                               \label{inst:0128}
\and Finnish Geospatial Research Institute FGI, Geodeetinrinne 2, 02430 Masala, Finland\relax                                                                                                                \label{inst:0129}
\and Isdefe for ESA/ESAC, Camino bajo del Castillo, s/n, Urbanizacion Villafranca del Castillo, Villanueva de la Ca\~{n}ada, 28692 Madrid, Spain\relax                                                       \label{inst:0130}
\and Institut UTINAM UMR6213, CNRS, OSU THETA Franche-Comt\'{e} Bourgogne, Universit\'{e} Bourgogne Franche-Comt\'{e}, 25000 Besan\c{c}on, France\relax                                                      \label{inst:0134}
\and STFC, Rutherford Appleton Laboratory, Harwell, Didcot, OX11 0QX, United Kingdom\relax                                                                                                                   \label{inst:0136}
\and Dpto. de Inteligencia Artificial, UNED, c/ Juan del Rosal 16, 28040 Madrid, Spain\relax                                                                                                                 \label{inst:0139}
\and Elecnor Deimos Space for ESA/ESAC, Camino bajo del Castillo, s/n, Urbanizacion Villafranca del Castillo, Villanueva de la Ca\~{n}ada, 28692 Madrid, Spain\relax                                         \label{inst:0147}
\and Thales Services for CNES Centre Spatial de Toulouse, 18 avenue Edouard Belin, 31401 Toulouse Cedex 9, France\relax                                                                                      \label{inst:0148}
\and Department of Astrophysics/IMAPP, Radboud University, P.O.Box 9010, 6500 GL Nijmegen, The Netherlands\relax                                                                                             \label{inst:0150}
\and European Southern Observatory, Karl-Schwarzschild-Str. 2, 85748 Garching, Germany\relax                                                                                                                 \label{inst:0157}
\and Department of Terrestrial Magnetism, Carnegie Institution for Science, 5241 Broad Branch Road, NW, Washington, DC 20015-1305, USA\relax                                                                 \label{inst:0165}
\and Universit\`{a} di Torino, Dipartimento di Fisica, via Pietro Giuria 1, 10125 Torino, Italy\relax                                                                                                        \label{inst:0171}
\and Departamento de Astrof\'{i}sica, Centro de Astrobiolog\'{i}a (CSIC-INTA), ESA-ESAC. Camino Bajo del Castillo s/n. 28692 Villanueva de la Ca\~{n}ada, Madrid, Spain\relax                                \label{inst:0175}
\and Leicester Institute of Space and Earth Observation and Department of Physics and Astronomy, University of Leicester, University Road, Leicester LE1 7RH, United Kingdom\relax                           \label{inst:0177}
\and Departamento de Estad\'{i}stica, Universidad de C\'{a}diz, Calle Rep\'{u}blica \'{A}rabe Saharawi s/n. 11510, Puerto Real, C\'{a}diz, Spain\relax                                                       \label{inst:0182}
\and Astronomical Institute Bern University, Sidlerstrasse 5, 3012 Bern, Switzerland (present address)\relax                                                                                                 \label{inst:0185}
\and EURIX S.r.l., Corso Vittorio Emanuele II 61, 10128, Torino, Italy\relax                                                                                                                                 \label{inst:0186}
\and Harvard-Smithsonian Center for Astrophysics, 60 Garden Street, Cambridge MA 02138, USA\relax                                                                                                            \label{inst:0190}
\and Kapteyn Astronomical Institute, University of Groningen, Landleven 12, 9747 AD Groningen, The Netherlands\relax                                                                                         \label{inst:0198}
\and SISSA - Scuola Internazionale Superiore di Studi Avanzati, via Bonomea 265, 34136 Trieste, Italy\relax                                                                                                  \label{inst:0199}
\and University of Turin, Department of Computer Sciences, Corso Svizzera 185, 10149 Torino, Italy\relax                                                                                                     \label{inst:0209}
\and SRON, Netherlands Institute for Space Research, Sorbonnelaan 2, 3584CA, Utrecht, The Netherlands\relax                                                                                                  \label{inst:0210}
\and Dpto. de Matem\'{a}tica Aplicada y Ciencias de la Computaci\'{o}n, Univ. de Cantabria, ETS Ingenieros de Caminos, Canales y Puertos, Avda. de los Castros s/n, 39005 Santander, Spain\relax             \label{inst:0214}
\and Unidad de Astronom\'ia, Universidad de Antofagasta, Avenida Angamos 601, Antofagasta 1270300, Chile\relax                                                                                               \label{inst:0220}
\and CRAAG - Centre de Recherche en Astronomie, Astrophysique et G\'{e}ophysique, Route de l'Observatoire Bp 63 Bouzareah 16340 Algiers, Algeria\relax                                                       \label{inst:0231}
\and University of Antwerp, Onderzoeksgroep Toegepaste Wiskunde, Middelheimlaan 1, 2020 Antwerp, Belgium\relax                                                                                               \label{inst:0234}
\and INAF - Osservatorio Astronomico d'Abruzzo, Via Mentore Maggini, 64100 Teramo, Italy\relax                                                                                                               \label{inst:0236}
\and Instituto de Astronomia, Geof\`{i}sica e Ci\^{e}ncias Atmosf\'{e}ricas, Universidade de S\~{a}o Paulo, Rua do Mat\~{a}o, 1226, Cidade Universitaria, 05508-900 S\~{a}o Paulo, SP, Brazil\relax          \label{inst:0238}
\and Department of Astrophysics, Astronomy and Mechanics, National and Kapodistrian University of Athens, Panepistimiopolis, Zografos, 15783 Athens, Greece\relax                                            \label{inst:0249}
\and Leibniz Institute for Astrophysics Potsdam (AIP), An der Sternwarte 16, 14482 Potsdam, Germany\relax                                                                                                    \label{inst:0252}
\and RHEA for ESA/ESAC, Camino bajo del Castillo, s/n, Urbanizacion Villafranca del Castillo, Villanueva de la Ca\~{n}ada, 28692 Madrid, Spain\relax                                                         \label{inst:0254}
\and ATOS for CNES Centre Spatial de Toulouse, 18 avenue Edouard Belin, 31401 Toulouse Cedex 9, France\relax                                                                                                 \label{inst:0256}
\and School of Physics and Astronomy, Tel Aviv University, Tel Aviv 6997801, Israel\relax                                                                                                                    \label{inst:0259}
\and UNINOVA - CTS, Campus FCT-UNL, Monte da Caparica, 2829-516 Caparica, Portugal\relax                                                                                                                     \label{inst:0260}
\and School of Physics, O'Brien Centre for Science North, University College Dublin, Belfield, Dublin 4, Ireland\relax                                                                                       \label{inst:0271}
\and Dipartimento di Fisica e Astronomia, Universit\`{a} di Bologna, Via Piero Gobetti 93/2, 40129 Bologna, Italy\relax                                                                                      \label{inst:0277}
\and Barcelona Supercomputing Center - Centro Nacional de Supercomputaci\'{o}n, c/ Jordi Girona 29, Ed. Nexus II, 08034 Barcelona, Spain\relax                                                               \label{inst:0287}
\and Department of Computer Science, Electrical and Space Engineering, Lule\aa{} University of Technology, Box 848, S-981 28 Kiruna, Sweden\relax                                                            \label{inst:0293}
\and Max Planck Institute for Extraterrestrial Physics, High Energy Group, Gie{\ss}enbachstra{\ss}e, 85741 Garching, Germany\relax                                                                           \label{inst:0295}
\and Astronomical Observatory Institute, Faculty of Physics, Adam Mickiewicz University, S{\l}oneczna 36, 60-286 Pozna{\'n}, Poland\relax                                                                    \label{inst:0312}
\and Konkoly Observatory, Research Centre for Astronomy and Earth Sciences, Hungarian Academy of Sciences, Konkoly Thege Mikl\'{o}s \'{u}t 15-17, 1121 Budapest, Hungary\relax                               \label{inst:0318}
\and E\"{ o}tv\"{ o}s Lor\'and University, Egyetem t\'{e}r 1-3, H-1053 Budapest, Hungary\relax                                                                                                               \label{inst:0319}
\and American Community Schools of Athens, 129 Aghias Paraskevis Ave. \& Kazantzaki Street, Halandri, 15234 Athens, Greece\relax                                                                             \label{inst:0322}
\and Faculty of Mathematics and Physics, University of Ljubljana, Jadranska ulica 19, 1000 Ljubljana, Slovenia\relax                                                                                         \label{inst:0325}
\and Villanova University, Department of Astrophysics and Planetary Science, 800 E Lancaster Avenue, Villanova PA 19085, USA\relax                                                                           \label{inst:0326}
\and Physics Department, University of Antwerp, Groenenborgerlaan 171, 2020 Antwerp, Belgium\relax                                                                                                           \label{inst:0328}
\and McWilliams Center for Cosmology, Department of Physics, Carnegie Mellon University, 5000 Forbes Avenue, Pittsburgh, PA 15213, USA\relax                                                                 \label{inst:0334}
\and Astronomical Institute, Academy of Sciences of the Czech Republic, Fri\v{c}ova 298, 25165 Ond\v{r}ejov, Czech Republic\relax                                                                            \label{inst:0338}
\and Telespazio for CNES Centre Spatial de Toulouse, 18 avenue Edouard Belin, 31401 Toulouse Cedex 9, France\relax                                                                                           \label{inst:0342}
\and Institut de Physique de Rennes, Universit{\'e} de Rennes 1, 35042 Rennes, France\relax                                                                                                                  \label{inst:0345}
\and INAF - Osservatorio Astronomico di Capodimonte, Via Moiariello 16, 80131, Napoli, Italy\relax                                                                                                           \label{inst:0346}
\and Shanghai Astronomical Observatory, Chinese Academy of Sciences, 80 Nandan Rd, 200030 Shanghai, China\relax                                                                                              \label{inst:0352}
\and School of Astronomy and Space Science, University of Chinese Academy of Sciences, Beijing 100049, China\relax                                                                                           \label{inst:0353}
\and Niels Bohr Institute, University of Copenhagen, Juliane Maries Vej 30, 2100 Copenhagen {\O}, Denmark\relax                                                                                              \label{inst:0355}
\and DXC Technology, Retortvej 8, 2500 Valby, Denmark\relax                                                                                                                                                  \label{inst:0356}
\and Las Cumbres Observatory, 6740 Cortona Drive Suite 102, Goleta, CA 93117, USA\relax                                                                                                                      \label{inst:0357}
\and Astrophysics Research Institute, Liverpool John Moores University, 146 Brownlow Hill, Liverpool L3 5RF, United Kingdom\relax                                                                            \label{inst:0365}
\and Baja Observatory of University of Szeged, Szegedi \'{u}t III/70, 6500 Baja, Hungary\relax                                                                                                               \label{inst:0370}
\and Laboratoire AIM, IRFU/Service d'Astrophysique - CEA/DSM - CNRS - Universit\'{e} Paris Diderot, B\^{a}t 709, CEA-Saclay, 91191 Gif-sur-Yvette Cedex, France\relax                                        \label{inst:0371}
\and Warsaw University Observatory, Al. Ujazdowskie 4, 00-478 Warszawa, Poland\relax                                                                                                                         \label{inst:0408}
\and Institute of Theoretical Physics, Faculty of Mathematics and Physics, Charles University in Prague, Czech Republic\relax                                                                                \label{inst:0409}
\and AKKA for CNES Centre Spatial de Toulouse, 18 avenue Edouard Belin, 31401 Toulouse Cedex 9, France\relax                                                                                                 \label{inst:0416}
\and HE Space Operations BV for ESA/ESTEC, Keplerlaan 1, 2201AZ, Noordwijk, The Netherlands\relax                                                                                                            \label{inst:0429}
\and Space Telescope Science Institute, 3700 San Martin Drive, Baltimore, MD 21218, USA\relax                                                                                                                \label{inst:0438}
\and QUASAR Science Resources for ESA/ESAC, Camino bajo del Castillo, s/n, Urbanizacion Villafranca del Castillo, Villanueva de la Ca\~{n}ada, 28692 Madrid, Spain\relax                                     \label{inst:0439}
\and Fork Research, Rua do Cruzado Osberno, Lt. 1, 9 esq., Lisboa, Portugal\relax                                                                                                                            \label{inst:0444}
\and APAVE SUDEUROPE SAS for CNES Centre Spatial de Toulouse, 18 avenue Edouard Belin, 31401 Toulouse Cedex 9, France\relax                                                                                  \label{inst:0447}
\and Nordic Optical Telescope, Rambla Jos\'{e} Ana Fern\'{a}ndez P\'{e}rez 7, 38711 Bre\~{n}a Baja, Spain\relax                                                                                              \label{inst:0452}
\and Spanish Virtual Observatory\relax                                                                                                                                                                       \label{inst:0457}
\and Fundaci\'{o}n Galileo Galilei - INAF, Rambla Jos\'{e} Ana Fern\'{a}ndez P\'{e}rez 7, 38712 Bre\~{n}a Baja, Santa Cruz de Tenerife, Spain\relax                                                          \label{inst:0468}
\and INSA for ESA/ESAC, Camino bajo del Castillo, s/n, Urbanizacion Villafranca del Castillo, Villanueva de la Ca\~{n}ada, 28692 Madrid, Spain\relax                                                         \label{inst:0473}
\and Dpto. Arquitectura de Computadores y Autom\'{a}tica, Facultad de Inform\'{a}tica, Universidad Complutense de Madrid, C/ Prof. Jos\'{e} Garc\'{i}a Santesmases s/n, 28040 Madrid, Spain\relax            \label{inst:0474}
\and H H Wills Physics Laboratory, University of Bristol, Tyndall Avenue, Bristol BS8 1TL, United Kingdom\relax                                                                                              \label{inst:0477}
\and Institut d'Estudis Espacials de Catalunya (IEEC), Gran Capita 2-4, 08034 Barcelona, Spain\relax                                                                                                         \label{inst:0483}
\and Applied Physics Department, Universidade de Vigo, 36310 Vigo, Spain\relax                                                                                                                               \label{inst:0485}
\and Stellar Astrophysics Centre, Aarhus University, Department of Physics and Astronomy, 120 Ny Munkegade, Building 1520, DK-8000 Aarhus C, Denmark\relax                                                   \label{inst:0500}
\and Argelander-Institut f\"{ ur} Astronomie, Universit\"{ a}t Bonn,  Auf dem H\"{ u}gel 71, 53121 Bonn, Germany\relax                                                                                       \label{inst:0506}
\and Research School of Astronomy and Astrophysics, Australian National University, Canberra, ACT 2611 Australia\relax                                                                                       \label{inst:0513}
\and Sorbonne Universit\'{e}s, UPMC Univ. Paris 6 et CNRS, UMR 7095, Institut d'Astrophysique de Paris, 98 bis bd. Arago, 75014 Paris, France\relax                                                          \label{inst:0515}
\and Department of Geosciences, Tel Aviv University, Tel Aviv 6997801, Israel\relax                                                                                                                          \label{inst:0517}
}

   \date{ }

\abstract
  % context heading (optional)
  {The second release of {\gaia} data ({\gdrtwo}) contains the astrometric parameters for more than half a million
quasars. This set defines a kinematically non-rotating reference frame in the optical domain. A subset of these quasars
have accurate VLBI positions that allow the axes of the reference frame to be aligned with the International Celestial Reference System (ICRF) radio frame.}  
  % aims heading (mandatory) 
  {We describe the astrometric and photometric properties of the quasars that were selected to represent{} 
the celestial reference frame of {\gdrtwo\ (\textit{Gaia}-CRF2)}, and to compare the optical and radio positions for sources 
with accurate VLBI positions.}
  % methods heading (mandatory)
  {Descriptive statistics are used to characterise the overall properties of the quasar sample. Residual rotation
  and orientation errors and large-scale systematics are quantified by means of expansions in vector spherical 
  harmonics. Positional differences are calculated relative to a prototype version of the forthcoming ICRF3.}  
  % results heading (mandatory)
  {{\gcrftwo} consists of the positions of a sample of 556\,869 sources in {\gdrtwo}, obtained from a 
positional cross-match with the {\icrf} and AllWISE AGN catalogues. The sample constitutes a clean, dense, and 
homogeneous set of extragalactic point sources in the magnitude range $G\simeq 16$ to 21~mag with
accurately known optical positions. The median positional uncertainty is 0.12~mas for $G<18$~mag and 
0.5~mas at $G=20$~mag. 
Large-scale systematics are estimated to be in the range 20 to 30~{\muas}. The accuracy claims are supported by 
the parallaxes and proper motions of the quasars in {\gdrtwo}. The optical positions for a
subset of 2820 sources in common with the {\icrf} show very good overall agreement with the radio positions, but several tens of sources have significantly discrepant positions.
  }  
  % conclusions heading (optional), leave it emp6ty if necessary 
  {Based on less than 40\% of the data expected from the nominal {\gaia} mission, {\gcrftwo} is the first 
realisation of a non-rotating global optical reference frame that meets the ICRS prescriptions, meaning that it is\ built only 
on extragalactic sources. Its accuracy matches the current radio frame of the ICRF, but the density of sources in all parts of the sky is much higher, except along the Galactic equator.}

   
%   \titlerunning{\textit{Gaia} reference frame and QSO astrometry} 
%   \authorrunning{F.~Mignard et al.}

   \maketitle

%
%________________________________________________________________

\section{Introduction}
One of the key science objectives of the European Space Agency's {\gaia}  mission
\citepads{2016A&A...595A...1G} is to build a rotation-free celestial reference frame 
in the visible wavelengths. This reference frame, which may be called the 
{\gaia}  Celestial Reference Frame ({\gcrf}), should meet the 
specifications of the International Celestial Reference System (ICRS; \citeads{1995A&A...303..604A}) 
in that its axes are fixed with respect to distant extragalactic objects, that is, to\ quasars. 
For continuity with existing reference frames and consistency across the electromagnetic 
spectrum, the orientation of the axes should moreover coincide with the 
International Celestial Reference Frame (ICRF; \citeads{1998AJ....116..516M}) 
that is established in the radio domain by means of VLBI observations of selected quasars.

The second release of data from  {\gaia}  ({\gdrtwo}; \citeads{2018Gaia36}) provides 
complete astrometric data (positions, parallaxes, and proper motions)  for more than 
550\,000 quasars. In the astrometric solution for {\gdrtwo} \citepads{2018Gaia51}, 
subsets of these objects were used to avoid the rotation and align the axes with a
prototype version of the forthcoming third realisation of the ICRF.
\footnote{ ``Rotation'' here refers exclusively to the kinematical rotation
of the spatial axes of the barycentric celestial reference system (BCRS), as used in
the {\gaia} catalogue, with respect to distant extragalactic objects (see e.g. \citeads{1998A&A...334.1123K}).
Similarly, ``orientation'' refers to the (non-)  alignment of the axes at the reference
epoch J2015.5.}
The purpose of this paper
is to characterise the resulting reference frame, {\gcrftwo}, by analysing 
the astrometric and photometric properties of quasars that are
identified in {\gdrtwo} from a positional
cross-match with existing catalogues, including the {\icrf}.

{\gcrftwo}  is the first optical realisation of a reference frame at sub-milliarcsecond (mas) precision,
using a large number of extragalactic objects. With a mean density of more than ten quasars per
square degree, it represents a more than 100-fold increase in the number of objects from the 
current realisation at radio wavelengths, the ICRF2 \citepads{2015AJ....150...58F}. The {\gcrftwo} is bound to replace the HCRF (\textsc{Hipparcos} Celestial Reference Frame) 
as the most accurate representation of the ICRS at optical wavelengths until the next 
release of {\gaia}  data. While the positions of the generally faint quasars constitute the primary realisation of
{\gcrftwo}, the positions and proper motions of the $\simeq\,$1.3~billion stars in {\gdrtwo} are
nominally in the same reference frame and thus provide a secondary realisation that covers
the magnitude range $G\simeq 6$ to 21~mag at similar precisions, which degrades with increasing distance from the reference epoch J2015.5. The properties of the stellar reference frame of{\gdrtwo} are not discussed here.

%The origin of the frame is at the barycentre of the solar system and the orientation of its axes is kinematically non-rotating with respect to the most distant sources in the universe. The  {\gcrf} ({\gaia} Celestial Reference Frame)  has been aligned  to the planned ICRF3 orientation using a set of optical and radio loud quasars found in both catalogues, while the space-fixed orientation is constrained by imposing that the proper motion vector field of the quasars has no global rotation.  

This paper explains in Sect.~\ref{sect:gaiaqsos} the selection
of the {\gaia} sources from which we built the reference frame. Section~\ref{sect:gaicrf} presents statistics summarising 
the overall properties of the reference frame in terms of the spatial distribution, accuracy, and magnitude 
distribution of the sources. The parallax and proper motion distributions are used as additional quality 
indicators and strengthen the confidence in the overall quality of the product. In Sect.~\ref{sect:gaiaicrf3} 
the optical positions in {\gdrtwo} are compared with the VLBI frame realised in the {\icrf}.
A brief discussion of other quasars in the data release (Sect.~\ref{sect:otherqsos}) is followed by 
the conclusions in Sect.~\ref{sec:concl}.

% for the conclusions:
%no systematic difference, but several sources exhibit  positional differences well above the normal statistical deviations and must be examined on a case by case basis with VLBI ICRF experts.

\section{Construction of {\gcrftwo}}\label{sect:gaiaqsos}

\subsection{Principles}\label{sect:principles}
Starting with {\gdrtwo,} the astrometric processing of the {\gaia}  data provides the parallax and the 
two proper motion components for most of the sources, in addition to the positions (see \citeads{2018Gaia51}). 
As a consequence of the {\gaia}  observing principle, the spin of the global reference frame 
must be constrained in some way in order to deliver stellar proper motions in a non-rotating frame. 
Less relevant for the underlying physics, but of great practical importance, is that the orientation of the 
resulting {\gaia}  frame should coincide with the current best realisation of the 
ICRS in the radio domain as well as possible, as implemented by the ICRF2 and soon by the ICRF3.

These two objectives were achieved in the course of the iterated astrometric solution by analysing the provisional 
positions and proper motions of a pre-defined set of sources, and by adjusting the source and attitude
parameters accordingly by means of the so-called frame rotator \citep{2012A&A...538A..78L}.
Two types of sources were used for this purpose: a few thousand sources identified as the optical
counterparts of ICRF sources were used to align the positions with the radio frame, and a much larger
set of probable quasars found by a cross-match with existing quasar catalogues were used,
together with the ICRF sources, to ensure that the set of quasar proper motions was globally non-rotating.
The resulting solution is then a physical realisation of the {\gaia} frame that is rotationally stabilised on the quasars.
The detailed procedure used for {\gdrtwo} is described in \citetads{2018Gaia51}.

\begin{figure}[ht]
\centering
      \resizebox{0.95\hsize}{!}{ \includegraphics{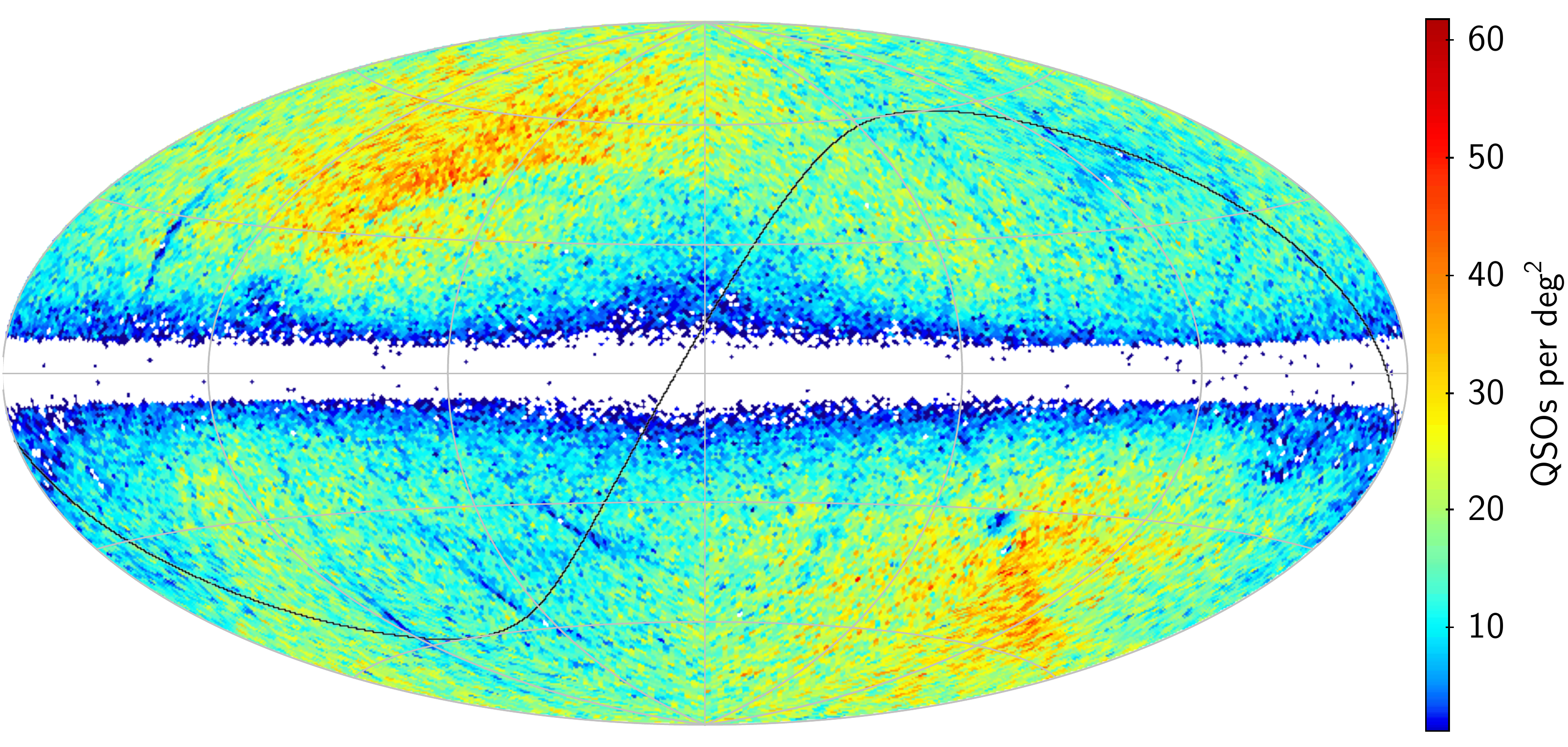}}     
        \caption{Sky density per square degree for the quasars of {\gcrftwo}  on an equal-area 
        Hammer--Aitoff projection in Galactic coordinates. The Galactic centre is at the origin of coordinates 
        (centre of the map), Galactic north is up, and Galactic longitude increases from right to left.
        The solid black line shows the ecliptic.
        The higher density areas are located around the ecliptic poles.
        \label{qso_density}}
\end{figure}

\subsection{Selection of quasars}\label{subsec:qsosel}
Although {\gaia}  is meant to be autonomous in terms of the recognition of quasars from their photometric 
properties (colours, variability), this functionality was not yet implemented for the first few releases. 
Therefore the sources that are currently identified as quasars are known objects drawn from available catalogues and 
cross-matched to {\gaia}  sources by retaining the nearest positional match. In {\gdrone,} 
quasars where flagged from a compilation made before the mission \citepads{2014jsrs.conf...84A},
and a subset of ICRF2 was used for the alignment. The heterogeneous spatial distribution of this 
compilation did not greatly affect the reference frame of {\gdrone} because of the special procedures 
that were used to link it to the HCRF \citepads{2016A&A...595A...4L,2016A&A...595A...5M}. 

For {\gdrtwo}, which is the first release that is completely independent of the earlier \textsc{Hipparcos}
and \textit{Tycho} catalogues, it was desirable to use the most recent VLBI positions for the 
orientation of the reference frame, and a large, homogeneous set of quasars for the rotation.
The {\gaia} data were therefore cross-matched with two different sets of known quasars:
\begin{itemize}
\item 
A prototype of the upcoming ICRF3, based on the VLBI solution of the  Goddard Space Flight Center (GSFC) that comprises 4262 
radio-loud quasars that are observed in the X (8.5~GHz) and S (2.3~GHz) bands. This catalogue,
referred to here as the {\icrf}, was kindly provided to the {\gaia} team  by 
the IAU Working Group on ICRF3 (see Sect.~\ref{sect:gaiaicrf3}) more than a year in advance of the scheduled release of the ICRF3.
The positional accuracy is comparable to that of {\gaia,} and this set was used to align the reference
frame of {\gdrtwo} to the radio frame.
\item  
The all-sky sample of 1.4~million active galactic nuclei (AGNs) of \citetads{2015ApJS..221...12S}, 
referred to below as the AllWISE AGN catalogue (AW in labels and captions). This catalogue 
resulted from observations by the Wide-field Infrared Survey Explorer (WISE; \citeads{2010AJ....140.1868W})
that operates in the mid-IR at 3.4, 4.6, 12, and 22~$\mu$m wavelength. The AllWISE AGN catalogue has a 
relatively homogenous sky coverage, except for the Galactic plane, where the coverage is less extensive because of 
Galactic extinction and confusion by stars, and at the ecliptic poles, which have a higher density because of the WISE 
scanning law. The sources are classified as AGNs from a two-colour infrared photometric criterion, 
and \citetads{2015ApJS..221...12S} estimated that the probability of stellar contamination is 
$\le 4.0\times 10^{-5}$ per source.  About half of the AllWISE AGN sources have an optical 
counterpart that is detected at least once by {\gaia} in its first two years. 
\end{itemize}
Cross-matching the two catalogues with a provisional {\gaia} solution and applying some filters 
based on the {\gaia} astrometry (see Sect.~5.1, Eq.13 in \citeads{2018Gaia51}) resulted in a list of 492\,007 
putative quasars, including 2844 {\icrf} objects. The filters select sources with good observation records, a parallax formal uncertainty $< 1$ mas,  a reliable level of significance in parallax and proper motion, and they avoid the Galactic plane by imposing $|\sin b| > 0.1$.
These sources were used by the frame rotator, 
as explained above, when calculating the final solution; in {\gdrtwo,} they are identified by means 
of the flag \texttt{frame\_rotator\_object\_type}. 
This subset of (presumed) quasars cannot, however, be regarded as a proper representation of {\gcrftwo} 
because of the provisional nature of the solution used for the cross-matching 
and the relatively coarse selection criteria. Several of these sources were indeed later found 
to be Galactic stars. 

A new selection of quasars was therefore made after {\gdrtwo}
was completed. This selection took advantage 
of the higher astrometric accuracy of \textit{Gaia} DR2 and applied better selection criteria that are detailed in 
Sect.~5.2, Eq. 14,  of \citetads{2018Gaia51}. In particular, this updated selection takes the  parallax zeropoint into account.
This resulted in a set of 555\,934 {\gdrtwo} sources that are
matched to the AllWISE AGN catalogue
and 2820 sources that are matched to the {\icrf}. The union of the two sets contains 556\,869 {\gdrtwo} 
sources. These sources and their positions in {\gdrtwo} are a
version of the {\gcrf} 
that we call {\gcrftwo}. 

The entire subsequent analysis in this paper (except in Sect.~\ref{sect:otherqsos}) is based on this sample
or on subsets of it.
For simplicity, we use the term quasar (QSO) for these objects, although other classifications 
(BL Lac object, Seyfert~1, etc.) may be more appropriate in many cases, and a very small number of them
may be distant ($> 1$~kpc) Galactic stars.

\begin{figure}[ht]
\centering
      \resizebox{0.85\hsize}{!}{ \includegraphics{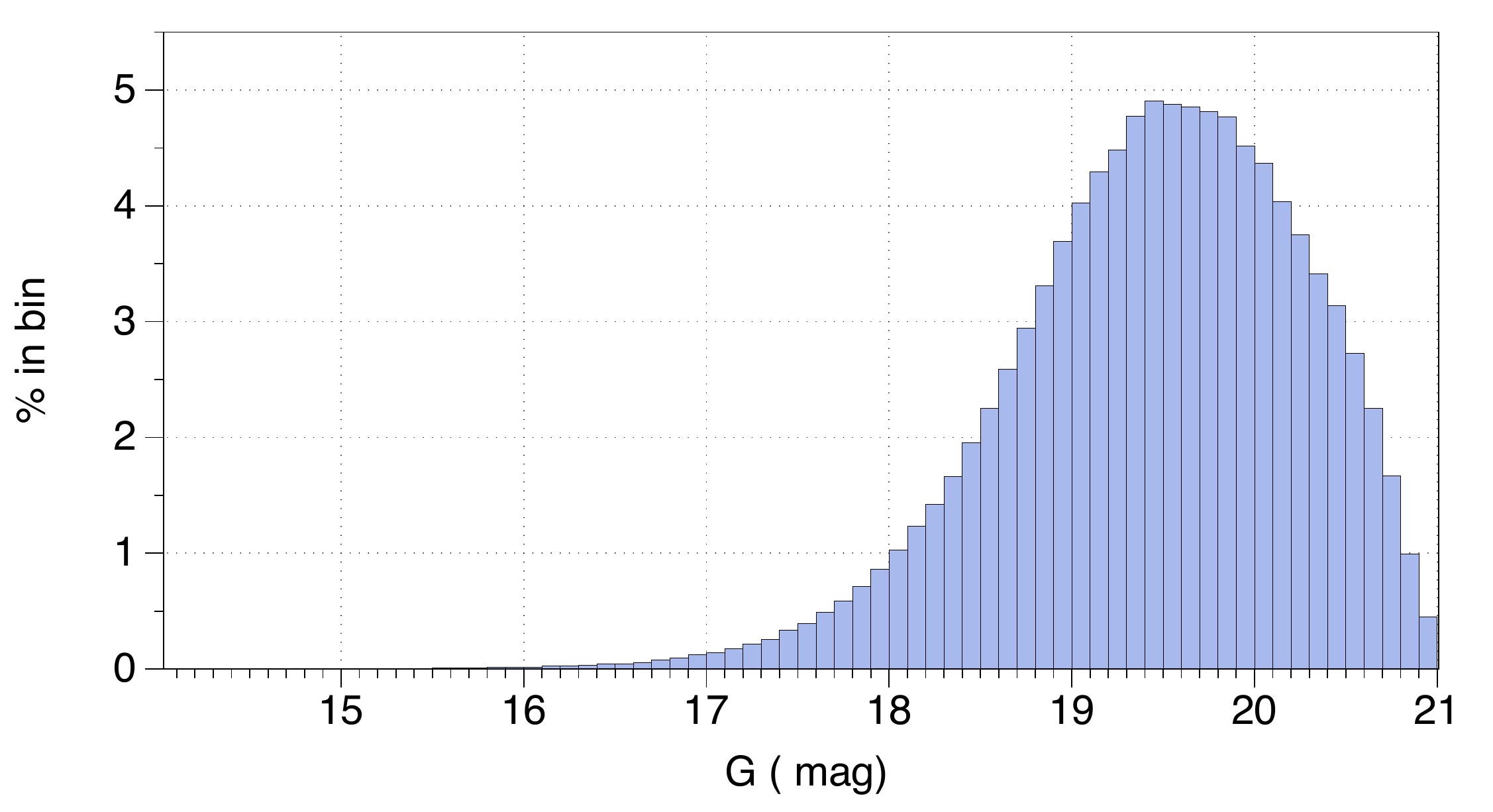}}
      \resizebox{0.85\hsize}{!}{ \includegraphics{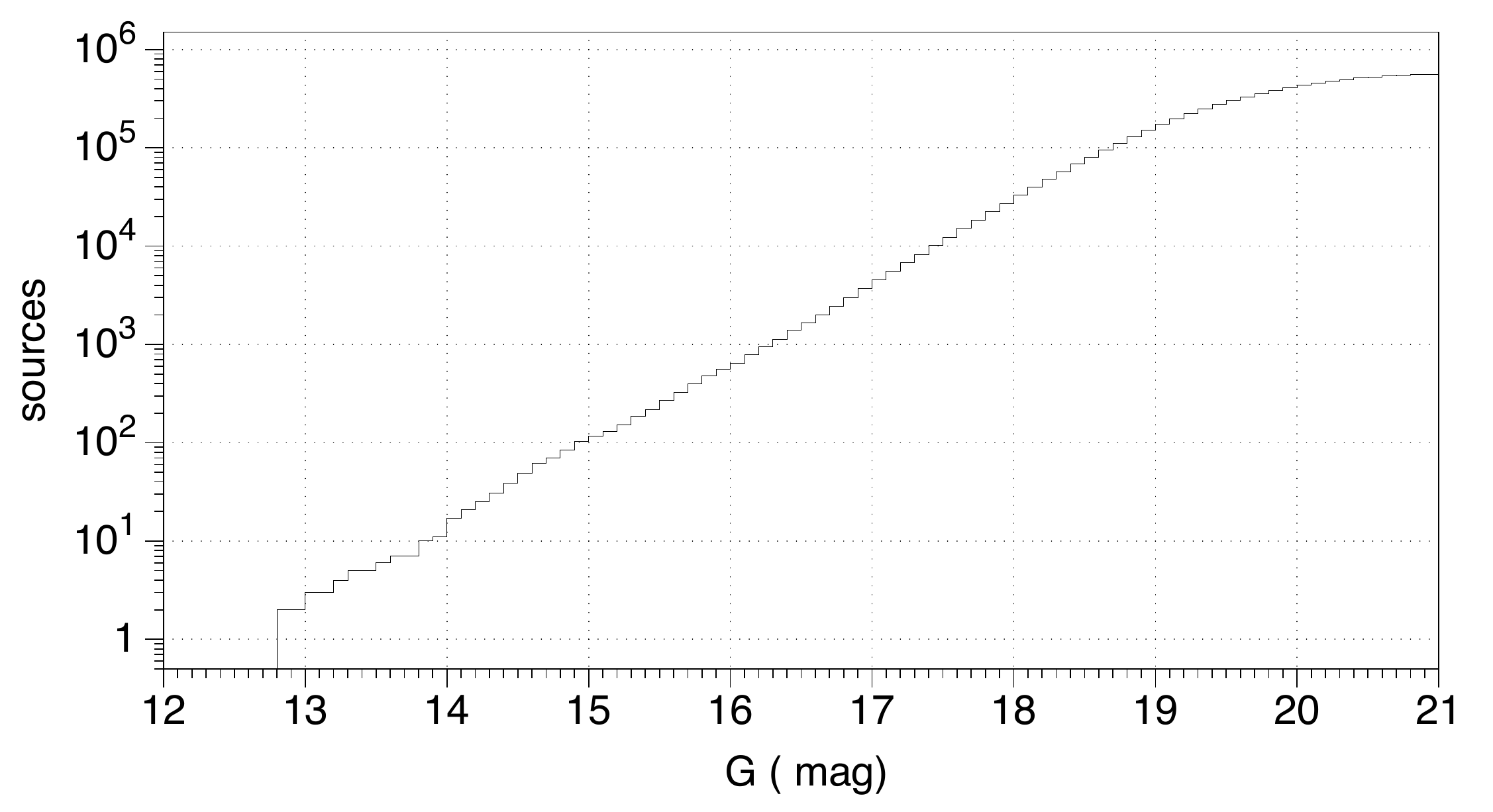} }      
        \caption{$G$ -magnitude distribution of the {\gcrftwo} quasars. 
        Percentage per bin of 0.1~mag (top) and cumulative distribution (bottom).
        \label{mag_G_distribution}}
\end{figure}

\section{Properties of {\gcrftwo}}\label{sect:gaicrf}

This section describes the overall astrometric and photometric properties in {\gdrtwo} of the 
sources of the {\gcrftwo}, that is,\ the 556\,869 quasars we obtained from the match to 
the AllWISE AGN catalogue and the {\icrf}.
Their sky density is displayed in \figref{qso_density}. The Galactic plane area is filtered out 
by the AllWISE AGN selection criteria, while areas around the ecliptic poles are higher than the average 
density, as noted above.  Lower density arcs from the WISE survey
are also visible, but as a rule, the whole-sky coverage outside the Galactic plane has an 
average density of about 14 sources per deg$^2$.
The few sources in the Galactic plane area come from the \icrf.

\subsection{Magnitude and colour}

Figure~\ref{mag_G_distribution} shows the magnitude distribution of the {\gcrftwo} sources. 
In rounded numbers, there are 27\,000 sources with $G < 18$~mag,  150\,000 with $G < 19$~mag, 
and 400\,000 with $G < 20$~mag. The average density of one source per square degree is reached at 
$G =18.2$~mag, where it is likely that the sample is nearly complete outside the Galactic plane. 

\begin{figure}[ht]
\centering
      \resizebox{0.85\hsize}{!}{ \includegraphics{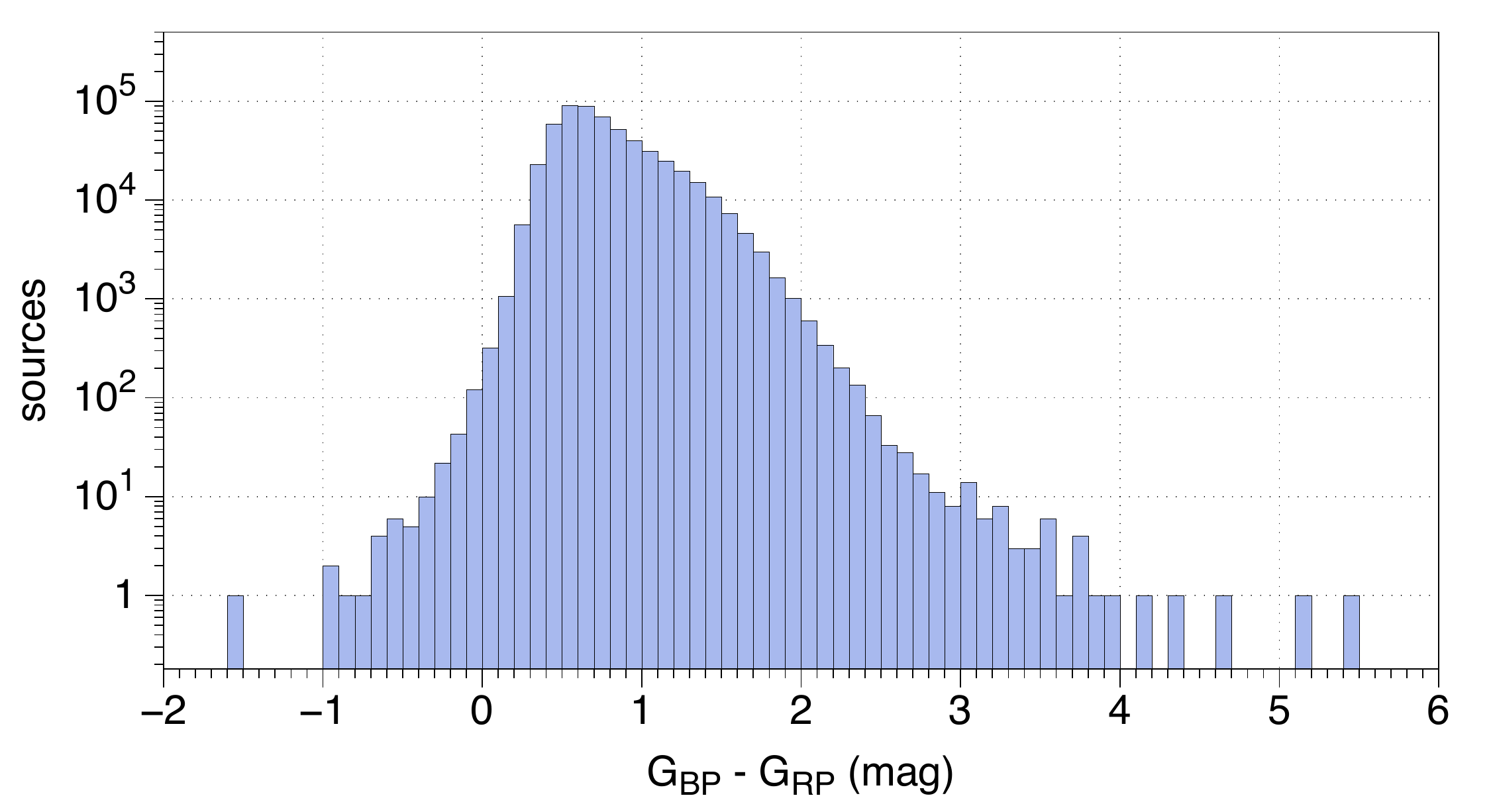}}     
        \caption{Colour distribution  of the {\gcrftwo} quasars (log scale for the
        number of sources per bin of 0.1 mag).
        \label{colour_distribution}}
\end{figure}

Figure~\ref{colour_distribution} shows the distribution in colour index $G_\text{BP}-G_\text{RP}$ 
(for the definition of the blue and red passbands, BP and RP, see \citeads{2018Gaia44}).
Of the sources, 2154 (0.4\%) have no colour index $G_\text{BP}-G_\text{RP}$ in {\gdrtwo}.  The distribution
is rather narrow with a median of 0.71~mag and only 1\% of the sources bluer than 0.28~mag or 
redder than 1.75~mag.

\begin{figure}[ht]
\centering
      \resizebox{0.95\hsize}{!}{ \includegraphics{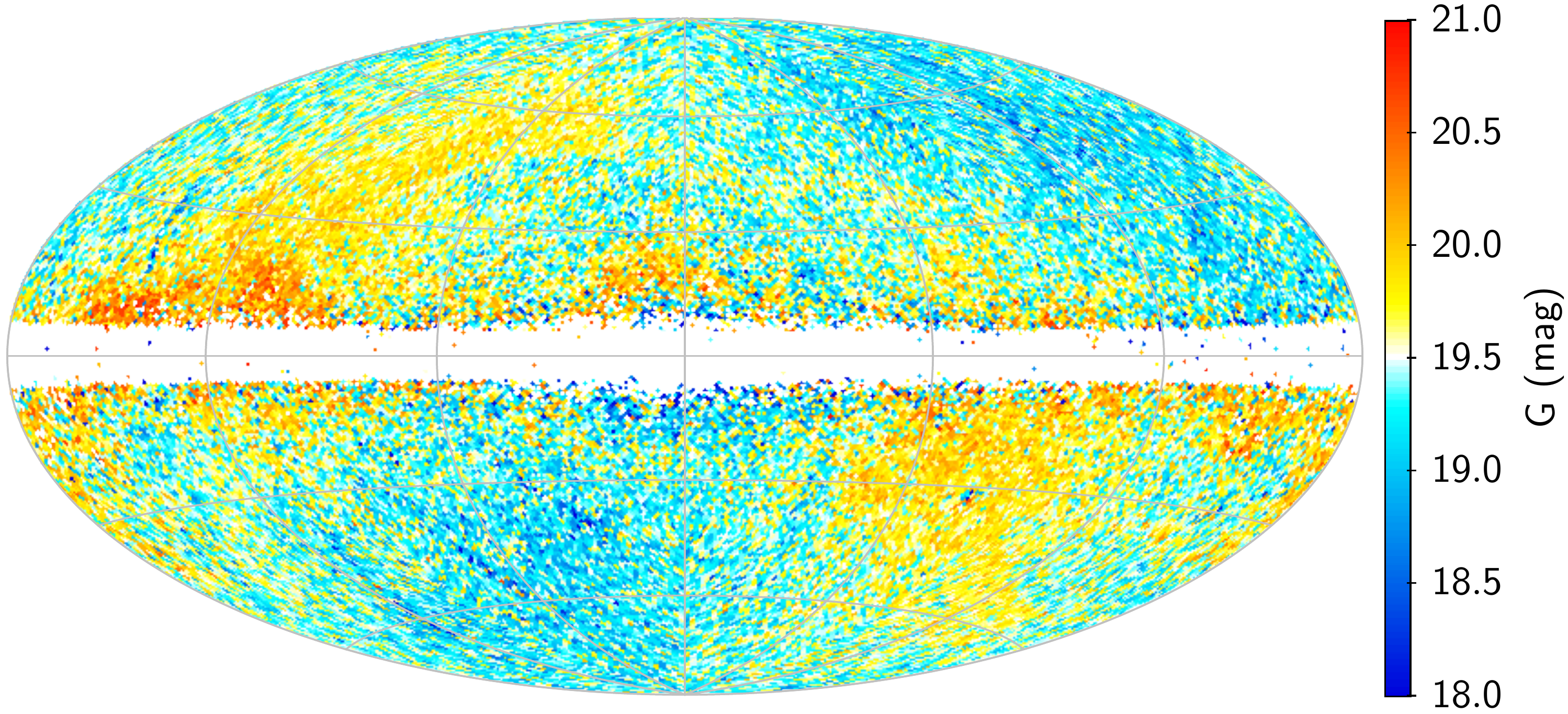}}     
      \caption{Sky distribution of the {\gcrftwo} source magnitudes.
        This map shows the median values of  the $G$ magnitude in cells of approximately 0.84 ${\rm deg}^2$ using an equal-area 
        Hammer--Aitoff projection in Galactic coordinates. The Galactic centre is at the origin of coordinates 
        (centre of the map), Galactic north is up, and Galactic longitude increases from right to left.
        \label{skymap_Gmag}}
\end{figure}

The magnitude  is not evenly distributed on the sky, as shown in \figref{skymap_Gmag}, with on 
the average fainter sources around the ecliptic poles, where the highest densities are found as well
(\figref{qso_density}). These two features result from a combination of the deeper survey of AllWISE in 
these areas and the more frequent {\gaia} observations from the scanning law.

\subsection{Astrometric quality}\label{subsect:astromqual}
In this section we describe the astrometric quality of the {\gcrftwo} quasars based on the formal 
positional uncertainties and on the distribution of observed parallaxes and proper motions, 
which are not expected to be measurable by {\gaia} at the level of individual sources. We defer a direct comparison of the Gaia
positions with VLBI astrometry to Sect.~\ref{sect:gaiaicrf3}.

\begin{figure}[ht]
\centering
      \resizebox{0.9\hsize}{!}{ \includegraphics{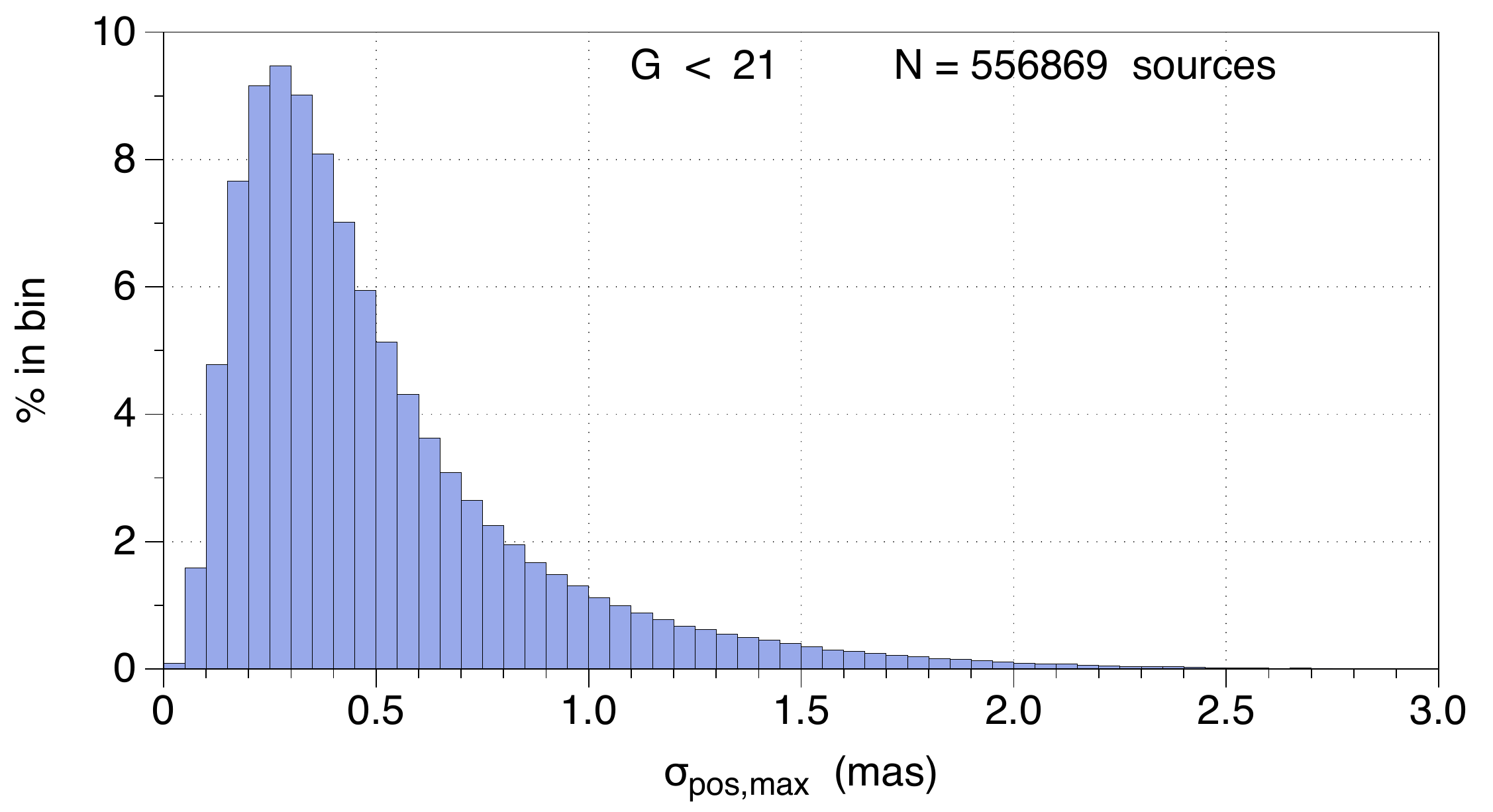}}     
      \resizebox{0.9\hsize}{!}{ \includegraphics{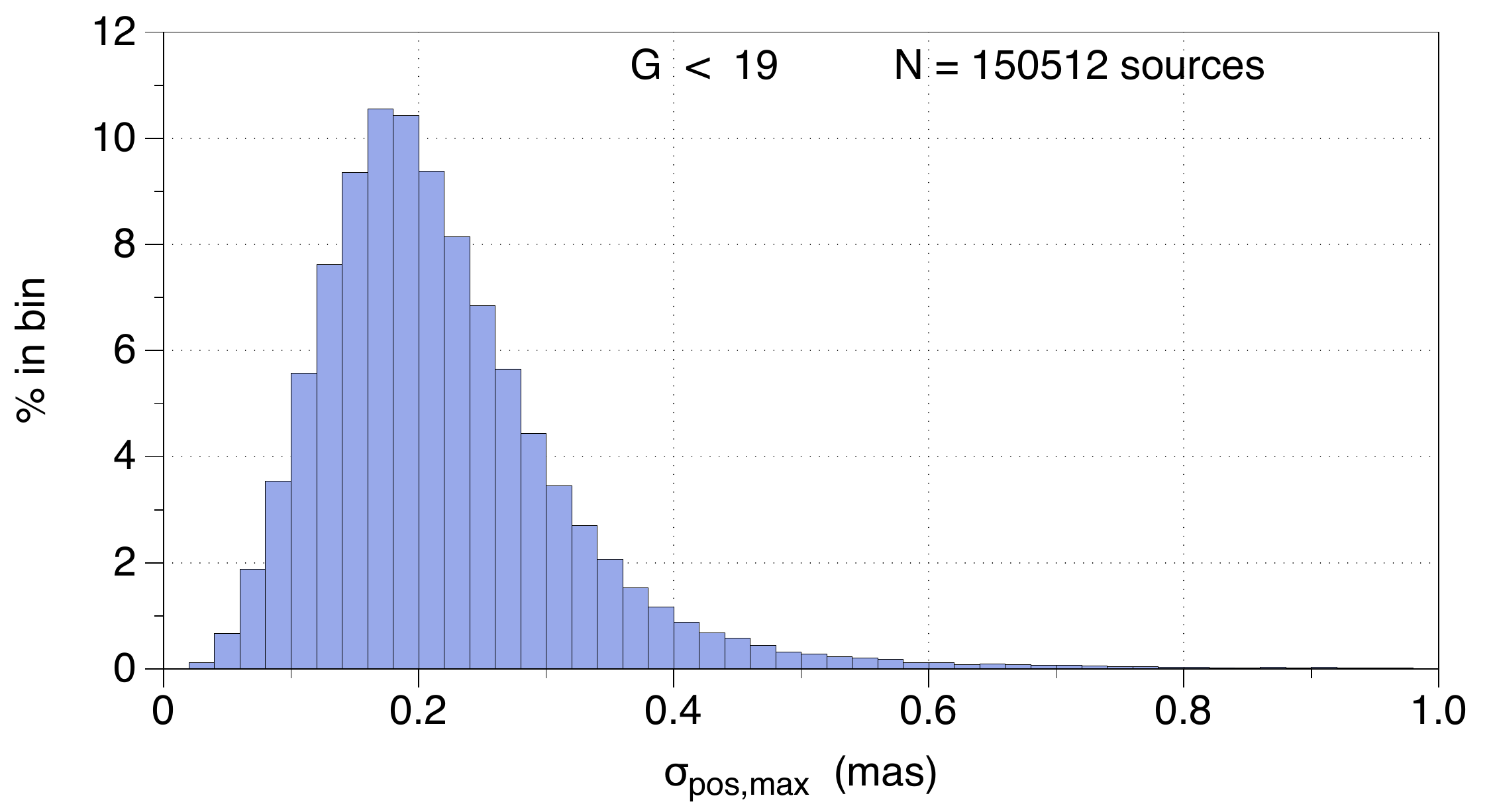}}     
        \caption{Distribution of positional uncertainties $\sigma_\text{pos,max}$ for the {\gcrftwo} quasars: all sources (top) and $G<19$~mag (bottom). 
        \label{sigpos_max}}
\end{figure}

\begin{figure}[ht]
\centering
      \resizebox{0.90\hsize}{!}{ \includegraphics{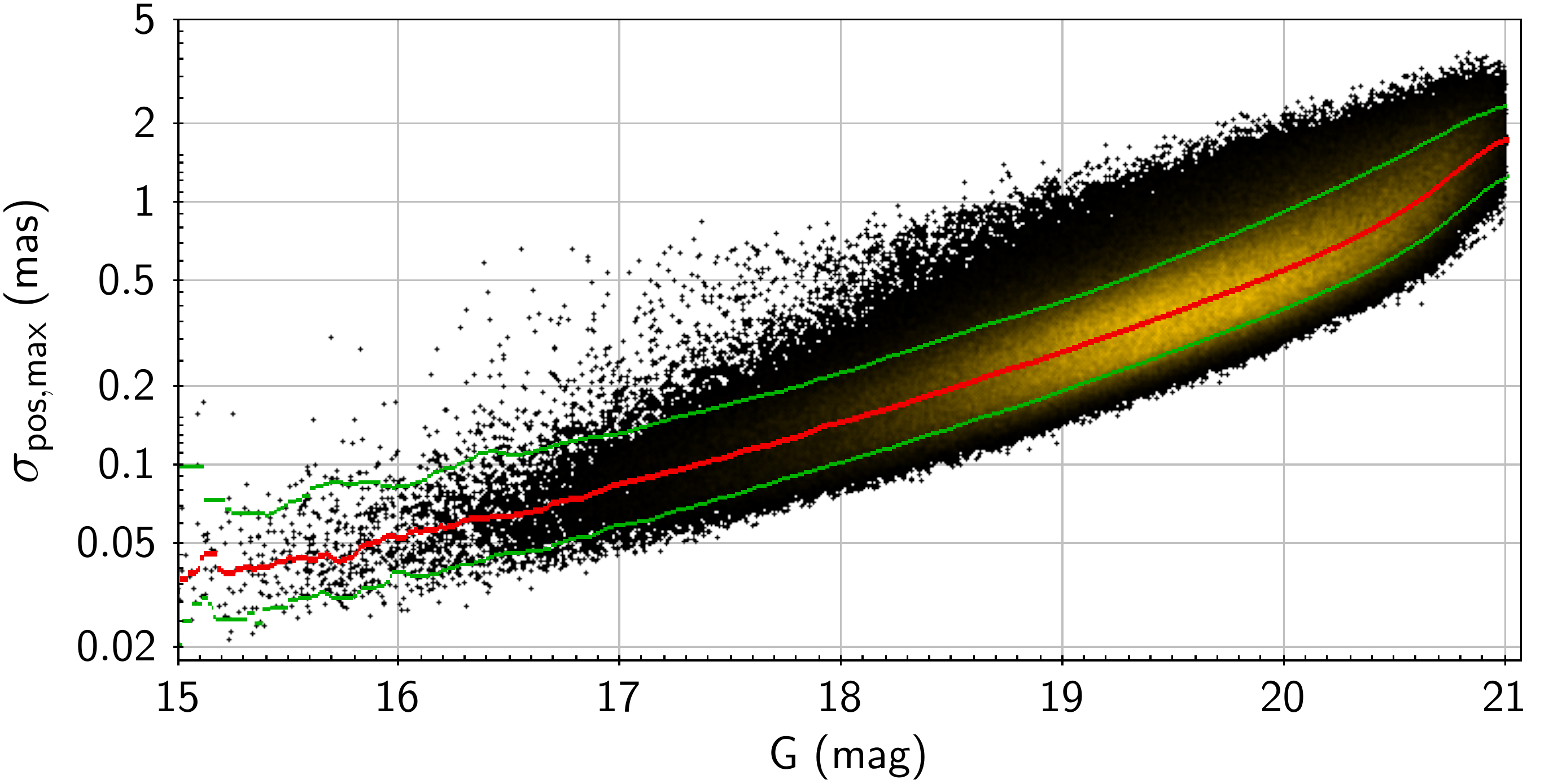}}          
        \caption{Positional uncertainties $\sigma_\text{pos,max}$ for the sources
        in the {\gcrftwo} as function of the $G$ magnitude. 
        The red solid line is the running median, and the two green lines are the first and ninth decile.
        \label{sigpos_max_G}}
\end{figure}

\begin{figure}[h]
\centering
      \resizebox{1.00\hsize}{!}{ \includegraphics{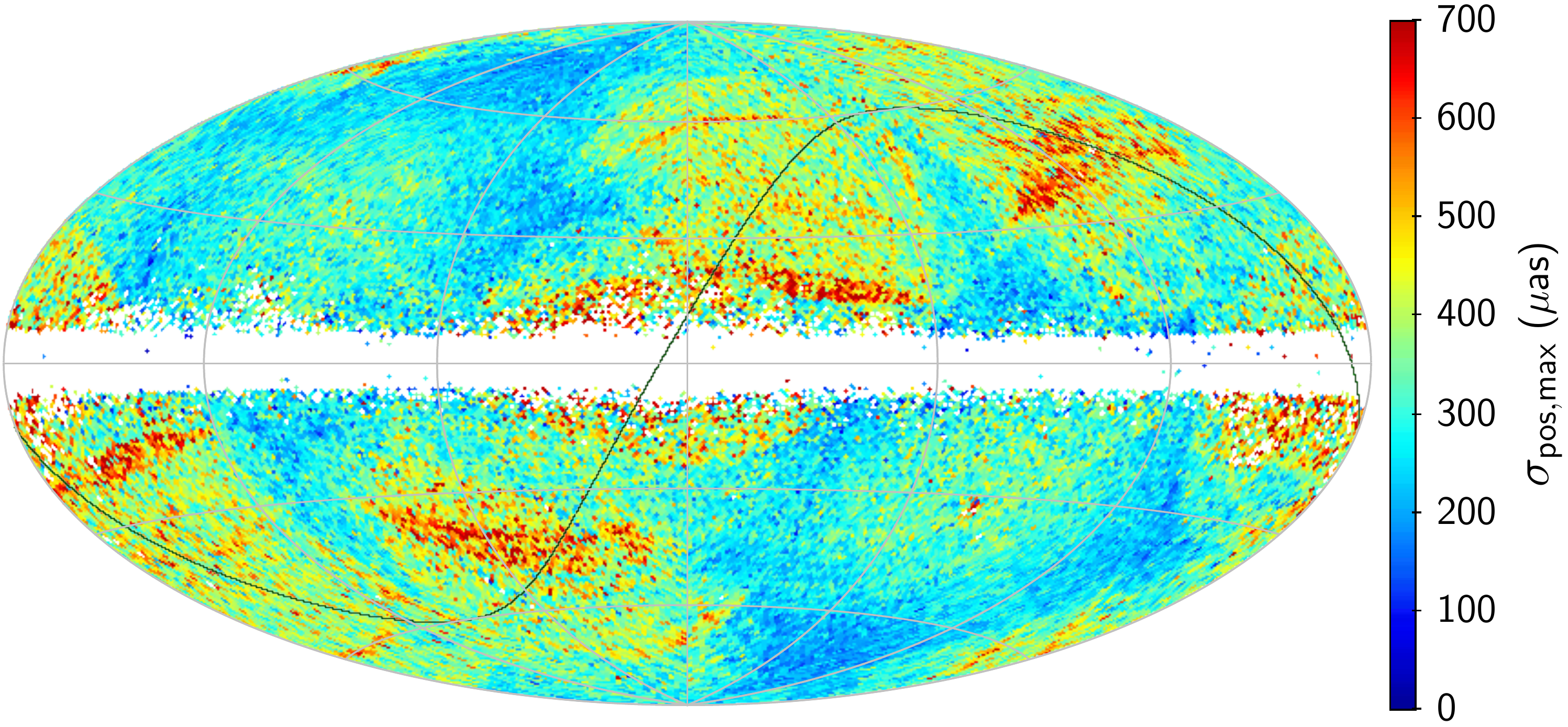}}      
        \caption{Spatial distribution of the formal position uncertainty in Eq.~(\ref{eq:sigpos}) for the 
        407\,959 sources of the {\gcrftwo} with $G<20$. The map shows the median value in each cell
        of approximately $0.84 \deg^2$, using a Hammer--Aitoff projection in Galactic coordinates with 
        zero longitude at the centre and increasing longitude from right to left. The solid black line shows the ecliptic.
        \label{skymap_sigpos}}
\end{figure}

\begin{table}[ht]
        \small
        \caption{Positional uncertainty $\sigma_\text{pos,max}$ of the {\gcrftwo}  
      quasars.\label{table:accuracy}}       
        \begin{tabular}{cccccc}
        \hline\hline
        \noalign{\smallskip}
        $G$ selection  & $N$ & 1st quartile & median  & 3rd quartile \\
         {[mag]}           &     & [mas] & [mas]  & [mas]   \\
        \noalign{\smallskip}\hline\noalign{\smallskip}         
         $< 18.0$        & $\phantom{1}$27\,275   &  0.09  & 0.12 & 0.15  \\
         $[18.0-19.0[$  & 123\,237  &  0.17  & 0.22 & 0.28  \\
         $[19.0-19.5[$   & 125\,029  &  0.27  & 0.33 & 0.41  \\
         $[19.5-20.0[$  & 132\,418  &  0.38  & 0.47 & 0.59  \\
         $\ge20.0$       & 148\,910  &  0.61  & 0.81 & 1.12  \\
        \noalign{\smallskip}\hline\noalign{\smallskip}         
           all            & 556\,869  &  0.26  & 0.40 & 0.64  \\
        \noalign{\smallskip}\hline
        \end{tabular}
\end{table}

\subsubsection{Formal uncertainty in position}
As a single number characterising the positional uncertainty of a source, $\sigma_\text{pos,max}$, 
we take the semi-major axis of the dispersion ellipse, computed from a combination 
$\sigma_{\alpha*}=\sigma_\alpha\cos\delta$, $\sigma_\delta$, and the correlation coefficient 
$\varrho_{\alpha,\delta}$:
\begin{equation}\label{eq:sigpos}
\sigma_\text{pos,max}^2 =  
\frac{1}{2}\left(\sigma_{\alpha*}^2 +\sigma_\delta^2+
\sqrt{(\sigma_{\alpha*}^2 -\sigma_\delta^2)^2+
(2\sigma_{\alpha*}\sigma_\delta\,\varrho_{\alpha,\delta})^2}\,\right)
%\frac{\sigma_{\alpha*}^2 +\sigma_\delta^2 + \left( (\sigma_{\alpha*}^2 -\sigma_\delta^2 )^2 
%+\left(2\sigma_{\alpha*} \sigma_\delta\, \varrho_{\alpha, \delta}\right)^2\right)^{1/2} }{2}
.\end{equation}
Because this is also the highest eigenvalue of the $2\times 2$ covariance matrix, it is invariant to a change 
of coordinates. These are formal uncertainties (see Sect.~\ref{subsec:plxpm} for a discussion of how real they are) for the reference epoch J2015.5 of {\gdrtwo}. % does the epoch matter?

The results for the whole sample of {\gcrftwo} quasars and the subset with $G<19$ 
are shown in \figref{sigpos_max}. The median accuracy is 0.40~mas for the full set and 0.20~mas 
for the brighter subset. Additional statistics are given in \tabref{table:accuracy}. 
%An important feature of the {\gaia} results is the relatively small scatter in the accuracy, compared 
%with for example the {\icrf} (see also below, \figref{ICRF3_rho_sigma}). 

The main factors governing the positional accuracy are the magnitude (\figref{sigpos_max_G})
and location on the sky (\figref{skymap_sigpos}). 
The larger-than-average uncertainty along the ecliptic in \figref{skymap_sigpos} is conspicuous;
this is a signature of the {\gaia} scanning law. This feature will also be present in future releases of {\gaia} 
astrometry and will remain an important characteristic of the \gcrf.

%The final plot of this section in \figref{rho_AW_Gaia} shows the differences in position between the  {\qsos} 
%as provided by the AllWISE survey and the final sub-mas positions of {\gaia}. The plot is by itself not relevant 
%for the properties of the reference frame, but illustrates dramatically the $\sim  200$-fold  improvement 
%in astrometry of faint sources achieved with {\gaia} and provides a good example of the benefit one draws 
%for the synergy between a photometric survey allowing to identify the quasars and a space mission like 
%{\gaia} able to locate them with an exquisite accuracy. 

%\begin{figure}[ht]
%\centering
%      \resizebox{0.8\hsize}{!}{ \includegraphics{Figures/rho_AW_Gaia.pdf}}          
%        \caption{Positional difference in mas between the AllWISE and Gaia positions for the $560,000$  {\qsos} used to materialise the {\gcrftwo}.
%        \label{rho_AW_Gaia}}
%\end{figure}

\subsubsection{Parallaxes and proper motions}\label{subsec:plxpm}
\begin{figure}[ht]
\centering
      \resizebox{0.9\hsize}{!}{ \includegraphics{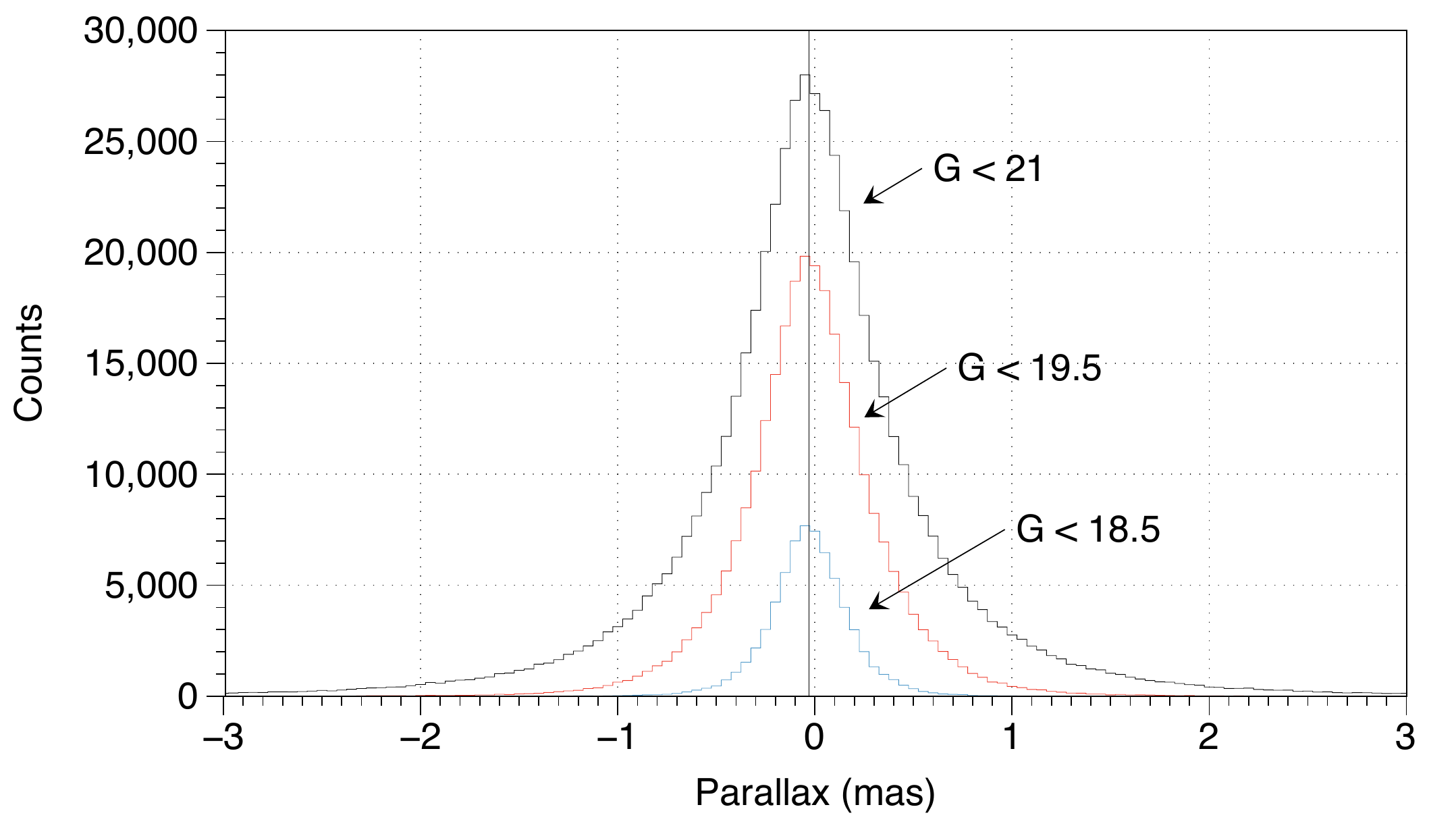}}         
        \caption{Distribution of parallaxes in the {\gaia} archive for the {\gcrftwo} quasars, subdivided by the
        maximum magnitude. The line at  $\varpi=-0.029$~mas shows the global zeropoint offset.
        \label{varpi_CRF}}
\end{figure}

Parallaxes and proper motions are nominally zero for the quasars that were selected for the reference frame
(we neglect here the expected global pattern from the Galactic acceleration, which is expected to have an amplitude of 4.5~{\muasyr}, see Sect.~\ref{subsec:vshqsos}). 
Their statistics are useful as complementary indicators of the global quality of the frame and support the accuracy claim. Here we consider the global statistics before investigating possible
systematics in Sect.~\ref{subsec:vshqsos}. Figure~\ref{varpi_CRF} shows the distribution of the parallaxes for 
different magnitude-limited subsets. As explained in \citet{2018Gaia51}, the {\gdrtwo} parallaxes have 
a global zeropoint 
error of $-0.029$~mas, which is not corrected for in the data available in the {\gaia} archive. This feature is well visible for the quasar sample and is a real instrumental effect that
is not yet eliminated by the calibration
models. Fortunately, the offset is similar for the different subsets. The shape of the distributions (best 
visible in the full set) is clearly non-Gaussian because of the mixture of normal distributions with a large 
spread in standard deviation, which is primarily linked to the source magnitude. The typical half-widths
of the distributions (0.4, 0.3, and 0.2~mas) are of a similar size as the median positional
uncertainties in \tabref{table:accuracy}. 

\begin{figure}[ht]
\centering
      \resizebox{0.85\hsize}{!}{ \includegraphics{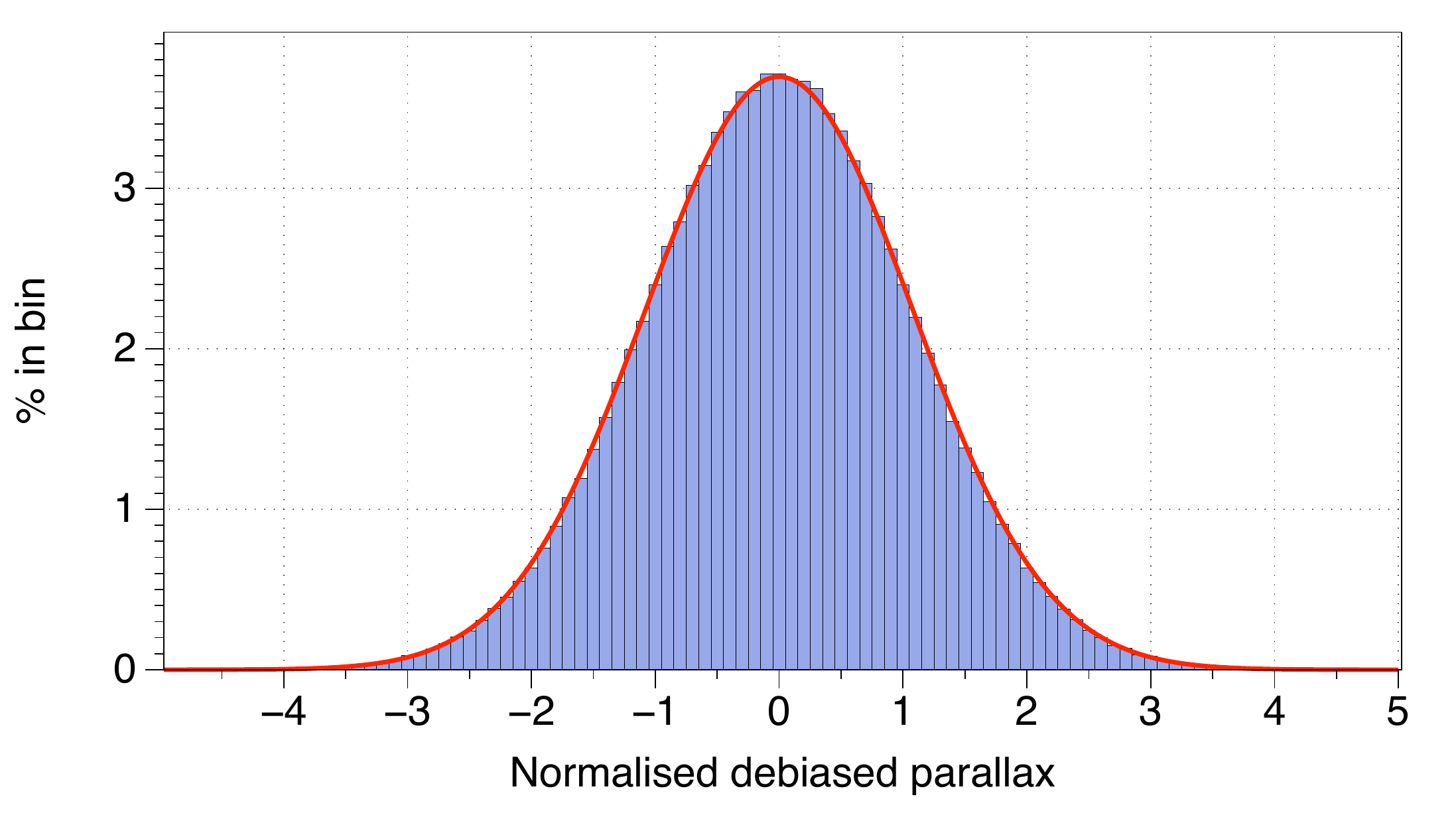}}     
      \resizebox{0.85\hsize}{!}{ \includegraphics{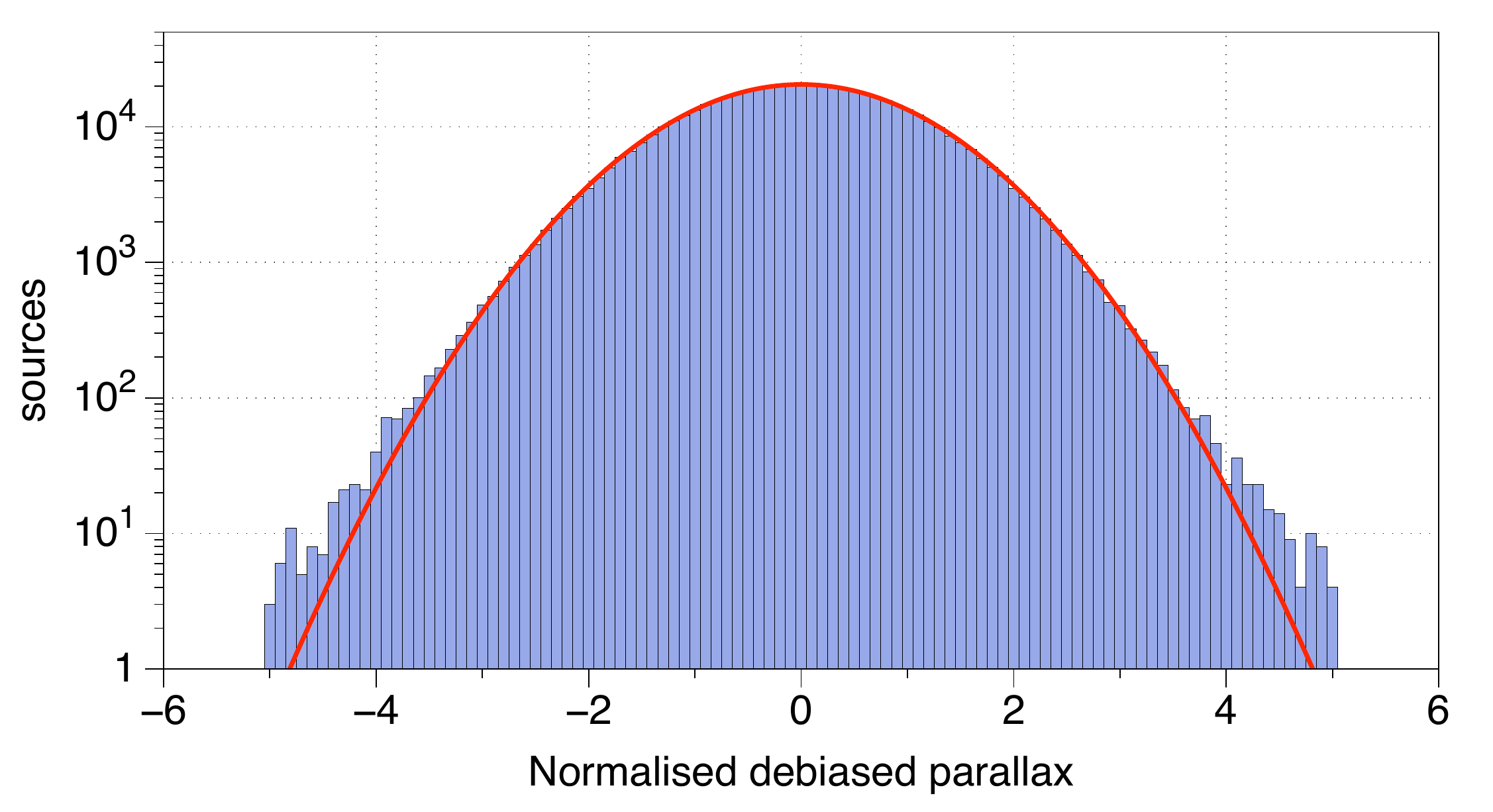}}     
        \caption{Distribution of the normalised debiased parallaxes, $(\varpi +0.029~\text{mas})/\sigma_\varpi$,  
        for the {\gcrftwo} quasars in linear scale (top) and logarithmic (bottom). The red curve is a normal 
        distribution with zero mean and standard deviation 1.08.  
        \label{norm_varpi_CRF}}
\end{figure}

The distribution of the normalised debiased parallaxes, computed as $(\varpi+0.029~\text{mas})/\sigma_\varpi$, 
should follow a standard normal distribution (zero mean and unit variance) if the errors are Gaussian
and the formal uncertainties $\sigma_\varpi$ are correctly estimated. The actual distribution for the full set of 
556\,869 quasars is plotted in \figref{norm_varpi_CRF}. The red continuous curve is a normal distribution 
with zero mean and standard deviation 1.08; that this very closely
follows the distribution up to normalised 
values of $\pm 3.5$ is an amazing feature for real data. The magnitude effect is then fully absorbed by 
the normalisation, indicating that the {\gaia} accuracy in this brightness range is dominated by the 
photon noise. The factor 1.08 means that the formal uncertainties of the parallaxes are too small by 8\%.

\begin{figure}[ht]
\centering
      \resizebox{0.8\hsize}{!}{ \includegraphics{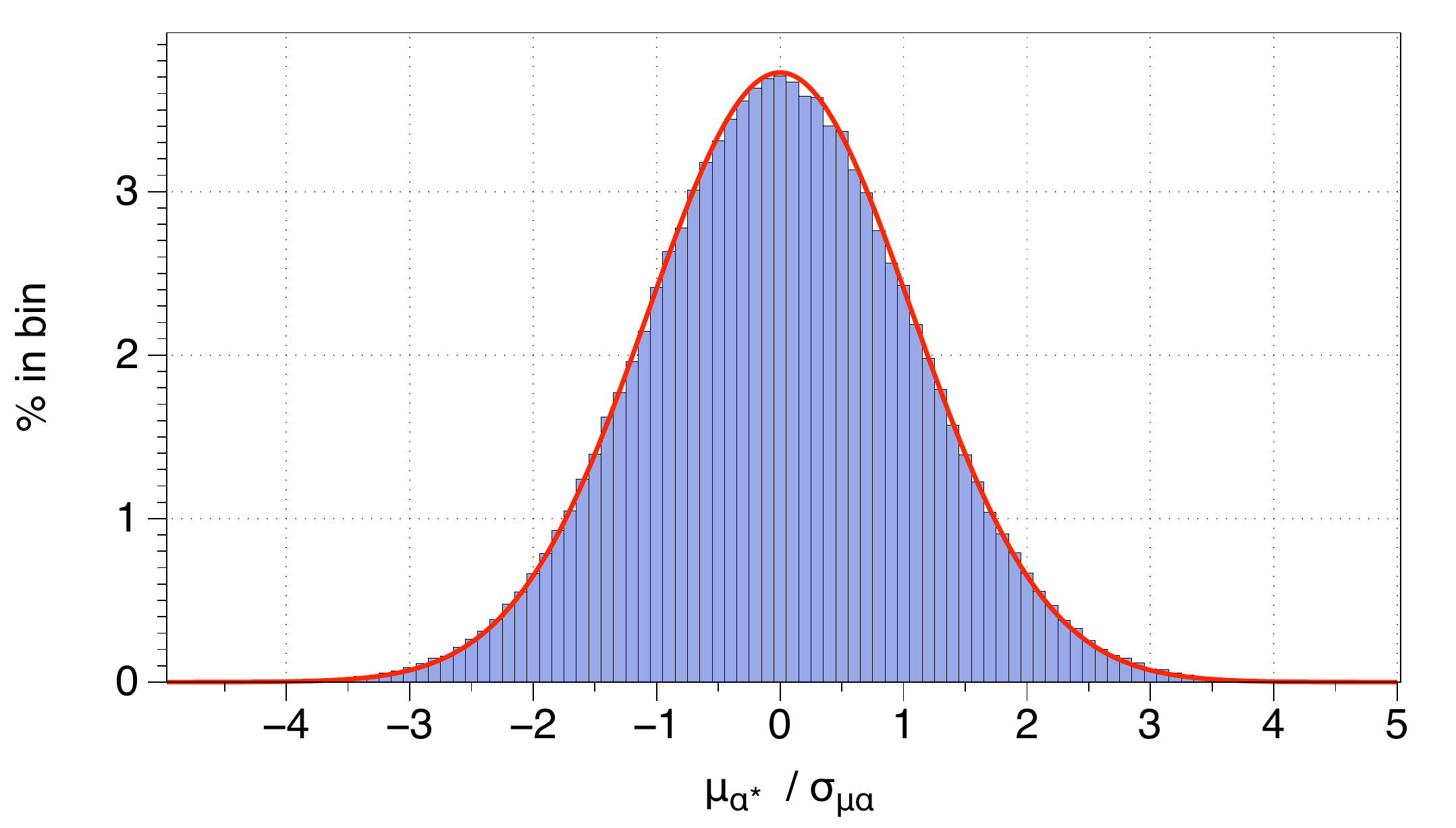}}     
      \resizebox{0.8\hsize}{!}{ \includegraphics{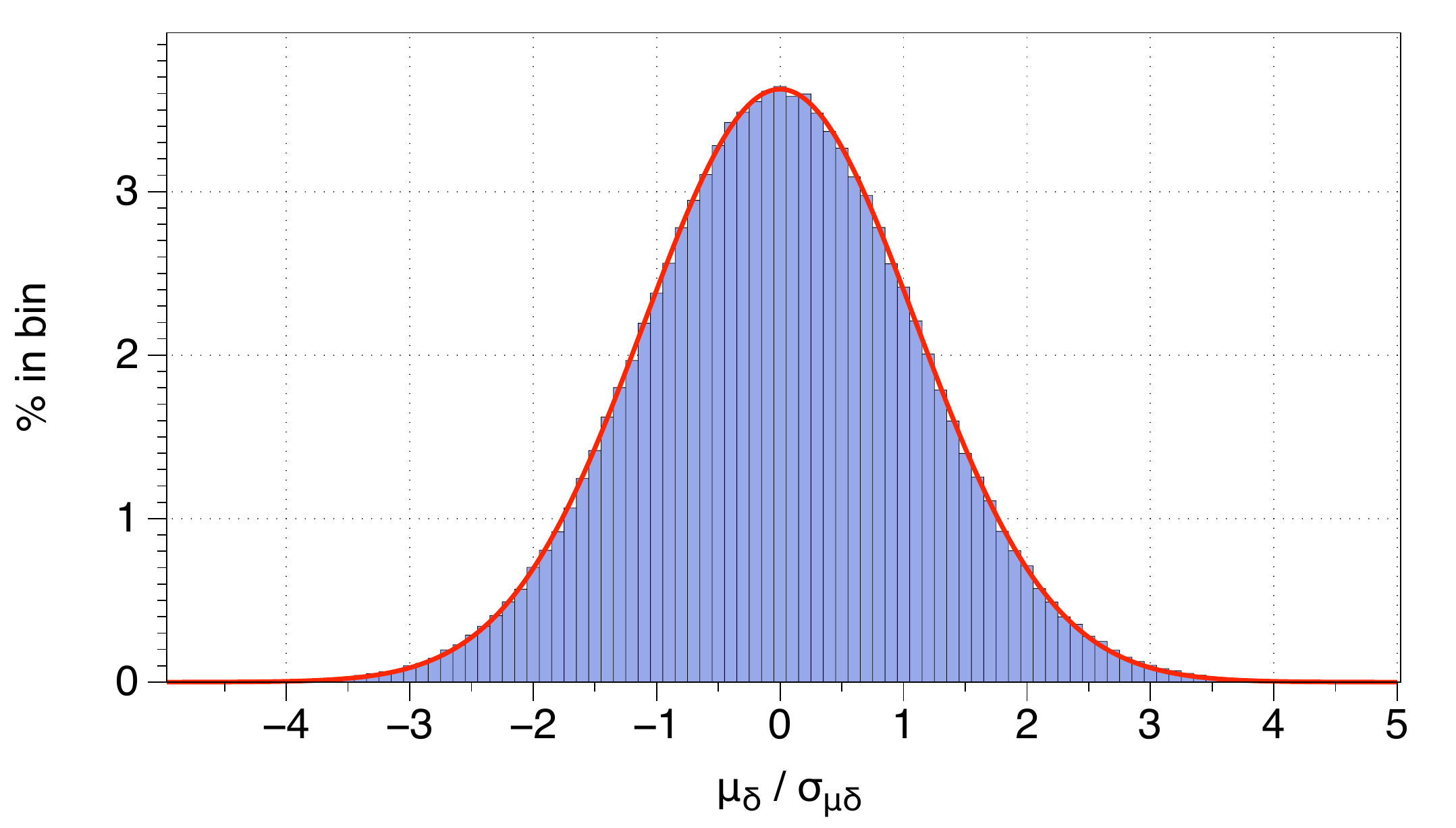}}     
        \caption{Distributions of the normalised components of proper motions of the {\qsos} found with Gaia data, with $\mu_{\alpha*}$ (top) and $\mu_{\delta}$ (bottom)  A  normal distribution with zero mean and standard deviation of 1.09 for $\mu_{\alpha*}$  (1.11 for $\mu_{\delta}$) is drawn in red.  
        \label{norm_mu_CRF}}
\end{figure}

Similarly, the distributions in \figref{norm_mu_CRF} for the normalised components of proper motions are very close to a normal distribution, with zero mean and standard deviations of 1.09 ($\mu_{\alpha*}$) and 1.11 ($\mu_{\delta}$). The extended distributions in log scale are very similar to the parallax and are not plotted.

\begin{figure}[ht]
\centering
      \resizebox{1.00\hsize}{!}{ \includegraphics[clip = true]{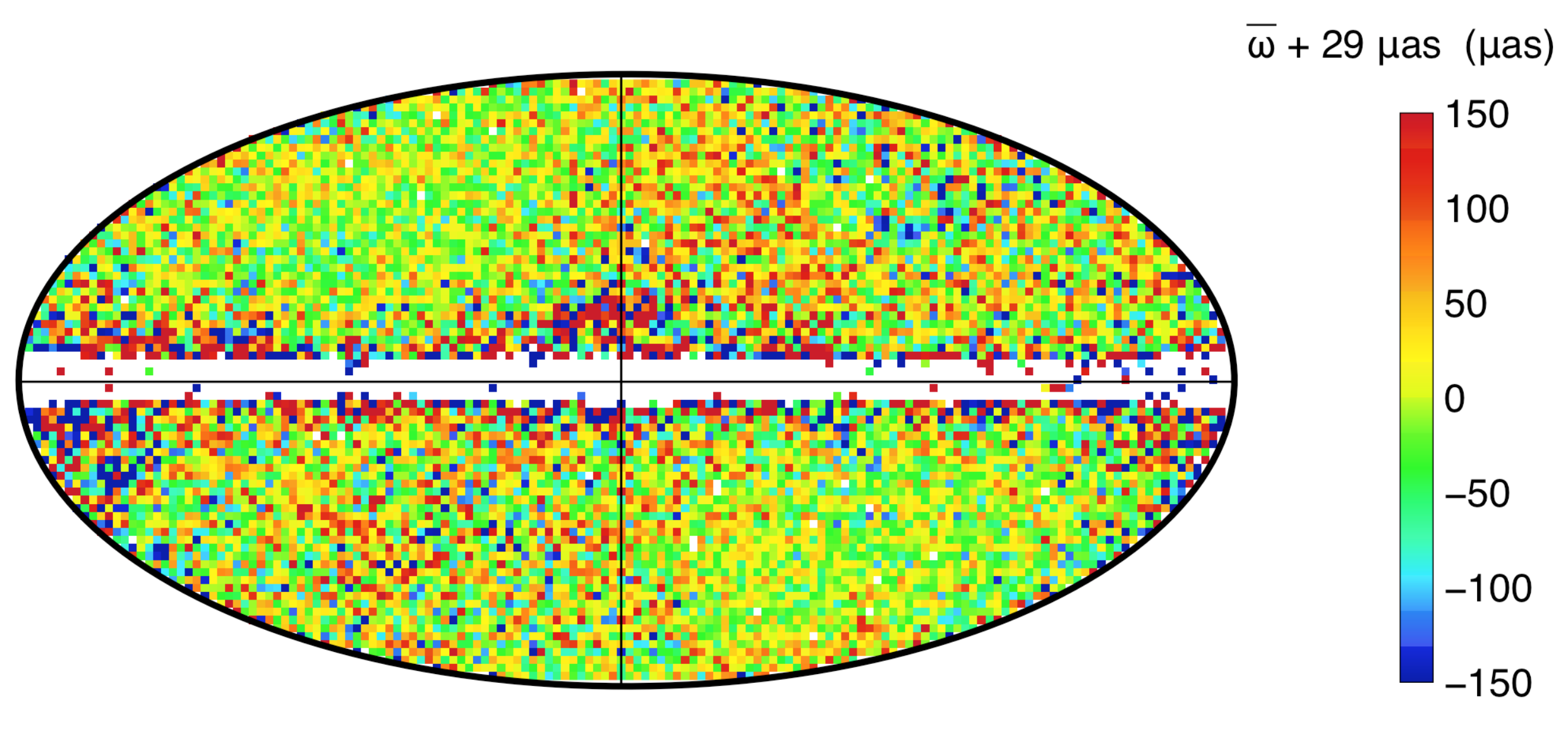}}     
      \resizebox{1.00\hsize}{!}{ \includegraphics[clip = true]{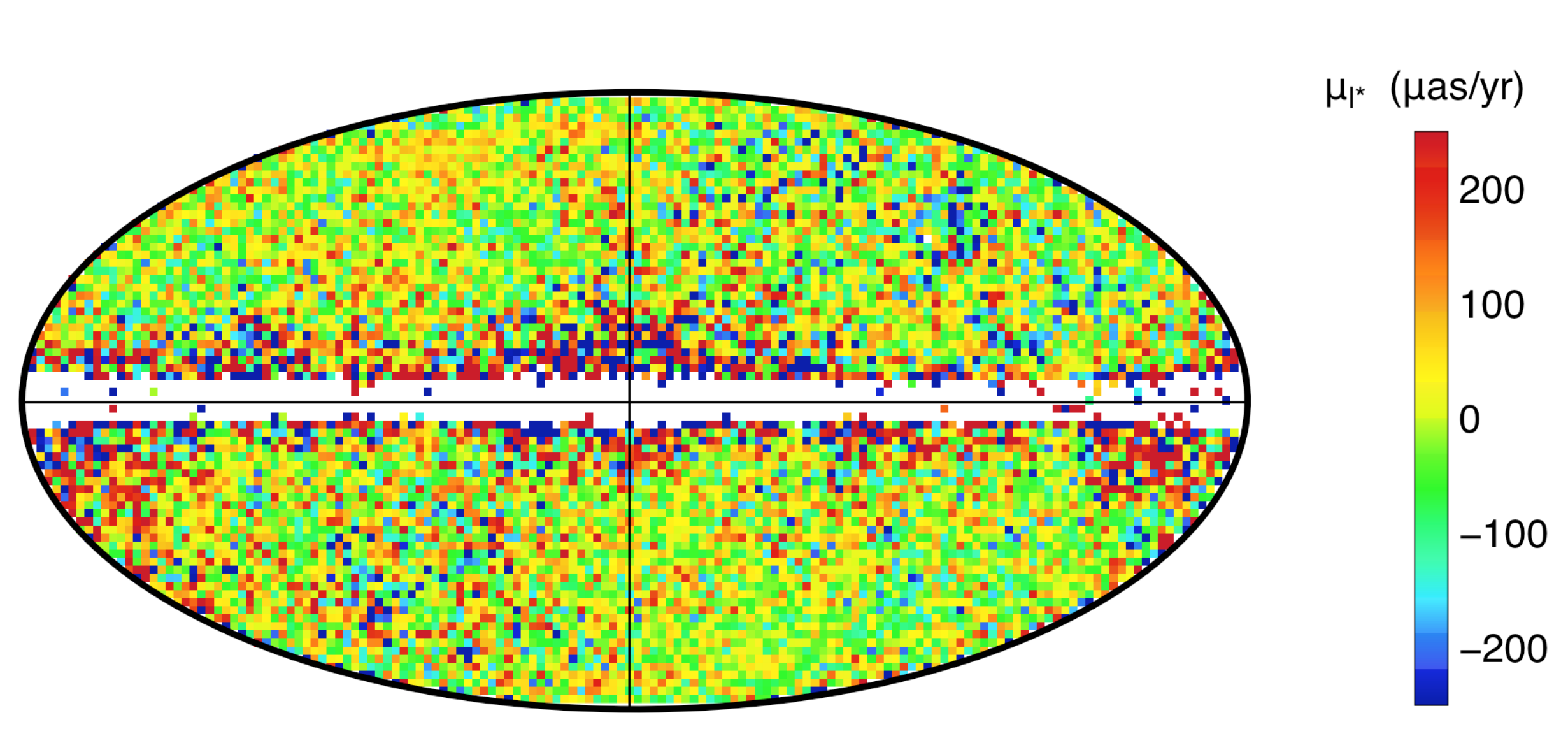}}     
      \resizebox{1.00\hsize}{!}{ \includegraphics[clip = true]{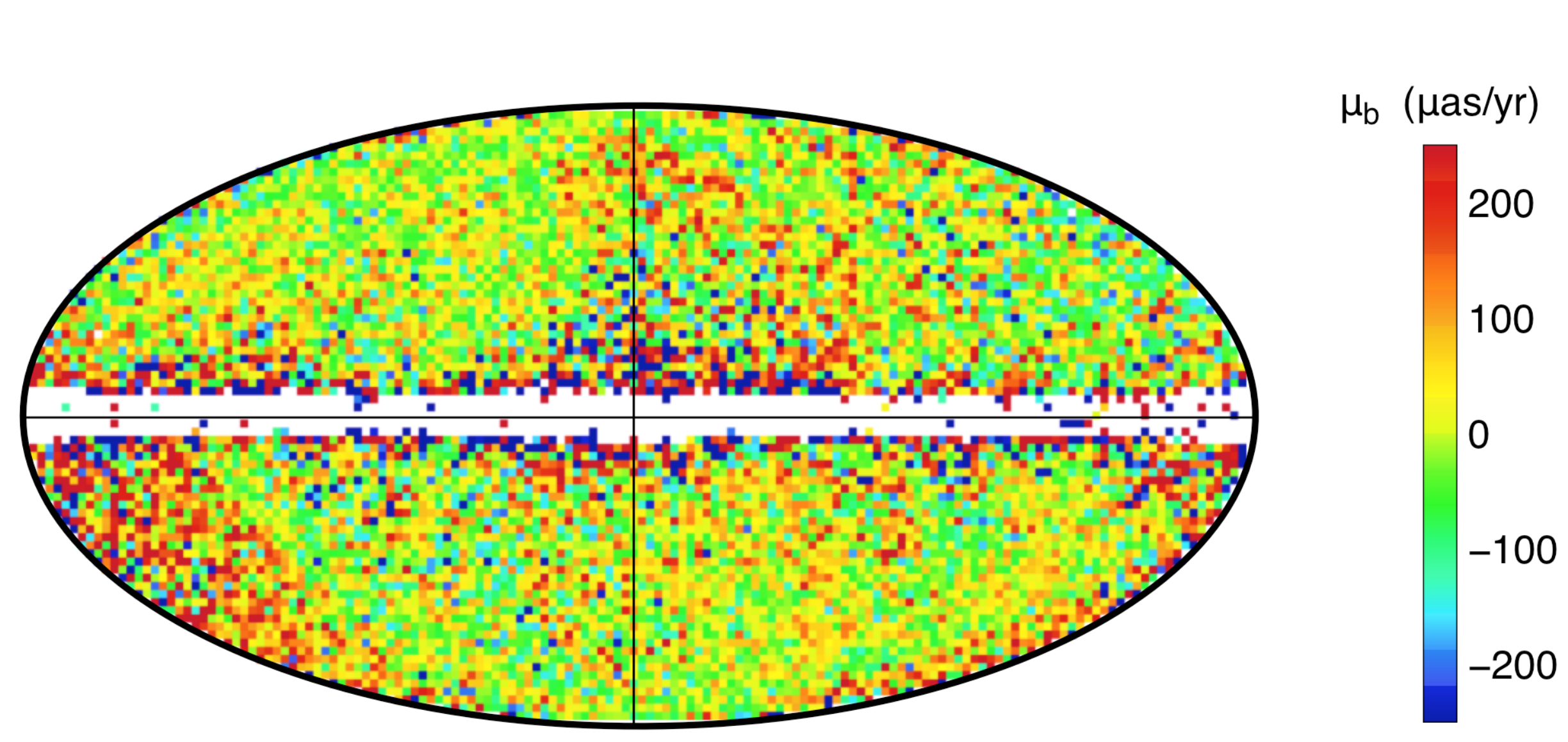}}     
        \caption{Spatial distributions (in Galactic coordinates) of the parallaxes and proper motions of the 
        {\gcrftwo} quasars. From top to bottom: parallax ($\varpi$), proper motion in Galactic longitude 
        ($\mu_{l*}$), and proper motion in Galactic latitude ($\mu_{b}$). Median values are computed in 
        cells of approximately 5~deg$^2$. Maps are in the Hammer--Aitoff projection with Galactic longitude 
        zero at the centre and increasing from right to left.
        \label{fig:spatial_galact}}
\end{figure}

\subsection{Systematic effects}\label{subsec:vshqsos}
\subsubsection{Spatial distributions}
In an ideal world, the errors in position, parallax, and proper motion should be purely random and not 
display any systematic patterns as function of position on the celestial sphere. While the non-uniform
sampling of the sky produced by the {\gaia} scanning law is reflected in the formal \textit{\textup{uncertainties}} 
of the quasar astrometry, as shown in \figref{skymap_sigpos} for the positions, this does not imply that
the \textit{\textup{errors}} (i.e.\ the deviations from the true values) show patterns of a similar nature. 
In the absence of a reliable external reference for the positions (except for the VLBI subset), the
possibility of investigating the true errors in position is limited. 
However, the positions are derived from the same set of observations as the other astrometric parameters, 
using the same solution. Since the errors in parallax and proper motion are found to be in good agreement 
with the formal uncertainties calculated from the solution, we expect this to be the case for the 
positional errors as well.

Figure~\ref{fig:spatial_galact} shows maps in Galactic coordinates of the median parallax and proper motion 
components of the {\gcrftwo} sources, calculated over cells of 4.669~deg$^2$. For cells of this size, the 
median number of sources per cell is 70, with the exception of low Galactic latitude, where the density is 
lower (see \figref{qso_density}), resulting in a larger scatter of the median from cell to cell than in other 
parts of the map. In \figref{fig:spatial_galact} this is visible as an increased number of cells with red and 
blue colours, instead of green and yellow, in the less populated areas. 

The median parallaxes shown in the top panel of \figref{fig:spatial_galact} were corrected for the 
global zeropoint of $-29$~{\muas} (Sect.~\ref{subsec:plxpm}). In all three maps, various large-scale 
patterns are seen for Galactic latitudes $|b|\gtrsim 10$--15~deg, while at small angles (cell size), 
only a mixture of positive or negative offsets is visible that
results from normal statistical scatter. 
The visual interpretation is complicated by large-scale patterns in the amplitude of the statistical 
scatter, in particular the smaller scatter in the second and fourth quadrants, that is,\ around the ecliptic poles. 
This is the result of a combination of the sky distribution of the sources (\figref{qso_density}),
their magnitudes (\figref{skymap_Gmag}), and the {\gaia} scanning law (\figref{skymap_sigpos}), 
which all exhibit similar patterns. Quantifying the large-scale systematics therefore requires a 
more detailed numerical analysis.

\begin{table*}[ht]
        \caption{Large-scale structure of the proper motion field of the {\gcrftwo} quasars analysed using 
        vector spherical harmonics.\label{table:pmfield}}
        \small
        \begin{tabular}{llccr r@{ $\pm$ }r r@{ $\pm$ }r r@{ $\pm$ }r c  r@{ $\pm$ }r r@{ $\pm$ }r r@{ $\pm$ }r}
        \hline\hline
        \noalign{\smallskip}
      &  & & & & \multicolumn{6}{c}{Rotation [{\muasyr}]} &&  \multicolumn{6}{c}{Glide [{\muasyr}]}  \\
      Fit & Source selection & $W$ &$l_\text{max}$  & \multicolumn{1}{c}{$N$} & \multicolumn{2}{c}{$x$}&\multicolumn{2}{c}{$y$}&\multicolumn{2}{c}{$z$}  &&  \multicolumn{2}{c}{$x$}&\multicolumn{2}{c}{$y$}&\multicolumn{2}{c}{$z$} \\
        \noalign{\smallskip} 
        \hline         
        \noalign{\smallskip} 
1 &all                          & y &1  &$556869$ &$  -3.1$&$   0.8$&$  -1.9$&$   0.7$&$  -1.0$&$   0.9$&& \multicolumn{2}{c}{--} & \multicolumn{2}{c}{--} & \multicolumn{2}{c}{--} \\
2 &all                          & y &1  &$556869$ &$  -3.6$&$   0.8$&$  -2.2$&$   0.7$&$  -0.9$&$   0.9$&&$  -7.0$&$   0.8$&$   4.7$&$   0.7$&$  12.1$&$   0.7$ \\          
3 &all                          & y &5  &$556869$ &$  -5.5$&$   1.1$&$  -7.4$&$   0.9$&$   5.6$&$   1.2$&&$  -9.2$&$   1.2$&$   4.7$&$   1.0$&$  11.6$&$   1.0$ \\[6pt]         
4 &$\mu<2$~mas yr$^{-1}$, $G<18$ & y &5  &$27189$  &$ -13.8$&$   2.0$&$ -13.2$&$   1.7$&$   4.0$&$   2.2$&&$  -7.9$&$   2.2$&$  4.7$&$   1.8$&$  10.3$&$   1.7$ \\  
5 &$\mu<2$~mas yr$^{-1}$, $G<18$ & n &5  &$27189$  &$  -8.9$&$   2.9$&$ -12.1$&$   2.4$&$   2.8$&$   2.5$&&$ -10.4$&$   2.9$&$  5.7$&$   2.4$&$  16.6$&$   2.5$ \\
6 &$\mu<2$~mas yr$^{-1}$, $G<19$ & y &5  &$149146$ &$ -11.2$&$   1.3$&$ -12.0$&$   1.1$&$   4.4$&$   1.4$&&$  -9.8$&$   1.5$&$   4.6$&$   1.2$&$  10.4$&$   1.1$ \\
7 &$\mu<3$~mas yr$^{-1}$, $G<20$ & y &5  &$400472$ &$  -5.9$&$   1.1$&$  -8.6$&$   0.9$&$   5.1$&$   1.2$&&$  -9.0$&$   1.2$&$   4.1$&$   1.0$&$  11.9$&$   1.0$ \\
8 &$\mu<3$~mas yr$^{-1}$         & y &5  &$513270$ &$  -5.7$&$   1.1$&$  -7.9$&$   0.9$&$   5.2$&$   1.2$&&$  -8.8$&$   1.2$&$   4.1$&$   1.0$&$  11.6$&$   0.9$ \\[6pt]
9a & $\lfloor{10^5\alpha}\rfloor\!\!\mod2=0$ & y &5  &$278170$ &$ -5.8$&$   1.6$&$  -8.9$&$   1.3$&$   6.4$&$   1.7$&&$  -8.5$&$   1.7$&$   3.0$&$   1.4$&$  12.5$&$   1.4$ \\      
9b & $\lfloor{10^5\alpha}\rfloor\!\!\mod2=1$ & y &5  &$278699$ &$ -5.1$&$   1.6$&$  -5.8$&$   1.3$&$   4.8$&$   1.7$&&$  -9.8$&$   1.7$&$   6.6$&$   1.4$&$  10.7$&$   1.4$ \\[6pt] 
10 &$G>19$                       & y &5  &$406356$ &$   9.8$&$   2.1$&$   6.2$&$   1.8$&$   7.0$&$   2.4$&&$  -8.3$&$   2.3$&$   3.3$&$   1.9$&$  15.6$&$   1.9$ \\[6pt]

        \hline
        \end{tabular}
        \tablefoot{$\mu=(\mu_{\alpha*}^2+\mu_\delta^2)^{1/2}$ is the modulus of the proper motion.  
          $N$ is the number of sources used in the solution. $W$ = ``y'' or ``n'' for weighted or unweighted solution.
          The weighted solutions use a block-diagonal weight matrix obtained from the $2\times 2$ 
          covariance matrix of each source.
        $l_\text{max}$ is the highest degree of the fitted VSH from which rotation and glide are extracted for 
        $l=1$. The columns headed $x$, $y$, $z$ give the components of the rotation and glide along the 
        principal axes of the ICRS. In rows 9a and 9b, two independent halves of the sample are selected 
        according to whether $\lfloor{10^5\,\alpha}\rfloor$ is even (9a) or odd (9b), with $\alpha$ in degrees.}
\end{table*}

\subsubsection{Spectral analysis}

The  vector field of the proper motions of the {\gcrftwo} quasars was analysed using expansions on 
a set of vector spherical harmonics (VSH), as explained in \citetads{2012A&A...547A..59M} or 
\citetads{2014MNRAS.442.1249V}. 

In this approach the components of proper motion are projected onto a set of orthogonal functions up 
to a certain degree $l_\text{max}$. The terms of lower degrees provide global signatures such as the 
rotation and other important physical effects (secular acceleration, gravitational wave signatures), while 
harmonics of higher degree hold information on local distortions at different scales. Given the patterns seen in 
\figref{fig:spatial_galact}, we expect to see a slow decrease in the power of harmonics with $l>1$. 
The harmonics of degree $l=1$ play a special role, since any global rotation of the system of proper 
motions will be observed in the form of a rotation vector directly extracted from the three components with
$(l,m)=(1,0)$, $(1,-1)$, and $(1,+1)$, where $m$ is the order of the harmonic ($|m|\le l$).

 \citetads{2012A&A...547A..59M} derived a second global term from $l=1\text{ that they called}$  \textit{\textup{glide}}.
This physically corresponds to a dipolar displacement originating at one point on a sphere and ending at 
the diametrically opposite point. For the quasar proper motions, this vector field is precisely the 
expected signature of the the galactocentric acceleration
(\citeads{1983jpl..rept.8339F}, % Fanselow
\citeads{1995ESASP.379...99B}, % Bastian
\citeads{1998RvMP...70.1393S}, % Sovers
\citeads{2003A&A...404..743K}, % Kovalevsky
\citeads{2013A&A...559A..95T}). % Titov & Lambert

As summarised in \tabref{table:pmfield}, several VSH fits were made using different selections of 
quasars or other configuration parameters. Fit~1 uses all the quasars and fits only the rotation, 
without glide or harmonics with $l>1$. This is very close to the conditions used to achieve the 
non-rotating frame in the astrometric solution for {\gdrtwo}. It is therefore not surprising that the rotation 
we find is much smaller than in the other experiments. The remaining rotation can be
explained by differences in the set of sources used, treatment of outliers, and so on. This also illustrates the 
difficulty of producing a non-rotating frame that is non-rotating for every reasonable subset that a user 
may wish to select: This is not possible, at least at the level of formal uncertainties. Experiment 2 fits both the rotation and glide to all the data. The very
small change in rotation compared with fit 1 shows the stability of the rotation resulting from the regular 
spatial distribution of the sources and the consequent near-orthogonality of the rotation and glide on this 
set. Fit~3 includes all harmonics of degree $l\le 5$, that is, 70 fitted parameters. Again the 
results do not change very much because of the good spatial distribution. The next five fits show the 
influence of the selection in magnitude and modulus of proper motion, and of not weighting the data
by the inverse formal variance. In the next two fits (9a and
9b), the data are divided into two independent subsets,
illustrating the statistical uncertainties. Most of these fits use fewer sources with a less regular 
distribution on the sky.  

The last fit, fit 10, uses only the faint sources and has a similar glide but a very different rotation ($x$ and $y$ components, primarily), although it comprises the majority ($73\%$) of the  {\gcrftwo} sources. This agrees with Figs.~3 and 4 in \citetads{2018Gaia51},
which show a slight dependency on colour and magnitude of the 
{\gaia} spin relative to quasars. 
Again, this illustrates the sensitivity of the determination of the residual spin to the source selection, and at this stage, we cannot offer a better explanation than that a single solid rotation is too simple a model to fit 
the entire range of magnitudes. No attempt was made to introduce a magnitude equation in the fits.

The formal uncertainty of all the fits using at least a few hundred thousand quasars is of the order of
1~{\muasyr}. It is tempting to conclude from this that the residual rotation of the frame with respect to 
the distant universe is of a similar magnitude. However, the scatter from one fit to the next is considerably
larger, with some values exceeding 10~{\muasyr}. 
Clearly, an overall solid rotation does not easily fit all the {\gaia} data, but gives results that vary 
with source selection well above the statistical noise. However, the degree of consistency between 
the various selections allows us to state that the residual rotation rate of the {\gcrf} is probably not 
much higher than $\pm 10$~{\muasyr} in each axis for any subset of sources.

The typical glide vector is about $(-8, +5, +12) \pm 1$~{\muasyr} for the components in the ICRS. The 
expected signature for the  galactocentric acceleration is a vector directed towards the Galactic centre 
with a magnitude of $\simeq\,$4.50~{\muasyr}, or $(-0.25, -3.93, -2.18)$~{\muasyr} in the ICRS components. 
Clearly, the large-scale systematic effects in the {\gaia} proper motions, being of the order of 
10~{\muasyr} at this stage of the data analysis, prevent a fruitful analysis of the quasar proper 
motion field in terms of the Galactic acceleration. For this purpose, an order-of-magnitude improvement 
is needed in the level of systematic errors, which may be achieved in future releases of {\gaia} data based
on better instrument calibrations and a longer observation time-span. A similar improvement is needed 
to achieve the expected estimate of the energy flux of the primordial gravitational waves 
(\citeads{1997ApJ...485...87G}; \citeads{2012A&A...547A..59M}; \citeads{2018CQGra..35d5005K}).

The overall stability of the fits in \tabref{table:pmfield} is partly due to the fairly uniform distribution
of the {\gcrftwo} sources over the celestial sphere, and it does not preclude the existence of significant 
large-scale distortions of the system of proper motions. Such systematics may be quantified
by means of the fitted VSH, however, and a convenient synthetic indicator of how much signal is found at different 
angular scales is given by the total power in each degree $l$ of the VSH expansion. This power ${\cal P}_l$
is invariant under orthogonal transformation (change of coordinate system) and therefore describes a 
more intrinsic, geometric feature than the individual components of the VSH expansion. The degree
$l$ corresponds to an angular scale of $\sim\! 180^\circ/l$.
 
\begin{figure}[ht]
\centering
      \resizebox{0.95\hsize}{!}{ \includegraphics{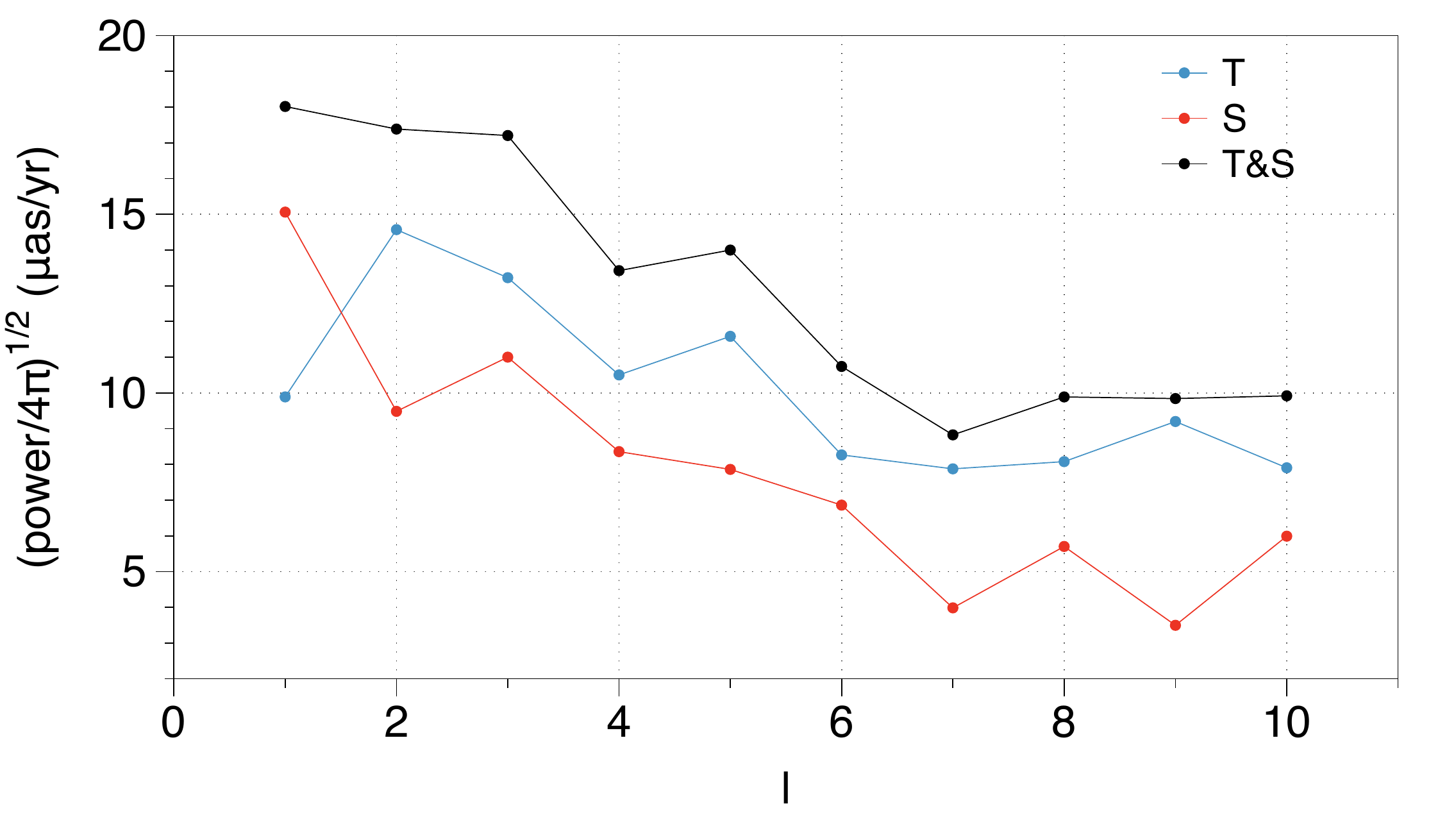}}     
      \resizebox{0.95\hsize}{!}{ \includegraphics{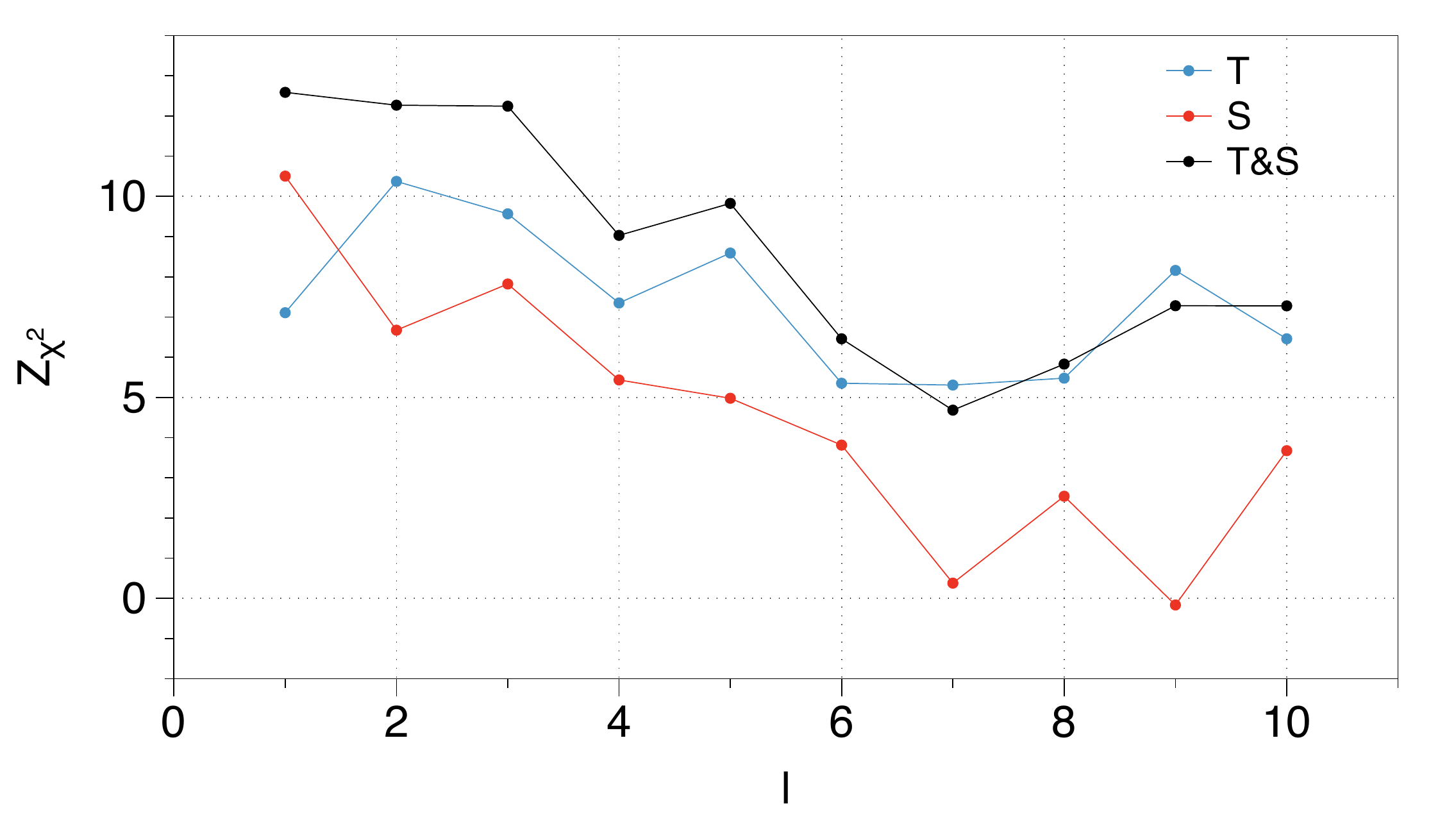}}     
        \caption{Distributions of the RMS values (top) and their statistical levels of significance (bottom) in 
        the VSH decomposition of the proper motion vector field of the {\gcrftwo} from $l=1$ to $10$. 
        $S$ and $T$ refer to the spheroidal and toroidal harmonics, and $T\&S$ signifies their quadratic combination. 
        \label{VSH_power}}
\end{figure}

In \figref{VSH_power} (top panel) we plot $({\cal P}_l/4\pi)^{1/2}$ in {\muasyr}, representing the RMS 
value of the vector field for the corresponding degree $l$. The lower panel in \figref{VSH_power} shows
the significance level of the power given as the equivalent standard normal variate derived from the 
asymptotic $\chi^2$ distribution; see \citetads{2012A&A...547A..59M} for details. The points labelled 
$S$ and $T$ correspond to the spheroidal and toroidal harmonics, with $T\&S$ for their quadratic 
combination. To illustrate the interpretation of the diagrams, for $T_1$ the RMS value is 
$({\cal P}_1/4\pi)^{1/2}\simeq 10$~{\muasyr}, which should be similar to the magnitude of the 
rotation vector for fit~3 in Table~\ref{table:pmfield}. The significance of this value is  
$Z_{\chi^2}\simeq 7$, corresponding to $7\sigma$ of a normal distribution, or a probability below $10^{-11}$.

For the low degrees plotted in \figref{VSH_power}, the power generally decreases with increasing $l$
(smaller angular scales). This indicates that the systematics are generally dominated by the large 
angular scales. The total RMS for $l\le 10$ (angular scales $\gtrsim 18$~deg) is 42~{\muasyr}.
%larger in the harmonics of low degrees and even become not statistically significant for the $S$ 
%harmonic when $l>6$. But at $l=10$ there is still a systematic signature of $\approx 10${ \muasyr} 
%over scale of $\approx 40\deg$. 

\citetads{2018Gaia51} analysed the large-scale systematics of the {\gdrtwo} proper motions of exactly 
the same quasar sample, using a very different spatial correlation technique. A characteristic angular
scale of 20~deg was found, with an RMS amplitude of 28~{\muasyr} per component of proper motion 
(their Eq.~18). Since this corresponds to 40~{\muasyr} for the total proper motion, their result is in 
good agreement with ours. They also found higher-amplitude oscillations with a spatial period of 
$\simeq\,$1~deg, which in the present context of {\gcrftwo} are almost indistinguishable 
from random noise, however. 

%\clearpage

\section{{\icrf} subset of {\gcrftwo}}\label{sect:gaiaicrf3}
This section describes the subset of 2820 {\gcrftwo} quasars matched to the {\icrf} (Sect.~\ref{subsec:qsosel}),
that is,\ the optical counterparts of compact radio sources with accurate VLBI positions. 
%The VLBI is the only alternative technique achieving a global positional accuracy comparable to {\gaia}, 
%and even better for the best-observed sources. The {\icrf} sources are therefore suitable for validating the 
%optical positions in {\gcrftwo}, even though they only constitute a very small subset (0.5\%). 
A comparison between the optical and VLBI positions is in fact a two-way exercise, as useful for 
understanding the radio frame as
it is to {\gaia}, since neither of the two datasets is significantly better than the other. A similar investigation of 
the reference frame for {\gdrone} \citepads{2016A&A...595A...5M} showed the limitations of ICRF2, the 
currently available realisation of the ICRS, for such a comparison. A subset ICRF2 sources also had a less extensive VLBI observation record, the accuracy was lower for the best sources, and it would have been only marginally useful for a comparison to the {\gdrtwo}.

In discussions with the IAU working group in charge of preparing the upcoming ICRF3, which is scheduled for 
mid-2018, it was agreed that the working group would provide a 
prototype version of ICRF3 in the form of their best current solution to the {\gaia}
team. This {\icrf} was officially delivered 
in July 2017 and is particularly relevant in the current context for two reasons.
\begin{itemize}
\item With the assumption that there is no  globally systematic difference between the radio and optical positions, the common sources allowed the axes of the two reference frames to be 
aligned, as explained in \citetads{2018Gaia51}. 
The existence of radio--optical offsets with random orientation for each source is not a great problem
for this purpose as it only adds white noise to the position differences. If large enough, it will be detected in 
the normalised position differences (Sect.~\ref{subsec:normpos}).
\item The VLBI sources included in this prototype, together with the associated sets worked out in the X/Ka and K band (not yet released), are the 
most accurate global astrometric solutions available today that
are fully independent of {\gaia}. The quoted 
uncertainties are very similar to what is formally achieved in {\gdrtwo,} and the best-observed VLBI sources 
have positions that are nominally better than those from {\gaia}. This is therefore the only dataset from which the true errors and possible systematics in the positions of either dataset can
be assessed and individual
cases of truly discrepant positions between the radio and optical domains can be identified. 
The VLBI positions are less homogeneous in accuracy than the corresponding {\gaia} data, but the 
$\simeq\,$1650 {\icrf} sources with a (formal) position uncertainty $<\,$0.2~mas match the
{\gaia} positions of the brighter ($G<18$~mag) sources well in
quality. 
\end{itemize}

\subsection{Properties of the {\gaia} sources in the {\icrf}}\label{subsec:icrfminusgaia}

\begin{figure}[ht]
\centering
      \resizebox{0.95\hsize}{!}{ \includegraphics{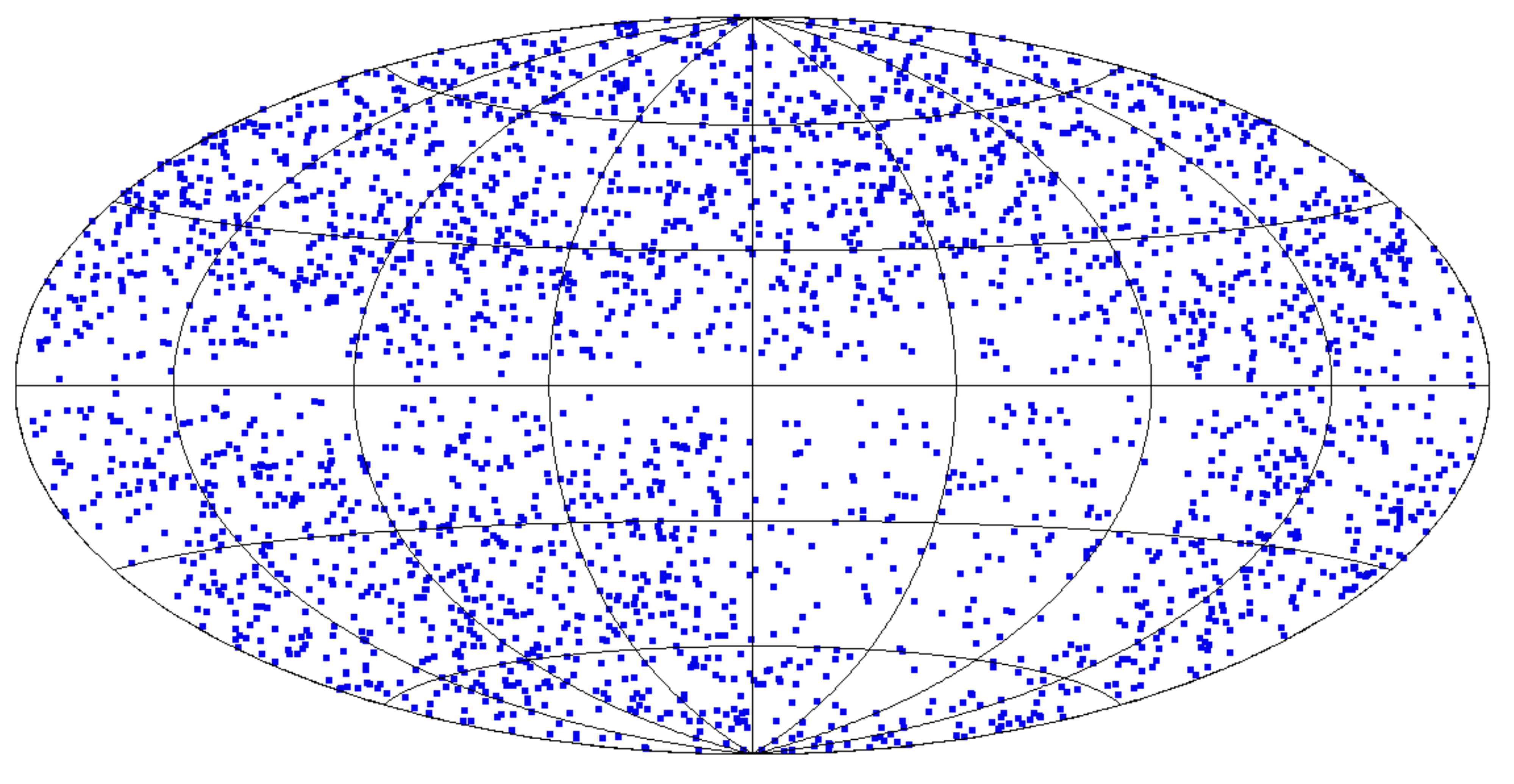}}      
        \caption{Sky distribution of the 2820 {\gaia} sources identified as most probable optical counterparts of quasars in
        the {\icrf}. Hammer--Aitoff projection in Galactic coordinates with origin at the centre of the map and
        longitude increasing from right to left.
        \label{ICRF3_space_distribution}}
\end{figure}

Figure~\ref{ICRF3_space_distribution} shows the spatial distribution of the 2820 optical counterparts 
of {\icrf} sources on the sky. The plot is in Galactic coordinates to facilitate comparison with \figref{qso_density},
showing the full {\gcrftwo} sample. The area in the lower right quadrant with low density corresponds to the 
region of the sky at $\delta<-40$~deg with less VLBI coverage. Otherwise the distribution is relatively uniform, 
but with a slight depletion along the Galactic plane, as expected for an instrument operating at optical 
wavelengths. 

\begin{figure}[ht]
\centering
      \resizebox{0.95\hsize}{!}{ \includegraphics{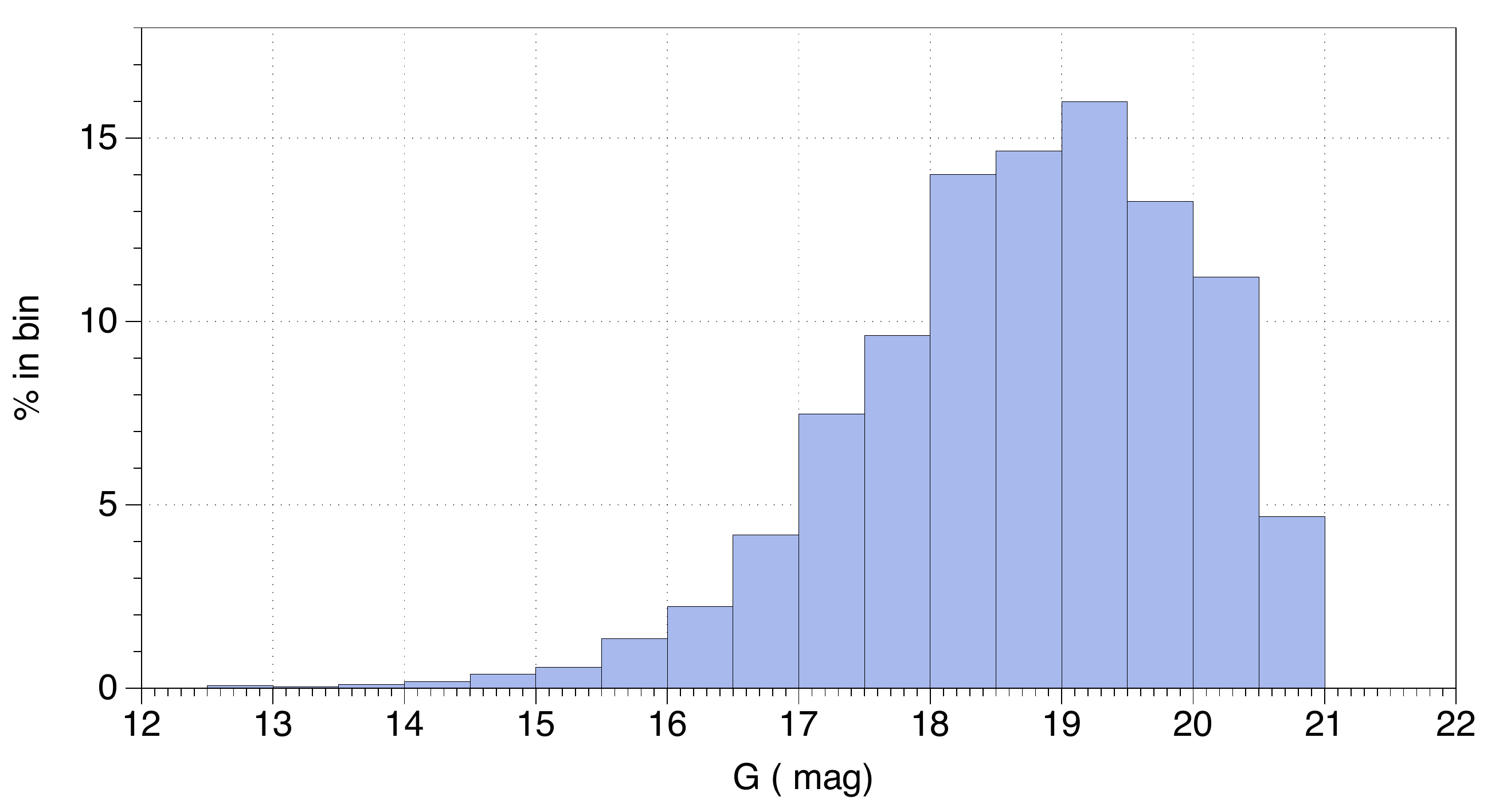}}      
        \caption{Magnitude distribution of the 2820 {\gaia} sources identified as likely optical counterparts of quasars in
        the {\icrf}.
        \label{ICRF3_mag_G_distribution}}
\end{figure}

The magnitude distribution of the {\icrf} sources is shown in \figref{ICRF3_mag_G_distribution}. 
The median is 18.8~mag, compared with 19.5~mag for the full {\gcrftwo} sample shown in 
\figref{mag_G_distribution}. 
%meaning that on the average a Gaia ICRF3 source has a more accurate position than an 
%average quasar of the {\gcrftwo}. 
The colour distribution (not shown) is similar to that of the full sample, shown in 
\figref{mag_G_distribution}, only slightly redder: the median $G_\text{BP}-G_\text{RP}$ is $\simeq 0.8$~mag
for the {\icrf} subset, compared with 0.7~mag for the full sample.

\begin{figure}[ht]
\centering
      \resizebox{0.95\hsize}{!}{ \includegraphics{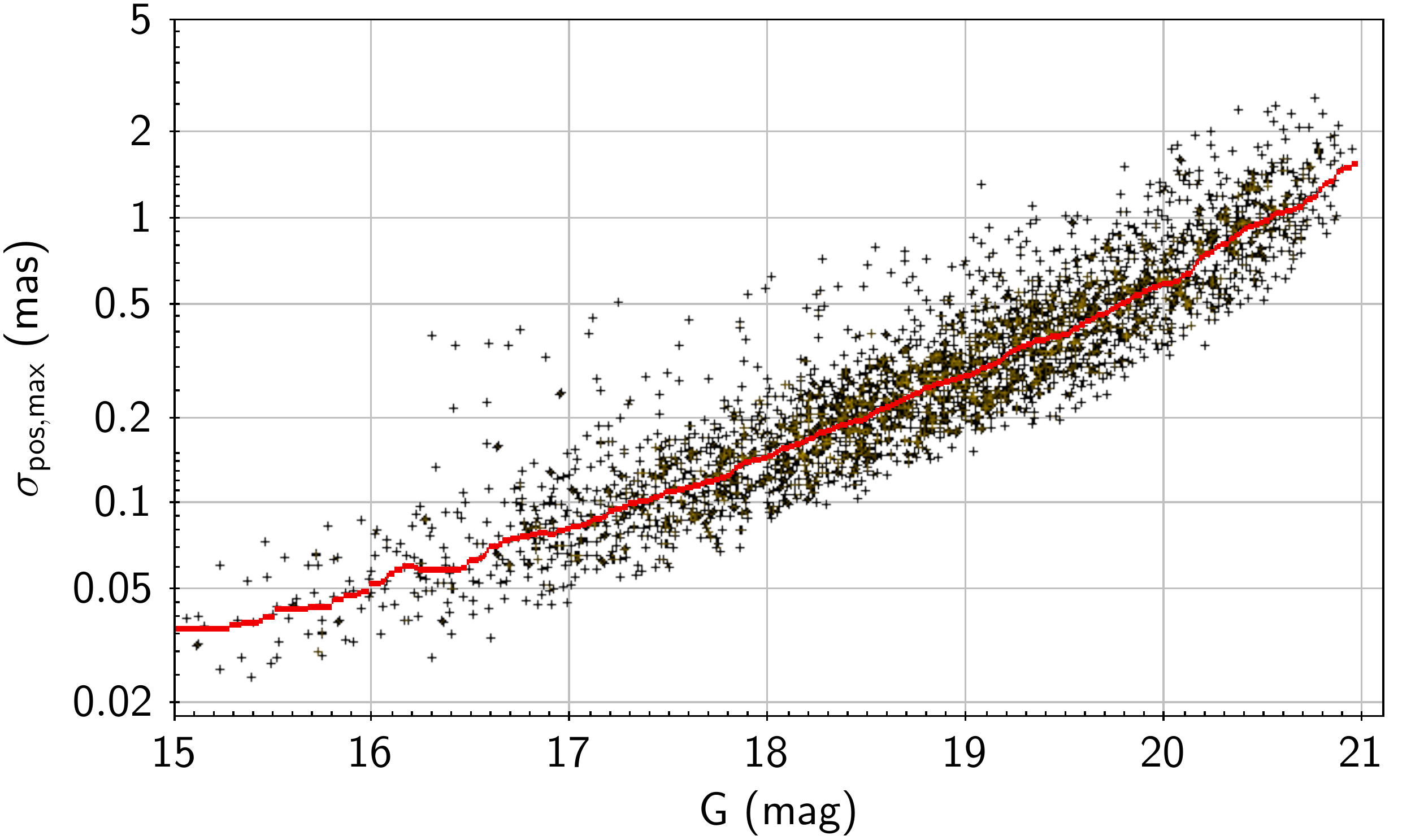}}      
        \caption{Formal position uncertainty as a function of magnitude for the 2820 {\gaia} sources identified as 
        optical counterparts of quasars in the {\icrf}. The solid red line is a running median through the data points.
        \label{ICRF3_sigpos_distribution}}
\end{figure}

In terms of astrometric quality, the {\gdrtwo} sources in the {\icrf} subset do not differ significantly
from other quasars in {\gcrftwo} at the same magnitude. 
Figure~\ref{ICRF3_sigpos_distribution} displays the formal uncertainty in position, computed with 
Eq.~(\ref{eq:sigpos}), as function of the $G$ magnitude. 
Both the median relation and the scatter about the median are virtually the same as for the general 
population of quasars in {\gcrftwo} shown in \figref{sigpos_max_G}.  
For $G\gtrsim 16.2,$ only few points in \figref{ICRF3_sigpos_distribution} lie clearly above the main relation.
This may be linked to the change in the onboard CCD observation window allocation that occurs at 
$G\simeq 16$ \citepads{2016A&A...595A...1G}. Four hundred and
nine sources are brighter than $G = 17.4,$  where the  median 
position uncertainty as shown on Figure ~\ref{ICRF3_sigpos_distribution} reaches $100$~{\muas}.

\subsection{Angular separations}

\begin{figure}[ht]
\centering
      \resizebox{0.95\hsize}{!}{ \includegraphics{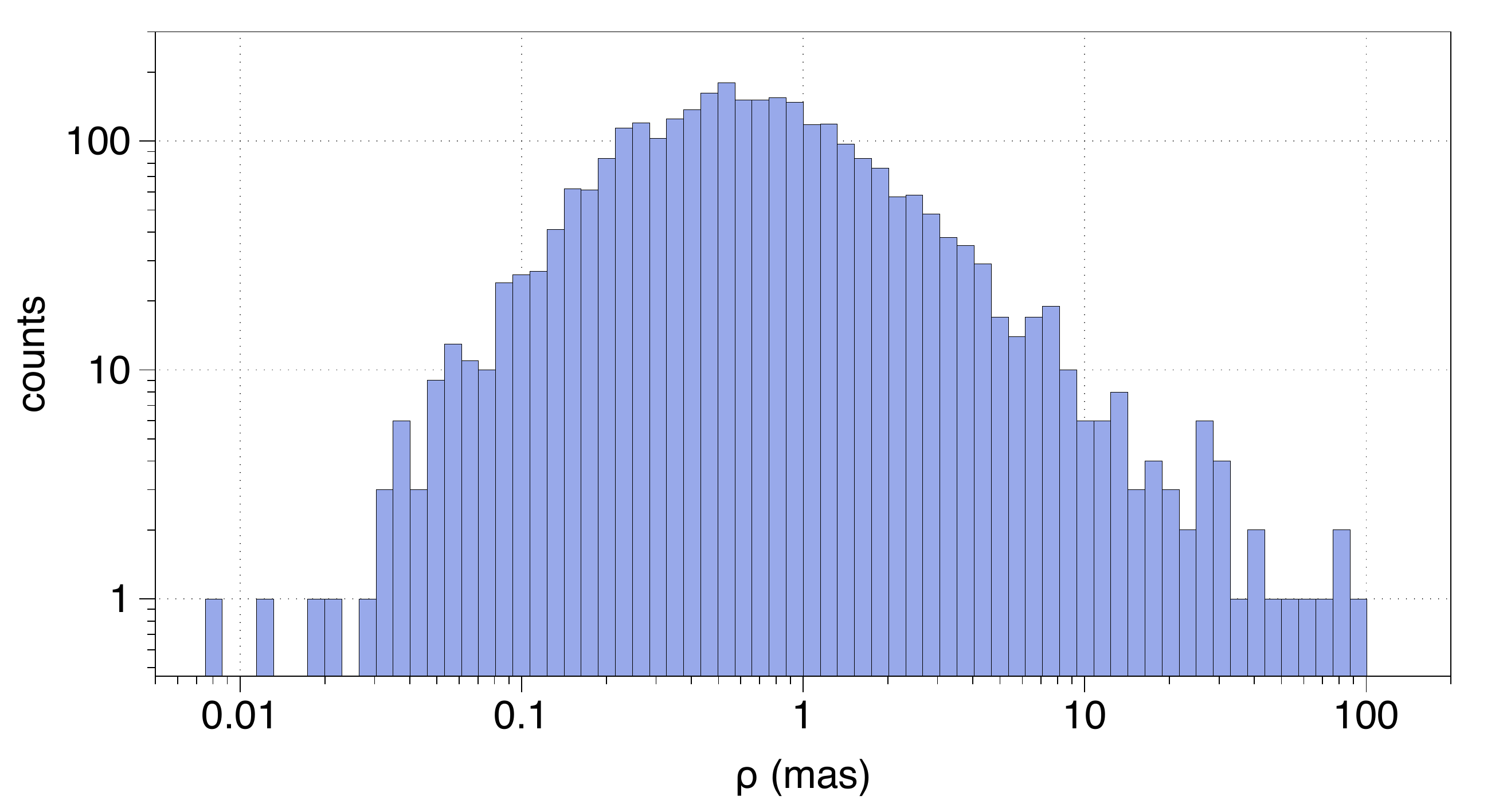}}      
        \caption{Distribution of the angular separation $\rho$ between {\gdrtwo} and the {\icrf} for the 2820 sources in common. Log-log scale plot  with bins in $\rho$ having a fractional  width of $2^{1/5}$.
        \label{ICRF3_rho_distribution}}
\end{figure}

We now compare the positions in {\gdrtwo} and {\icrf\ directly} for the 2820 quasars in common.
Figure~\ref{ICRF3_rho_distribution} gives in log-scale the distribution of the angular distances computed as
\begin{equation}\label{eq:rho}
\rho = (\Delta\alpha_*^2 +\Delta\delta^2)^{1/2}\, ,
\end{equation}
where $\Delta\alpha_*=(\alpha_{\gaia}-\alpha_\text{VLBI})\cos\delta$.
While for most of the sources, $\rho$ is lower than 1~mas and very often much below this level, the
number of  discrepant sources is significant, and a few even
have a position difference higher than 10~mas that would require 
individual examination. 

\begin{figure}[ht]
\centering
      \resizebox{0.95\hsize}{!}{ \includegraphics{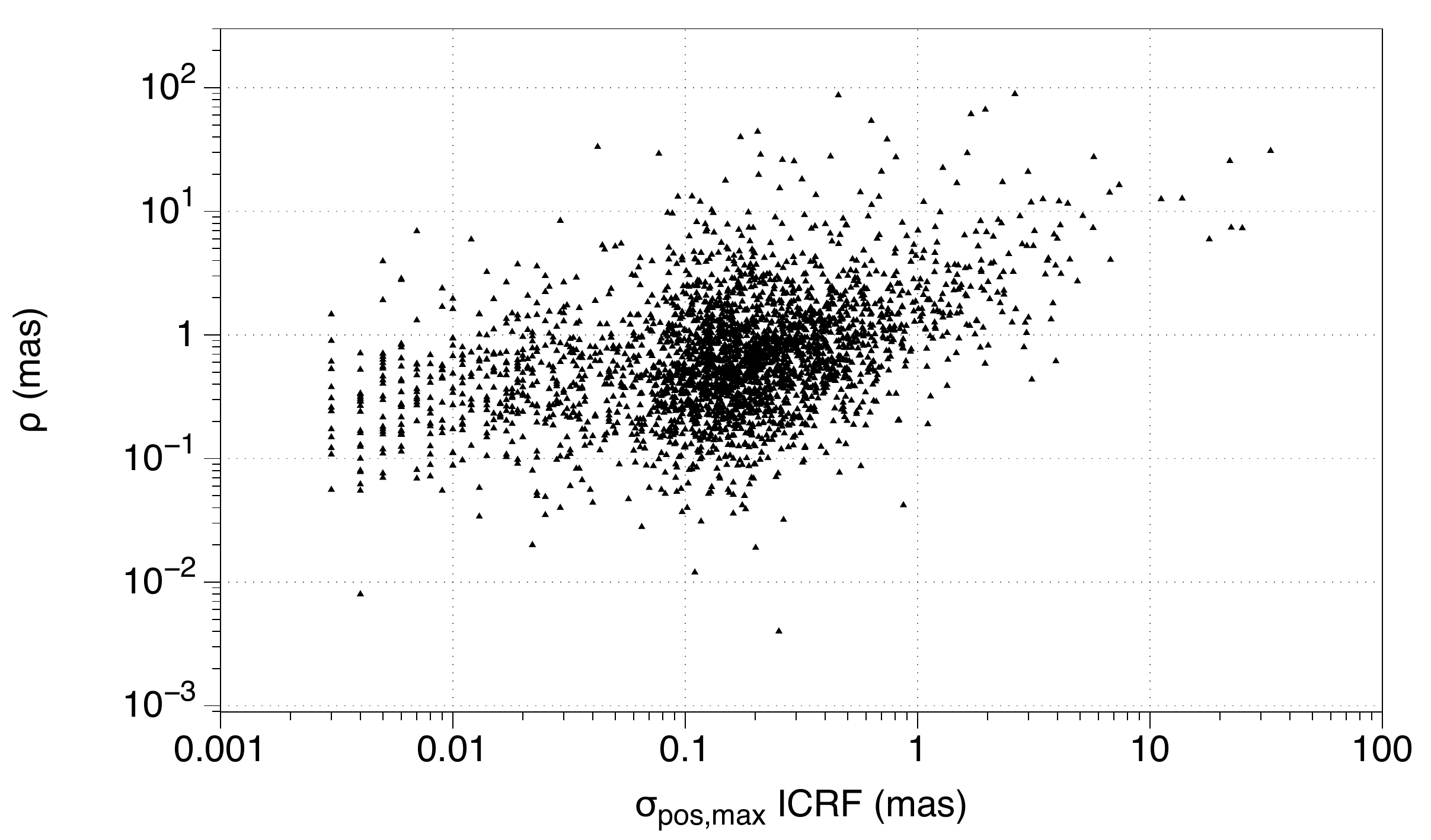}}
      \resizebox{0.95\hsize}{!}{ \includegraphics{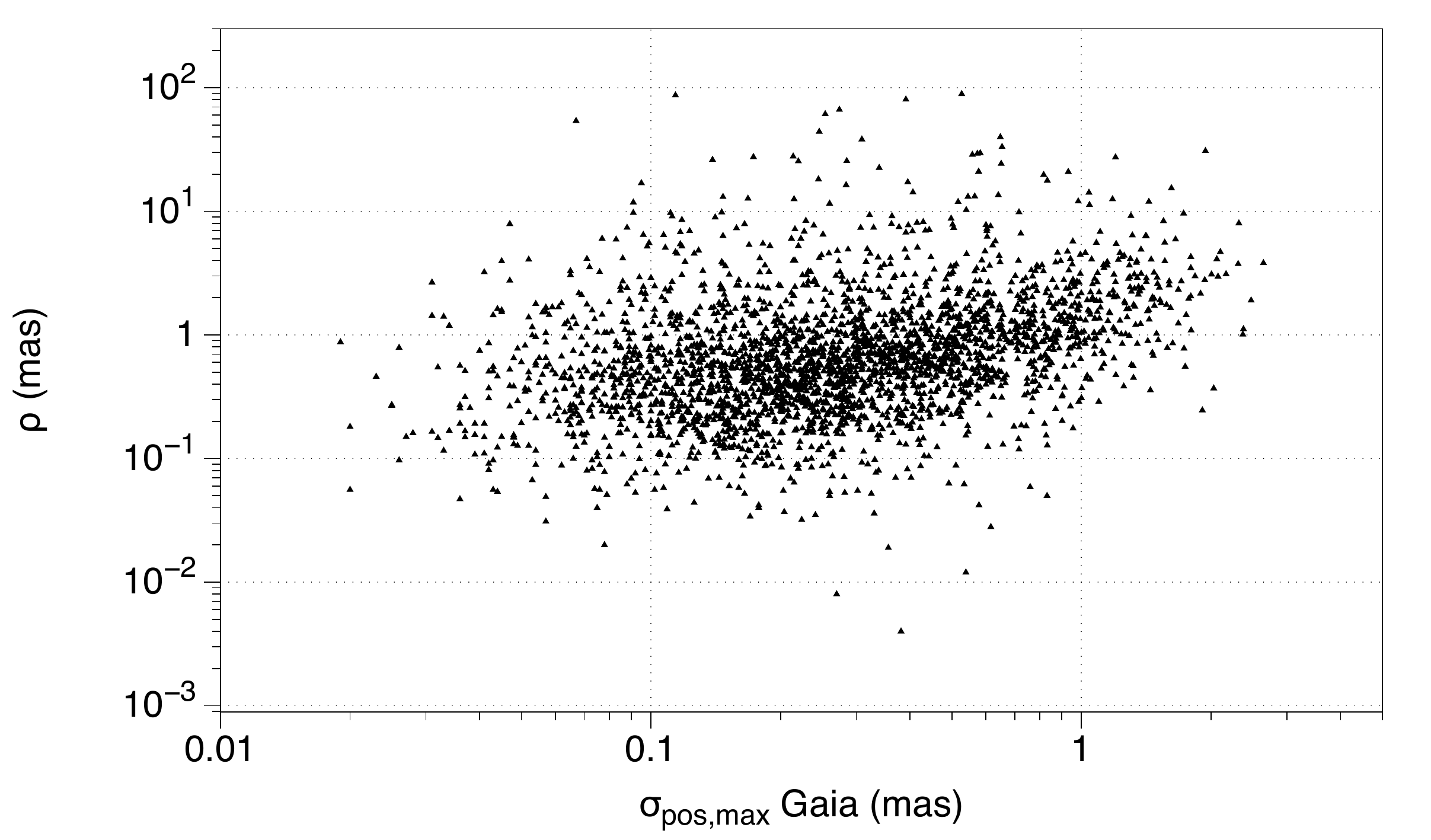}}      
        \caption{Angular position differences $\rho$ between {\gdrtwo} and the {\icrf} as function of 
        the formal uncertainties $\sigma_\text{pos,max}$ of the {\icrf}  (top) and {\gdrtwo} (bottom).       
        \label{ICRF3_rho_sigma}}
\end{figure}

To illustrate the dependence on the solution accuracies, \figref{ICRF3_rho_sigma} 
shows scatter plots of $\rho$ versus the formal uncertainty in the {\icrf} (top) and {\gcrftwo} (bottom). 
Several of the most extreme distances in the top diagram are for sources with a large uncertainty in the
{\icrf}. However, some sources with nominally good solutions in both datasets exhibit large positional 
differences. These deserve more attention as the differences could represent real offsets between the centres 
of emission at optical and radio wavelengths. This is not further investigated in this paper, which is devoted to present 
the main properties of the {\gcrftwo}. Other explanations for the large differences can be put forward, such as a 
mismatch on the {\gaia} side when the optical counterpart is too faint and a distant star happens to be matched 
instead (unlikely at $<\,$10~mas distance); an extended galaxy around the quasar that is misinterpreted by the {\gaia} detector (should
in general produce a poor solution); double or lensed quasars; or simply statistical outliers from the possibly 
extended tails of random errors.
Although the {\icrf} data in \figref{ICRF3_rho_sigma} cover a wider range in $\sigma_\text{pos,max}$ 
than the {\gaia} data, the cores of both distributions extend from $\simeq\,$0.1 to 0.5~mas.
%The best accuracy is achieved by the VLBI solution with a significant population having an uncertainty 
%$<0.05$~mas, below the {\gdrtwo} floor. However it must be stressed than  that uncertainties here are 
%formal errors and it is doubtful that the true uncertainty of the best {\icrf}  sources is at the level of 5 {\muas}.
%\tabref{table:offset} lists the most conspicuous anomalous sources with good astrometry both in {\icrf} and in {\gcrftwo}. The selection is defined by $\rho>3$~mas and $\sigma_\text{pos,max} < 0.5$~mas in  both catalogues. 

\begin{figure}[ht]
\centering
      \resizebox{0.95\hsize}{!}{ \includegraphics{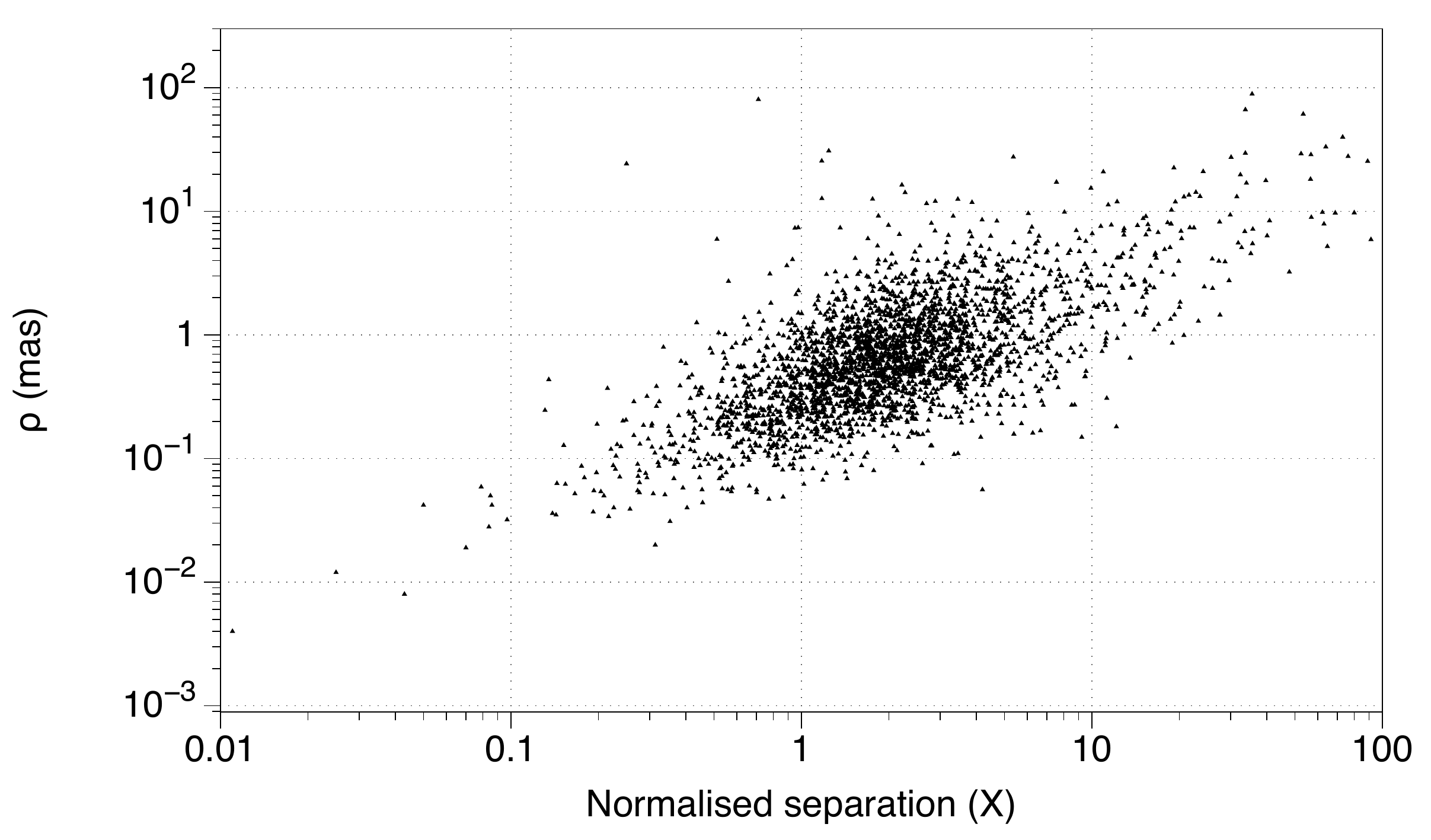}}
        \caption{Angular separation ($\rho$) vs. normalised separation ($X$) for the {\icrf} subset
        of {\gcrftwo}.       
        \label{rho_norm_scatter}}
\end{figure}

\subsection{Normalised separations}\label{subsec:normpos}

The angular separations $\rho$ become statistically more meaningful when scaled with the combined 
standard uncertainties. In the case of correlated variables, \citetads{2016A&A...595A...5M} have shown 
how to compute a dimensionless statistic $X$, called the normalised separation (their Eq.~4). If the positional
errors in both catalogues are Gaussian with the given covariances, then $X$ is expected to follow a standard 
Rayleigh distribution, and values of $X>3$ should be rare (probability $\simeq\,$0.01). We caution that the normalisation used in this section depends on the reliability of the reported position uncertainties from the {\gdrtwo} on one hand and from the {\icrf} on the other hand. The latter are still provisional and purely formal, without noise floor and other overall adjustment, which will be introduced in the final release of the ICRF3.

Figure~\ref{rho_norm_scatter} is a scatter plot of $\rho$ versus $X$, showing a fairly large subset of sources 
with $X>3$ and even much larger. The most anomalous cases are found in the upper
right part of the diagram. Some of the sources with the largest $\rho$ are located
in the upper centre of the diagram, with unremarkable $X$, meaning that their large angular separations are not significant in view of the formal uncertainties. 

\begin{figure}[ht]
\centering
      \resizebox{0.95\hsize}{!}{ \includegraphics{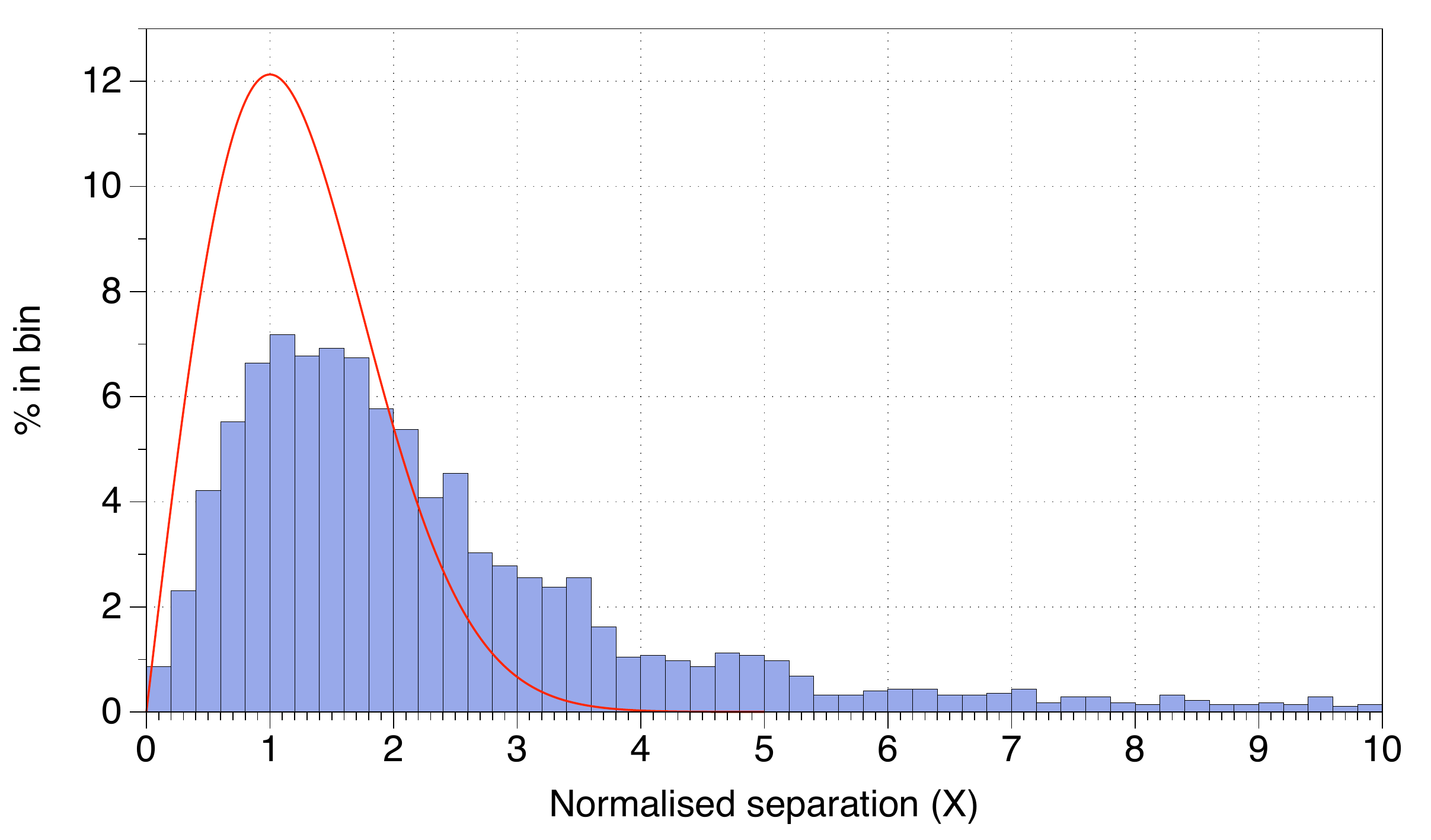}}      
        \caption{Distribution of the normalised separations $X$ between {\gdrtwo} and the {\icrf}.
        148 sources have a normalised separation $> 10$.
        The red curves show the (standard) Rayleigh distributions for unit standard deviation.      
        \label{norm_rho_histo}}
\end{figure}

The  diagram in \figref{norm_rho_histo} shows the frequency distribution of the normalised separations 
$X$ with the standard Rayleigh probability density function superimposed as a solid red line. The frequency diagram includes all the sources, although 148 sources have a normalised separation $> 10$ and would be outside the frame. The distribution cannot be represented by a standard Rayleigh distribution, even though its mode is not very far from one, but the spread at large normalised separations is much too large. The departure from a pure Rayleigh distribution between VLBI positions and the {\gdrone} has previously
been noted in \citeads{2017MNRAS.467L..71P} in a comparison using more than 6000 sources with VLBI positions. However, at this stage with the {\icrf}, we cannot draw definite conclusions, and this issue will have to be reconsidered with the official release of ICRF3.

%The next two plots in \figref{norm_rho_histo_twomas} compare a similar diagram with the 2423 sources with angular separation $< 2$ mas, and  the same data when  an extra uncertainty of 0.20~mas  is added quadratically to the formal standard uncertainties of the position differences in 
%$\alpha$ and $\delta$. The comparison to a standard Rayleigh is more satisfactory, although there is still a tail of very large $X$ values. But the core fits reasonably well the standard Rayleigh.

%The top panel uses a cut-off at 
%angular separation $\rho<10$~mas, which includes nearly all the sources, while the bottom panel is restricted 
%to the better half with $\rho <0.6$~mas. Not unexpectedly, the deviation from a Rayleigh distribution is
%more conspicuous when the larger angular separations are included. However, we have found that the 
%core of the distribution in the upper panel can be made much closer to a Rayleigh distribution by quadratically 
%adding an extra uncertainty of 0.18~mas to the formal standard uncertainties of the position differences in 
%\alpha$ and $\delta$ (a tail of very large $X$ values is however still present).
%The departure from the standard Rayleigh distribution is also found more conspicuous for bright sources than 
%faint sources, in agreement with the fact that small systematic differences should be more clearly seen on 
%the best solutions.

\begin{table*}[ht]
        \caption{Global differences between the {\gcrftwo} positions of ICRF sources and their 
        positions in the {\icrf}, expressed by the orientation and glide parameters.\label{table:alignment}}
        \small
        \begin{tabular}{rlcccc c@{ $\pm$ }c r@{ $\pm$ }r r@{ $\pm$ }r c  r@{ $\pm$ }r r@{ $\pm$ }r r@{ $\pm$ }r}
        \hline\hline
        \noalign{\smallskip}
      &  & & & && \multicolumn{6}{c}{Orientation ({\muas})} &&  \multicolumn{6}{c}{Glide ({\muas})}  \\
       Fit & Source selection & $W$ &$l_\text{max}$  & $N$ &&  \multicolumn{2}{c}{$x$}&\multicolumn{2}{c}{$y$}&\multicolumn{2}{c}{$z$}  &&  \multicolumn{2}{c}{$x$}&\multicolumn{2}{c}{$y$}&\multicolumn{2}{c}{$z$} \\
        \noalign{\smallskip} 
         \hline         
        \noalign{\smallskip} 
 1&all                         & y &1  &2820  &&$   -9$&$  29$&$    4$&$  27$&$   3$&$  28$&&\multicolumn{2}{c}{--}&\multicolumn{2}{c}{--}&\multicolumn{2}{c}{--} \\  
 2& all                        & y &1  &2820  &&$  -28$&$  31$&$   -8$&$  29$&$   10$&$  28$&&$   47$&$  29$&$  -69$&$  28$&$  -72$&$  29$ \\[6pt]
 3&$\rho<10$~mas               & y &1  &2773  &&$  -17$&$  16$&$   22$&$  15$&$  -23$&$  16$&&\multicolumn{2}{c}{--}&\multicolumn{2}{c}{--}&\multicolumn{2}{c}{--} \\ 
 4&$\rho<2$~mas                & y &1  &2423  &&$  -35$&$   9$&$   21$&$   8$&$  -24$&$   9$&&\multicolumn{2}{c}{--}&\multicolumn{2}{c}{--}&\multicolumn{2}{c}{--} \\ 
 5&$\rho<2$~mas                & n &1  &2423  &&$  -13$&$  14$&$    5$&$  14$&$   -5$&$  13$&&\multicolumn{2}{c}{--}&\multicolumn{2}{c}{--}&\multicolumn{2}{c}{--} \\  
 6&$\rho<2$~mas                & y &5  &2423  &&$  -47$&$  12$&$   30$&$  10$&$    0$&$  11$&&$    2$&$  12$&$  -40$&$  10$&$  -25$&$  11$ \\
 7&$\rho<1$~mas                & y &5  &1932  &&$  -47$&$  10$&$   12$&$   9$&$   -10$&$   9$&&$   -2$&$  10$&$  -42$&$   9$&$  -18$&$   9$ \\ 
 8&$\rho<1$~mas                & n &5  &1932  &&$  -15$&$  12$&$    2$&$  12$&$  -14$&$  11$&&$   -6$&$  12$&$    1$&$  12$&$   11$&$  11$ \\[6pt]
9&$\rho<2$~mas, $ G<19$                   & y &5  &1382  &&$  -57$&$  16$&$   33$&$  13$&$   9$&$  14$&&$    3$&$  15$&$  -48$&$  13$&$  -24$&$  14$ \\ 
10&$\rho<2$~mas, $ G<19$                   & n &5  &1382  &&$  -65$&$  20$&$    0$&$  18$&$   22$&$  17$&&$    5$&$  20$&$  -30$&$  18$&$   24$&$  17$ \\[6pt] 
 11a&$\rho<2$~mas,  $\lfloor{10^5\alpha}\rfloor\!\!\mod2=0$ & y &5  &1255  &&$  -19$&$  18$&$   34$&$  15$&$   -10$&$  16$&&$   28$&$  17$&$  -10$&$  15$&$  -22$&$  16$ \\
 11b&$\rho<2$~mas,  $\lfloor{10^5\alpha}\rfloor\!\!\mod2=1$ &  y &5  &1168  &&$ -61$&$  17$&$   33$&$  15$&$   17$&$  15$&&$  -31$&$  17$&$  -64$&$  15$&$  -18$&$  15$ \\[6pt]  
        \noalign{\smallskip} 
        \hline
        \end{tabular}
        \tablefoot{$\rho$ is the angular separation between the optical and radio positions.  $N$ is the number of sources
          used in the fit. $W$ = ``y'' or ``n'' for weighted or unweighted solution.
          The weighted solutions use a non-diagonal weight matrix resulting from the combination of {\gaia} covariances
          and the covariances from the ICRF3-prototype.
        $l_\text{max}$ is the highest degree of the fit from which 
        orientation and glide are extracted for $l=1$. The columns headed $x$, $y$, $z$ give the components of the 
        orientation and glide along the principal axes of the ICRS.
        }
\end{table*}

\subsection{Large-scale systematics}\label{subsec:vshicrf}
In this section we analyse the positional difference between {\gdrtwo} and the {\icrf} in terms of large-scale 
spatial patterns. As in Sect.~\ref{subsec:vshqsos}, the vector field of position differences is decomposed 
using VSH, where in particular the coefficients for degree $l=1$ give the orientation difference of the two
frames and a glide in position. Several fits were made to assess the stability of the orientation rotation 
against various selections of sources. Nominally, {\gdrtwo} has been aligned to the {\icrf} and no significant
orientation difference should remain. However, stating that the two frames have been aligned is not the 
complete story, since the final alignment depends on many details of the fit: weighting scheme, outlier filtering, 
magnitude selection, and the model used for the fit. Furthermore, as explained in Sect.~\ref{subsec:qsosel},
the alignment was made using a slightly different set of {\icrf} sources than currently considered.
As a consequence of these differences, we often find statistically significant non-zero orientation errors in our fits. 
The amplitude of these errors provides the best answer to the question of how precisely the two frames share 
the same axes.

The results of the various fits are summarised in \tabref{table:alignment}. The first fit is similar to the alignment
procedure in the astrometric solution for {\gdrtwo} in that only the three orientation parameters 
(otherwise denoted $\epsilon_x$, $\epsilon_y$, $\epsilon_z$) are fitted without a glide component. Of all the 
fits in the table, this has the overall smallest, statistically most insignificant orientation parameters. It gives a formal
uncertainty in the alignment of about 30~{\muas} per axis. Fit 2, using the same data set, but fitting the 
glide as well, reveals a different picture. The orientation parameters remain negligible, but not as close to zero as in
fit 1, and the glide components have a significant amplitude. The uncertainty is unchanged at about 30~{\muas}.
This is a good illustration of the ambiguity in the alignment when the procedure is not fully implemented. 

In fits 3 to 5, only the orientation parameters are estimated, but with different filtering of the data, with or without 
statistical weighting of the differences. We showed in Sect.~\ref{subsec:icrfminusgaia} that a subset of 
sources has good astrometric quality in both catalogues, but the position differences are not compatible with the 
formal uncertainties. Removing these sources from the fit greatly improves the formal precision of the fit, 
while the orientation parameters are changed by a few tens of {\muas}, which is still only marginally significant. 
More significant changes result from including the glide and higher degrees of VSH (fits 6 to 8), or restricting 
the sample to the brighter subset (fits 9 and 10) with or without weighting. In these fits particularly the orientation error in $x$ and the 
glide in $y$ become significant. Finally, cases 11a and 11b are run on two independent halves of the data to ascertain the sensitivity of the solution to the selection.

Based on these experiments, we state that the axes of the {\gcrftwo} and the {\icrf }  are aligned with an 
uncertainty of 20 to 30~{\muas}, but no precise value can be provided without agreeing on the detailed 
model and numerical procedures for determining the orientation errors.

\section{Other quasars in {\gdrtwo}}\label{sect:otherqsos}

\begin{figure}[ht]
\centering
      \resizebox{0.95\hsize}{!}{ \includegraphics{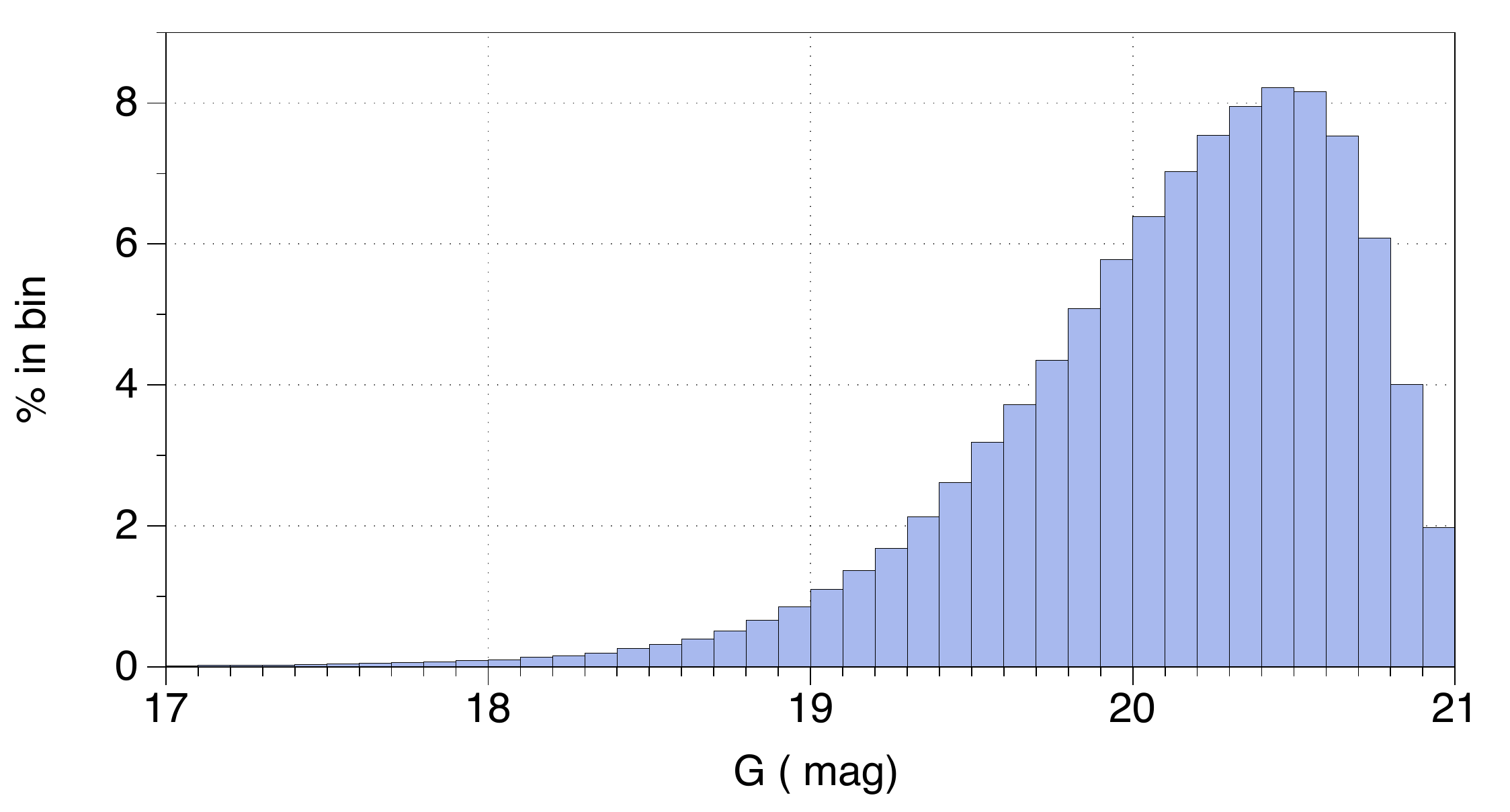}}    
        \caption{Magnitude distribution for $\sim\,$507\,000 {\gaia} sources that are not included in {\gcrftwo,}
        but are tentatively identified as quasars through a cross-match with the MILLIQUAS catalogue.
        \label{mag_G_histo_MQ}}
\end{figure}

The cross-match of {\gdrtwo} with the AllWISE AGN catalogue provided a very clean and homogeneous 
sample of quasars that is suitable for the definition of the {\gcrftwo} and systematic investigation of its properties.
However, other catalogues exist that will enlarge the sample of known or probable quasars in {\gdrtwo}
for other purposes. The Million Quasars Catalogue (MILLIQUAS; \citeads{2015PASA...32...10F})
is a compilation of quasars and AGNs from the literature, including the release of SDSS-DR14 
and AllWISE. We have cross-matched MILLIQUAS%
\footnote{\url{http://quasars.org/milliquas.htm}, version of August 2017, containing 1\,998\,464 entries.}
to {\gdrtwo} using a matching radius of 5~arcsec, but otherwise applying the
same selection criteria as for {\gcrftwo}. This yielded 1\,007\,920 sources with good five-parameter solutions
in {\gdrtwo}, of which 501\,204 are in common with the AllWISE selection in {\gcrftwo}. The magnitude 
distribution of the 506\,716 additional sources is shown in \figref{mag_G_histo_MQ}. With a median 
$G\simeq 20.2$~mag, these sources are typically one magnitude fainter than the AllWISE AGNs in {\gcrftwo},
with positional uncertainties of about 1~mas.

Obviously, the {\gdrtwo} release  contains even more quasars. They can be found by cross-matching with 
other QSO catalogues such as the LQAC (\citeads{2015A&A...583A..75S}) and various VLBI catalogues. 
Ultimately, a self-consistent identification of quasars from photometric 
and astrometric data of \textit{Gaia} will be possible in a future release.

%The position differences with the astrometry given in the MILLIQUAS catalogue are for the large majority of the sources concentrated within a distance (Eq.~\ref{eq:rho}) $ \rho  < 200$ mas, but there are also two small populations with systematic offset of 200 to 300 mas from the origin.

\section{Conclusions}\label{sec:concl}
With {\gdrtwo,} a long-awaited promise of {\gaia} has come to fruition: the publication of the first full-fledged 
optical realisation of the ICRS, that is to say, an optical reference frame built only on extragalactic sources.
Comprising more than half a million extragalactic sources that
are globally positioned on the sky with a median 
uncertainty of 0.4~mas on average, this represents a major step in the history of non-rotating celestial 
reference frames built over the centuries by generations of astronomers. 
The brighter subset with $G < 18$~mag, comprising nearly 30\,000 quasars with 
$\simeq\,$0.12~mas astrometric accuracy, is the best reference frame available today and within relatively 
easy reach for telescopes of moderate size. 

We have summarised the detailed content and mapped the main properties of {\gcrftwo} as functions of 
magnitude and position. 
The quality claims regarding positional accuracy are supported by independent indicators such as the 
distribution of parallaxes or proper motions. Large-scale systematics are characterised by means of
expansions in vector spherical harmonics. Comparison with VLBI positions in a prototype version of the 
forthcoming ICRF3 shows a globally satisfactory agreement at the level of 20 to 30~{\muas}. Several sources with significant radio--optical differences of several mas require further investigation
on a case-by-case basis.

\begin{acknowledgements}

This work has made use of data from the ESA space mission Gaia, 
processed by the Gaia Data Processing and Analysis Consortium
(DPAC).
We are grateful to the developers of TOPCAT (\citeads{2005ASPC..347...29T}) for their software.
Funding for the DPAC has been provided by national institutions, in
particular the institutions participating in the Gaia Multilateral Agreement.
The Gaia mission website is \url{http://www.cosmos.esa.int/gaia}.

The authors are members of the Gaia DPAC.  
This work has been supported by

the European Space Agency in the framework of the Gaia project; 

the Centre National d'Etudes Spatiales (CNES);

the French Centre National de la Recherche Scientifique (CNRS) and the Programme National GRAM of CNRS/INSU with INP and IN2P3 co-funded by CNES ;

the German Aerospace Agency DLR under grants 50QG0501, 50QG1401 50QG0601, 50QG0901, and 50QG1402; 

and 

the Swedish National Space Board.

We gratefully acknowledge the IAU Working Group on ICRF3 for their cooperation during the 
preparation of this work and for their willingness to let us use an unpublished working version of 
the ICRF3 (solution from the GSFC).

\end{acknowledgements}

\bibliographystyle{aa} % style aa.bst
\bibliography{BiblioICRF} % your references Yourfile.bib

\end{document}